\documentclass[11pt,english,onecolumn]{IEEEtran}
\usepackage[T1]{fontenc}
\usepackage[latin9]{inputenc}
\usepackage{babel}
\usepackage{float}
\usepackage{amsmath}
\usepackage{amsthm}
\usepackage{amssymb}
\usepackage{stackrel}
\usepackage{graphicx}
\usepackage{setspace}
\usepackage[unicode=true,pdfusetitle,
 bookmarks=true,bookmarksnumbered=false,bookmarksopen=false,
 breaklinks=false,pdfborder={0 0 1},backref=false,colorlinks=false]
 {hyperref}

\makeatletter

\providecommand{\tabularnewline}{\\}
\floatstyle{ruled}
\newfloat{algorithm}{tbp}{loa}
\providecommand{\algorithmname}{Algorithm}
\floatname{algorithm}{\protect\algorithmname}

\theoremstyle{plain}
\newtheorem{thm}{\protect\theoremname}
\theoremstyle{definition}
\newtheorem{defn}[thm]{\protect\definitionname}
\theoremstyle{remark}
\newtheorem{claim}[thm]{\protect\claimname}
\theoremstyle{plain}
\newtheorem{lem}[thm]{\protect\lemmaname}
\theoremstyle{plain}
\newtheorem{cor}[thm]{\protect\corollaryname}
\theoremstyle{plain}
\newtheorem{prop}[thm]{\protect\propositionname}

\DeclareMathOperator{\Tr}{Tr}
\DeclareMathOperator*{\argmax}{arg\,max}
\DeclareMathOperator*{\argmin}{arg\,min}

\DeclareMathOperator{\Span}{Span}
\DeclareMathOperator{\st}{subject\;to}
\DeclareMathOperator{\p}{\mathbb{P}}
\DeclareMathOperator*{\diag}{diag}
\DeclareMathOperator*{\dsym}{dsym}

\usepackage{cancel}
\renewcommand\[{\begin{equation}}
\renewcommand\]{\end{equation}}
\allowdisplaybreaks

\usepackage{cite}

\usepackage{algorithm,algpseudocode}

\global\long\def\d{\mathrm{d}}

\makeatother

\providecommand{\claimname}{Claim}
\providecommand{\corollaryname}{Corollary}
\providecommand{\definitionname}{Definition}
\providecommand{\lemmaname}{Lemma}
\providecommand{\propositionname}{Proposition}
\providecommand{\theoremname}{Theorem}

\begin{document}
\title{Learning Maximum Margin Channel Decoders}
\author{Amit Tsvieli and Nir Weinberger\\
The Viterbi Faculty of Electrical and Computer Engineering\\
Technion - Israel Institute of Technology\\
Technion City, Haifa 3200004, Israel\\
amit.tsvieli@campus.technion.ac.il, nirwein@technion.ac.il}
\maketitle
\begin{abstract}
The problem of learning a channel decoder is considered for two channel
models. The first model is an additive noise channel whose noise distribution
is unknown and nonparametric. The learner is provided with a fixed
codebook and a dataset comprised of independent samples of the noise,
and is required to select a precision matrix for a nearest neighbor
decoder in terms of the Mahalanobis distance. The second model is
a non-linear channel with additive white Gaussian noise and unknown
channel transformation. The learner is provided with a fixed codebook
and a dataset comprised of independent input-output samples of the
channel, and is required to select a matrix for a nearest neighbor
decoder with a linear kernel. For both models, the objective of maximizing
the margin of the decoder is addressed. Accordingly, for each channel
model, a regularized loss minimization problem with a codebook-related
regularization term and hinge-like loss function is developed, which
is inspired by the support vector machine paradigm for classification
problems. Expected generalization error bounds for the error probability
loss function are provided for both models, under optimal choice of
the regularization parameter. For the additive noise channel, a theoretical
guidance for choosing the training signal-to-noise ratio is proposed
based on this bound. In addition, for the non-linear channel, a high
probability uniform generalization error bound is provided for the
hypothesis class. For each channel, a stochastic sub-gradient descent
algorithm for solving the regularized loss minimization problem is
proposed, and an optimization error bound is stated. The performance
of the proposed algorithms is demonstrated through several examples.
\end{abstract}


\section{Introduction\label{sec:Introduction}}

\thispagestyle{empty}The success of machine learning (ML) based methods
in various domains have spurred a great contemporary interest in the
application of ML algorithms to communication problems \cite{o2017introduction,ye2019deep,nachmani2018deep,sahai2019learning,park2019meta,shlezinger2019viterbinet,liao2021doubly,gruber2017deep,shlezinger2020data,watanabe2021deep,lian2019learned,askri2019dnn,wadayama2019deep,jiang2019mind,kim2018communication}.
However, state-of-the art communication systems heavily rely on expert-based
design, and so obtaining a significantly improved performance with
ML algorithms typically necessitates the use of the most advanced
ML algorithms, most notably, deep neural networks (DNN). In accordance,
this also typically prevents a theoretical justification for the developed
algorithm or theoretical performance guarantees. For instance, despite
a considerable effort from the statistical-learning community, the
generalization properties of DNN are still considered a theoretical
puzzle \cite[Sec. 2]{berner2021modern}.

In order to address the theoretical aspects of learning in communication
systems, we take an alternative route in this paper, and focus on
two basic channel models: 1)\emph{ An additive noise channel model},
whose noise distribution is unknown and does not even known to belong
to a parametric family; and 2)\emph{ A non-linear channel model} with
additive white Gaussian noise, whose deterministic channel transformation
is unknown. Both of these models can be seen as extreme points of
a general model which contains both non-linearity and additive noise
from a general distribution. We focus on these two extreme cases because
they represent prior knowledge on the structure of the communication
problem, which is prevalent in various scenarios. More importantly,
this prior knowledge leads to learning problems which are markedly
different from standard classification problems in ML. Previously,
in \cite{weinberger2021generalization},\emph{ empirical risk minimization}
(ERM) algorithms for the the additive noise channel model were considered.
In this work, we focus on a different learning approach \textendash{}
\emph{maximum margin}. Our choice of channel models and the maximum
margin approach enables us to develop learning algorithms from first
principles, as well as theoretical performance bounds. We thus next
review the problem of learning channel decoders, and then justify
the maximum margin approach.

\subsection{Decoder Learning}

The choice of a proper channel decoder is a key element in the design
of a communication system, and is typically based on rich expert knowledge.
This knowledge is reflected in a statistical model of the channel
operation, which then leads to optimal decoder selection and theoretical
performance guarantees. The approach considered here follows the common
practice of partitioning the communication epoch to a \emph{training
phase} \textendash{} in which no data is transmitted, and the received
signal is used to properly select a decoder, and a \emph{data phase}
\textendash{} in which the decoder is fixed to its chosen value (or
only tracks slight changes in the channel statistics). However, our
approach here deviates from the common one to the training phase.
For models with unknown noise distribution a parametric model for
the noise distribution is typically assumed, the training phase is
used to estimate the parameter, and an optimal decoder matched to
the estimated parameter is used in the data phase. In various scenarios
of interest \textendash{} such as massive multiple-input multiple-output
(MIMO) systems \cite{marzetta2015massive} or in ultra low-latency
\cite{sybis2016channel} communication \textendash{} parameter estimation,
or even the parametric modeling itself, may be inaccurate. Thus, parameter
estimation during the training phase is excluded. In the same spirit,
for the non-linear channel, estimation of the transformation value
for each codeword may suffice. However, we target the regime in which
the codebook is too large with respect to (w.r.t.) the training phase,
i.e., there are not enough training samples for each codeword in order
to estimate its channel-transformed value. This is similar to the
common practice in linear frequency-selective channels, in which an
equalizer is learned to roughly invert the channel (while accounting
for noise enhancement) \cite[Ch. 8 and 9]{barry2012digital}.

These settings naturally motivate the use of ML methods, as they are
typically \emph{distribution-free}, that is, do not make any assumptions
on the data statistics. In ML methods, assumptions are made, instead,
on the structure of the hypothesis class. We thus follow here the
typical setting in ML, in which the hypothesis class is more restricted
than the true data. The choice of hypothesis class results an \emph{inductive
bia}s and its use is then justified by its low variance, compared
to richer classes of decoders. For example, a linear binary classifier
can be learned even when the optimal separator between the two classes
is not an hyperplane \cite[Ch. 9]{shalev2014understanding}, because
it is assumed to have a proper bias-variance tradeoff. The learning
process we propose, however, does not ignore the channel model, and
is strongly based on both the given codebook and the structure of
the channel.

For the additive noise channel, we consider the class of nearest neighbor
(NN) decoders, w.r.t. the Mahalanobis distance, that are parameterized
by a precision matrix. A decoder from this hypothesis class is optimal
in the sense of minimizing the error probability only if the additive
noise is Gaussian. The learner is provided with a fixed codebook and
a dataset comprised of independent samples of the noise, and is required
to select a precision matrix for a NN decoder in terms of the Mahalanobis
distance. For the non-linear channel, we consider the class of NN
decoders with a \emph{linear kernel}, that is, decoders parameterized
by a matrix. Similarly, here, a decoder based on linear kernel is
optimal only if the channel transformation itself is linear. The learner
is provided with a fixed codebook and a dataset comprised of independent
input-output samples of the channel, and is required to select a matrix
for a NN decoder with a linear kernel.

\subsection{Maximum Margin Learning}

At first glance, the decoder learning problem is akin to multiclass
classification problem, in which the class represents the index of
the codeword. However these problems are different due to the prior
knowledge about the channel model and the codebook. Therefore, \cite{weinberger2021generalization}
considered an ERM approach to the decoder learning problem for the
additive noise channel model. However, in general, the ERM might be
difficult to find, and more importantly, the ERM approach does not
fully capture the structure of the codebook. For example, there could
be many ERM rules which achieve zero empirical error. One of the consequences
of this, is that the generalization bounds derived in \cite{weinberger2021generalization}
scale as $O(m/\sqrt{n})$, where $m$ is the number of codewords in
the codebook, and $n$ is the number of noise samples provided to
the learner. Therefore, in this paper we take a different approach,
and derive learning rules which attempt to maximize the \emph{margin}
of the decoder.

The maximum margin approach is common in ML, and naturally matches
the communication problem. Indeed, a NN decoder partitions the output
space $\mathbb{R}^{d_{y}}$ into $m$ decision regions, where the
boundaries of each such decision region (also called a Voronoi region)
are $d$-dimensional hyperplanes. In the additive noise channel, this
NN decoding rule maximizes the minimal Mahalanobis distance between
each pair of codewords, w.r.t. the noise covariance matrix. In the
non-linear Gaussian channel, the NN decoding rule maximizes the minimal
Euclidean distance $d_{\text{min}}$ between each pair of transformed
codewords, w.r.t. the linear kernel. As is well known, for Gaussian
channels at high signal-to-noise ratio (SNR), this minimal distance
is the dominant parameter in determining the error probability, or
more specifically, its decay rate w.r.t. the SNR. Specifically, a
union bound (over all pairs of codewords), and a Bhattacharyya-based
pairwise error bound, leads to the upper bound $(m-1)\cdot\exp(-d_{\text{min}}^{2}/8\sigma^{2})$
on the error probability \cite[Sec. 5.2]{viterbi2013principles} ($\sigma^{2}$
is the noise variance). Naturally, such a bound does not necessarily
hold for non-Gaussian noise distributions. However, we adopt this
criterion here for general noise distributions since in the lack of
any other knowledge, maximizing the minimum distance appears to be
a plausible criterion for the quality of the decoder. This approach
is again common in ML. In classification, support vector machines
(SVM) learn linear separating hyperplanes which aim to maximize the
\emph{margin} between the classes, even if the true probabilistic
law results a non-linear decision boundary between the classes.

The resulting learned decoder is codebook-dependent, in the sense
that it is chosen to maximize the \emph{margin}, or \emph{minimal
distance}, between the codewords. For the non-linear channel, ideally
the margin would correspond to the non-linearly transformed codewords,
but as we make no assumptions on this non-linearity, the best that
can be hoped for is to maximize the margin of a linear transformation
of the codewords \textendash{} as follows from the perspective of
the assumed class of decoders.

In \emph{binary} classification, the maximum margin problem is solved
by transforming it to an equivalent regularized loss minimization
(RLM) problem with a surrogate convex loss function, known as the
\emph{hinge loss} \cite[Sec. 15.2]{shalev2014understanding}. However,
in \emph{multiclass} classification, the transformation from maximum
margin to RLM is not immediate and requires heuristics for choosing
the regularization function \cite[Sec. 7]{crammer2002learnability},
reductions to binary classification \cite[Sec. 17.1]{shalev2014understanding},
or other approximations. The maximum margin decoder learning problem
studied here naturally resembles multiclass classification (when there
are more than two codewords in the codebook), and accordingly, it
also necessitates several steps of approximation in order to transform
it to an RLM problem. Nonetheless, following these steps, the result
is an RLM problem that uses a convex surrogate loss function that
is specifically tailored to the decoder learning problem.

We conclude by further illuminating the difference between the decoder
learning problem and standard multiclass classification. As we shall
see, for the additive noise channel model, the dataset for our learning
algorithm is synthesized by adding $n$ measured noise samples to
each of the $m$ codewords in the codebook. This results a dataset
of size $mn$, but one which has specific structure. In this synthesis
of the dataset, the scaling of the codebook (henceforth referred to
as ``training SNR'') is a design parameter that can be chosen by
the learner. One consequence of this possibility is that unlike standard
classification problems, in which the margin prevailing in the dataset
determines the sample complexity of the problem \cite[Th. 15.4]{shalev2014understanding},
the margin in the decoder-learning problem for the additive noise
channel is a parameter to be tuned. This can be used in order to achieve
the best generalization possible.\footnote{See also a discussion in \cite[Sec. II]{weinberger2021generalization}.}
Due to these differences, standard SVM learning algorithms cannot
be applied, and we develop learning algorithms specifically designed
for the decoder learning problem.

\subsection{Contributions\label{subsec:Contributions}}

Our contributions and the outline of the rest of the paper are as
follows. In Sec. \ref{sec:Problem-Formulation} we establish notation
conventions, formulate the learning problem for the general channel,
and then specify it for the two discussed extreme special cases. In
Sec. \ref{sec:Maximum-Margin-by}, we formulate a maximum margin optimization
problem for the decoder learning problem, and relax it in several
principled steps to obtain a tractable optimization problem in the
form of an RLM rule. It should be stressed, however, that the resulting
optimization problem is directly designed for the decoder learning
problem, and does not involve a reduction or a modification of a multiclass
classification problem. In Sec. \ref{sec:Generalization-Error-Bounds},
we prove that the expected error probability of the learned decoder
in the additive noise channel model is bounded by $m$ times the expected
empirical hinge loss of the decoder plus an estimation error term
of $\tilde{O}(m/(\lambda n))$, where $m$ is the number of codewords
in the codebook, $\lambda$ is the regularization parameter, and $n$
is the number of available samples. Then, we constrain the precision
matrix of the learned decoder to a properly chosen set and optimize
the regularization parameter. The resulting generalization bound is
$m$ times the optimal hinge loss plus an estimation error term of
$\tilde{O}(m/\sqrt{n}))$. The proof of this generalization bound
is based on establishing an on-average-replace-one-stability property.
We then use this generalization bound to offer a theoretical guideline
for choosing the training SNR. We then prove analogous expected generalization
bounds for the non-linear channel. In addition, we prove a $O(1/\sqrt{n})$
uniform high-probability generalization error bound for the hinge-type
loss function, for any decoder from the chosen decoder class for the
non-linear channel. 

The RLM mentioned above is tractable, yet still suffers from large
complexity, mainly due to a $O(n^{2})$ dependence on the number of
samples $n$, similarly to the computational cost associated with
SVM problems \cite{bottou2007support}. For classification SVM, this
problem was addressed in \cite{pegasos}, which proposed a stochastic
sub-gradient descent algorithm for solving the RLM problem called
PEGASOS (primal estimated sub-gradient solver for SVM). In Sec. \ref{sec:SGD},
we develop an algorithm in that spirit for the decoder learning problem,
and prove that $\tilde{O}(1/\epsilon)$ iterations suffice in order
to obtain a solution of accuracy $\epsilon$. We stress that this
bound does not depend on the problem's dimension or the number of
samples, which can be large in many communication problems. On top
of that, each iteration of the algorithm requires low computational
power, and thus suitable to learning on low-complexity devices. In
Sec. \ref{sec:Examples}, we exemplify the operation of the algorithm
through simulation for several noise distributions, channel transformations,
and codebooks. In Sec. \ref{sec:Future-Research}, we summarize the
paper and propose several directions for further research. All the
proofs are deferred to the appendixes.

\subsection{Other Related Work\label{subsec:Related-Work}}

As said, the recent success of ML algorithms (most notably DNNs) in
various problem domains \cite{Goodfellow-et-al-2016} has spurred
interest in applying machine-learning algorithms as a part of a communication
system \cite{o2017introduction,wang2017deep}. These applications
can be roughly divided to several approaches, and here we briefly
mention a sample of them. One approach is to use a machine-learning
algorithm as a replacement to one or more components of the system.
Examples include \cite{caciularu2020unsupervised} for channel equalization,
\cite{kim2018communication} for channel encoder and decoder, \cite{gruber2017deep,jiang2019mind,liao2021doubly}
for channel decoder, and \cite{li2021knowledge,ma2021model,zhao2021deep}
for channel estimation. A second approach is to modify an existing
algorithm by incorporating DNNs. In \cite{nachmani2018deep} DNNs
are used in the belief propagation algorithm, in \cite{shlezinger2019viterbinet}
for the Viterbi algorithm, in \cite{carpi2019reinforcement} for iterative
decoding algorithms, such as bit-flipping and residual belief propagation.
Among the classical ML algorithms, SVMs gained high popularity and
enjoy strong theoretical guarantees. A SVM based receiver that combines
the pilot-based channel estimation, data demodulation and decoding
processes in one joint operation was proposed in \cite{akin2020joint}.
They considered first-order Gauss-Markov fading process and additive
Gaussian noise with $N$ dimensional encoding vector. The system was
composed from $N$ classifiers, one per bit, or $2^{N}$ classifiers
in the one-vs.-one or one-vs.-rest techniques. SVM regression (SVR)
was used in \cite{charrada2017analyzing,sanchez2004svm} for channel
estimation. We note in passing, that similar theoretical studies have
been performed for the source coding problem of learning vector quantizers
\cite{lee2019learning,levrard2013fast,antos2005improved,antos2005individual,linder2000training}.

Channel decoding in non-linear channels was studied earlier in \cite{chen1990adaptive,kaleh1994joint}.
The non-linear channel can practically model the nonlinear effects
in wireless communication systems, the Kerr non-linearity in optical
fibers, and the saturation non-linearity of amplifies \cite{xenoulis2012new,roberts2016convex}.
Information theoretic limits on the capacity of non-linear fiber-optic
channels were studied in \cite{essiambre2010capacity,mecozzi2012nonlinear,mitra2001nonlinear},
and the shaping gain in these channels was studied in \cite{dar2014shaping}.

\section{Problem Formulation\label{sec:Problem-Formulation}}

\subsection{Notation Conventions\label{subsec:Notation-Conventions}}

Random variables or vectors are denoted by capital letters and specific
values they take are denoted by the corresponding lower case letters.
The expectation operator is denoted by $\mathbb{E}_{\mu}[\cdot]$
where $\mu$ is the underlying probability measure, which is omitted
if understood from context. The tail function of the standard normal
distribution is denoted by $Q(x)$. The indicator of an event $\mathcal{A}$
is denoted by $\mathbb{I}\{\mathcal{A}\}$. The probability simplex
is denoted by $\Delta^{d}\triangleq\{v\in\mathbb{R}_{+}^{d}:\sum_{i\in\left[d\right]}v_{i}=1\}$.
All vectors are taken as column vectors. The standard Euclidean norm
for $x\in\mathbb{R}^{d}$ is denoted by $\lVert x\rVert$ and the
inner product by either $\langle x_{1},x_{2}\rangle$ or $x_{1}^{T}x_{2}$,
interchangeably. The Frobenius norm for a matrix $A\in\mathbb{R}^{d\times d}$
is denoted by $\Vert A\Vert_{F}$. The operator norm for a matrix
$A\in\mathbb{R}^{d_{y}\times d_{x}}$ is denoted by $\Vert A\Vert_{\text{op}}\triangleq\sup_{v\in\mathbb{R}^{d_{x}}\colon\|v\|\le1}\|Av\|$.
The positive semidefinite (PSD) cone is denoted by $\mathbb{S}_{+}$.
The minimal (resp. maximal) eigenvalue of a symmetric matrix $A$
is denoted by $\lambda_{\text{min}}(A)$ (resp. $\lambda_{\text{max}}(A)$)
. For $n\in\mathbb{N}^{+}$, the set $\{1,2,\dots,n\}$ is denoted
by $[n]$. Standard Bachmann-Landau asymptotic notation will be used,
where specifically, $\tilde{O}(\cdot)$ is such that the logarithmic
factors are hidden, namely, $f(n)\in\tilde{O}(h(n))\iff\exists k:f(n)\in O(h(n)\log^{k}(h(n)))$.

\subsection{Channel Models\label{subsec:Channel-Models}}

Consider the problem of communication over a general channel 
\begin{equation}
Y=f(X)+Z,\label{eq: general model}
\end{equation}
where, $Y\in\mathbb{R}^{d_{y}}$ is the channel output, $X\in\mathbb{R}^{d_{x}}$
is a codeword that is chosen from a fixed given codebook $C=\{x_{j}\}_{j\in\left[m\right]}$
with a uniform probability, $f:\mathbb{R}^{d_{x}}\rightarrow\mathbb{R}^{d_{y}}$
is a deterministic channel transformation, and $Z\in\mathbb{R}^{d_{y}}$
is a noise statistically independent of the input $X$. The unknown
channel operation (\ref{eq: general model}), is denoted by $\mu$
and is thus comprised of the deterministic $f$ and the distribution
of $Z$, both which are unknown to the designer of the decoder. It
is further assumed that the decoder is chosen from the class of NN
decoders with Mahalanobis distance and linear kernel given by

\begin{equation}
\hat{j}\left(y\right)\in\argmin_{j\in\left[m\right]}\left\Vert Hx_{j}-y\right\Vert _{S}\triangleq\argmin_{j\in\left[m\right]}\sqrt{\left(Hx_{j}-y\right)^{T}S\left(Hx_{j}-y\right)},\label{eq: class of decoders}
\end{equation}
where $S\in\mathbb{S}_{+}^{d_{x}}$ is a precision matrix (the inverse
of the covariance matrix) and $H\in\mathbb{R}^{d_{y}\times d_{x}}$
is a linear kernel. As well known, this class is optimal if the channel
transformation is linear $f(x)=Hx$ and the noise is Gaussian with
inverse covariance matrix $S$, but it is not assumed here that the
channel model obeys this model.

For a decoder $(H,S)$, the expected error probability conditioned
that the $j$th codeword was transmitted is given by
\[
\boldsymbol{p}_{\mu}\left(H,S\mid j\right)=\mathbb{E}_{\mu}\left[\mathbb{I}\left\{ \min_{j^{'}\in\left[m\right]\backslash\left\{ j\right\} }\left\Vert Hx_{j'}-Y\right\Vert _{S}<\left\Vert Hx_{j}-Y\right\Vert _{S}\right\} \;\middle|\;Y=f\left(x_{j}\right)+Z\right],
\]
and the expected error probability, averaged over all codewords, is
given by
\[
\boldsymbol{p}_{\mu}\left(H,S\right)=\frac{1}{m}\sum_{j\in\left[m\right]}\boldsymbol{p}_{\mu}\left(H,S|j\right).
\]
A learner, which does not know $\mu$, is provided with $n$ input-output
samples $\boldsymbol{D}=\{X_{i},Y_{i}\}_{i\in[n]}$ where $X_{i}\in C$,
and $Y_{i}$ is the corresponding channel output (\ref{eq: general model})
for input $X_{i}$, $i\in[n]$. Based on this dataset, and the given
codebook $C$, the learner is required to find $(H,S)$ which minimize
the expected error probability. A common learning approach is ERM,
which aims to minimize the empirical average error probability of
the noise samples, given by 
\[
\boldsymbol{p}_{\boldsymbol{D}}\left(H,S\right)=\frac{1}{n}\sum_{i\in[n]}\mathbb{I}\left\{ \min_{x'\in C\backslash\{x_{i}\}}\left\Vert Hx'-y_{i}\right\Vert _{S}<\left\Vert Hx_{i}-y_{i}\right\Vert _{S}\right\} .
\]

In this paper, we focus on two structured extremes of this model,
which both result decoder classes which are substantially different
from multiclass classifiers, and are also relevant in practical communication
scenarios in which further prior knowledge on the model exists. The
first one is an additive noise channel model that assumes that $f$
is the identity transformation, and the second one is a non-linear
channel model that assumes that the noise is Gaussian. These models
best illuminate the affect of the structure of the communication channel
model on the decoder learning problem. The general model (\ref{eq: general model})
is not more challenging than these two extreme models, and bounds
and algorithms can be easily obtained for the general model by combining
the results obtained in this paper for the two extreme models. 

\subsubsection{The Additive Noise Channel\label{subsec:The-Additive-Model}}

In the first specified channel model that we consider, the channel
transformation is assumed to be the identity function, and so (\ref{eq: general model})
becomes $Y=X+Z$. We thus henceforth denote for this model $d\triangleq d_{x}=d_{y}$.
For this channel model, we restrict the general class (\ref{eq: class of decoders})
to have $H=I$, and this results a class of NN decoders with Mahalanobis
distance, given by
\begin{equation}
\hat{j}\left(y\right)\in\argmin_{j\in\left[m\right]}\left\Vert x_{j}-y\right\Vert _{S}\triangleq\argmin_{j\in\left[m\right]}\sqrt{\left(x_{j}-y\right)^{T}S\left(x_{j}-y\right)},\label{eq:NN decoder family}
\end{equation}
parameterized by a precision matrix $S\in\mathbb{S}_{+}^{d}$. In
what follows, we will identify a decoder from this class by its precision
matrix $S$. Since the channel law only depends on the noise distribution,
learning the noise distribution can be made with arbitrary inputs,
which we take for simplicity to be $X\equiv0$. Thus, we assume that
the learner is equipped with $n$ noise samples $\boldsymbol{Z}=\{Z_{i}\}_{i\in\left[n\right]}$
drawn independent and identically distributed (i.i.d.) from the distribution
of $Z$. The learner synthesize a new dataset from these noise samples
comprised of $mn$ labeled samples, in which each of the scaled codewords
in the codebook is perturbed by one of the noise samples $\{Z_{i}\}_{i\in[n]}$,
namely 
\[
\boldsymbol{D}\left(\boldsymbol{Z}\right)\triangleq\left\{ Y_{k},l_{k}\right\} _{k=1}^{mn}=\bigcup_{j\in\left[m\right]}\boldsymbol{D}_{j}\left(\boldsymbol{Z}\right),
\]
where $\boldsymbol{D}_{j}(\boldsymbol{Z})\triangleq\{\Gamma\cdot x_{j}+Z_{i},j\}_{i\in\left[n\right]}$,
and where $\Gamma>0$ is a scaling constant which determines the \emph{training
SNR. }We note that this dataset has $nm$ points, but it is based
only on $n$ random noise samples. For the sake of brevity, we will
omit from now the explicit dependence of $\boldsymbol{D}$ in $\boldsymbol{Z}$.
Henceforth, we refer to this model as the \emph{additive noise channel
model. }A learner is then required to find $S$ which minimizes the
expected error probability based on $\boldsymbol{D}$.

\subsubsection{The Non-linear White Gaussian Noise Channel\label{subsec:The-Non-linear-Model}}

In the second specified channel model that we consider, the noise
is assumed to be white and Gaussian, and so (\ref{eq: general model})
becomes $Y=f(X)+W$, where $W\in\mathbb{R}^{d_{y}}$ is a white Gaussian
noise, statistically independent of the input $X$. For this channel
model, we restrict the general class (\ref{eq: class of decoders})
to have $S=I$, and this results a class of NN decoders, given by
\begin{equation}
\hat{j}(y)\in\argmin_{j\in[m]}\left\Vert y-Hx_{j}\right\Vert ,\label{eq:hypothesis class}
\end{equation}
parameterized by a linear-kernel matrix $H\in\mathcal{H}\triangleq\mathbb{R}^{d_{y}\times d_{x}}$.
In what follows, we will identify a decoder from this class by its
matrix $H$. We assume that the learner is equipped with $n$ input-output
samples $\boldsymbol{D}=\{J_{i},Y_{i}\}_{i\in[n]}$, where $J_{i}$
is chosen i.i.d. and uniformly over $[m]$, and $Y_{i}$ is the corresponding
channel output for input $X_{J_{i}}$. Henceforth, we refer to this
model as the \emph{non-linear channel model.} A learner is then required
to find $H$ which minimizes the expected error probability based
on $\boldsymbol{D}$.

The matrix $H$ serves as a linear approximation of $f(\cdot)$, which
is non-linear in general. Note that knowing $f(x_{j})$ for any $x_{j}\in C$
suffices for optimal decoding, but we target the regime in which the
codebook is large w.r.t. the dataset, and so there are not enough
training samples for each codeword $x_{j}$ in order to accurately
estimate $f(x_{j})$. In addition, although this class does not have
optimality guarantees for a non-linear channel, it is possible to
enrich its expressive power by first mapping the codewords into a
high dimensional feature space \cite[Ch. 16]{shalev2014understanding}.

\section{Maximum Margin by RLM\label{sec:Maximum-Margin-by}}

In this section, we develop RLM problems for the decoder learning
problem that are strongly connected to the original maximum margin
objective. In Sec. \ref{subsec:The-Additive-RLM} we present the corresponding
RLM problem for the additive noise channel model. We then state the
steps required for its development, with proofs appearing in Appendix
\ref{subsec:Additive proofs}. Then, in Sec. \ref{subsec:The-Non-linear-RLM},
we present the RLM problem for the non-linear channel model. The steps
required for its development and the corresponding proofs both appear
in Appendix \ref{subsec:Non linear Dev}.

For brevity, we will henceforth denote $\delta_{pq}\triangleq x_{p}-x_{q}$
as the difference between a pair of codewords in the codebook. For
notational convenience, we will henceforth also use a single index
in the set $[\frac{1}{2}m(m-1)]$ to specify a pair of codewords (instead
of double indices $\{(p,q)\}_{1\le p<q\le m}$ ). In addition, we
define the following property of a partition of the codeword-pairs.
\begin{defn}
\label{def:proper partition def}A partition $P=\bigcup_{j=1}^{d+1}P_{j}$
of $\{(p,q)\}_{1\le p<q\le m}$ is called \emph{proper} if $\Span[\{\delta_{p_{j},q_{j}}\}_{j=1}^{d+1}]=\mathbb{R}^{d}$
for any set of representatives $\{\delta_{p_{j},q_{j}}\}_{j=1}^{d+1}$,
such that $(p_{j},q_{j})\in P_{j}$ for all $j\in[d+1]$.
\end{defn}
The RLM problems that we will next present will use proper partitions
for their regularization terms. We assume, without loss of generality
(w.l.o.g.),\footnote{If it is not the case we can project the codebook and samples to a
lower dimension spanned by the codebook.} that $\Span\{\delta_{pq}\}_{1\le p<q\le m}=\mathbb{R}^{d}$. Then,
a simple way of finding a proper partition is by first finding a basis
of $\mathbb{R}^{d}$: $\{\delta_{p_{j},q_{j}}\}_{j=1}^{d}\subset\{\delta_{pq}\}_{1\le p<q\le m}$
and then setting $P_{j}=\{\delta_{p_{j},q_{j}}\}$ for all $j\in[d]$
and $P_{d+1}=\{\delta_{pq}\}_{1\le p<q\le m}\setminus\{\delta_{p_{j},q_{j}}\}_{j=1}^{d}$.
Nonetheless, the RLM problem (and subsequent results in what follows)
holds for any arbitrary proper partition.

\subsection{The Additive Noise Channel RLM\label{subsec:The-Additive-RLM}}

The RLM problem for the additive noise channel model requires a few
additional definitions, as follows. For a given ordered pair of codeword
indices $(p,q)$, and a sample $y_{i}\in\boldsymbol{D}_{p}\bigcup\boldsymbol{D}_{p}$,
we denote the following transformation of the sample and codewords
$a_{pqi}\triangleq(-1)^{\mathbb{I}(i\in\boldsymbol{D}_{q})}(y_{i}-\frac{1}{2}(x_{p}+x_{q}))$.
Then, we denote by $\mathring{\ell}^{\text{hinge}}(S,p,q,i)\triangleq\max\{0,1-a_{pqi}^{T}S\delta_{pq}\}$,
a hinge-type surrogate loss function, and by

\begin{equation}
\mathring{L}_{\boldsymbol{D}}^{\text{hinge}}\left(S\right)\triangleq\frac{2}{m\left(m-1\right)}\sum_{1\le p<q<m}\frac{1}{2n}\sum_{i\in\boldsymbol{D}_{p}\bigcup\boldsymbol{D}_{q}}\mathring{\ell}^{\text{hinge}}\left(S,p,q,i\right)\label{eq:hinge as margin}
\end{equation}
the average \emph{hinge loss} of the induced binary linear classifiers
$\{S\delta_{pq}\}$, over the transformed noise samples $\{a_{pqi}\}$.
Let $P=\bigcup_{i=1}^{d+1}P_{i}$ be a proper partition according
to Definition \ref{def:proper partition def}, let $\{\eta_{i}\}_{i\in\left[d+1\right]}$
be positive parameters which sum to $1$, and let $\lambda>0$ be
a regularization parameter that controls the tradeoff between the
loss and the regularization. We show in this section that the RLM
learning rule 
\begin{equation}
\min_{S\in\mathbb{S}_{+}}\mathring{L}_{\boldsymbol{D}}^{\text{hinge}}\left(S\right)+\lambda\sum_{i=1}^{d+1}\eta_{i}\max_{j\in P_{i}}\left\Vert S\delta_{j}\right\Vert ^{2},\label{eq:stable RLM}
\end{equation}
learns a precision matrix which maximizes a lower bound on the margin
among all codeword-pairs. 

At this point, we may compare the surrogate loss function obtained
here to a \emph{different} surrogate hinge-type upper bound for the
average error probability loss over $\boldsymbol{Z}$ proposed in
\cite{weinberger2021generalization}. There, the hinge-type surrogate
loss function was defined as
\begin{equation}
\bar{L}_{\boldsymbol{Z}}^{\text{hinge}}\left(S\right)\triangleq\frac{1}{n}\sum_{i\in\left[n\right]}\frac{1}{m}\sum_{p\in\left[m\right]}\max\left\{ 0,1-\min_{q\in\left[m\right]\backslash\left\{ p\right\} }\left\Vert x_{p}+z_{i}-x_{q}\right\Vert _{S}-\left\Vert z_{i}\right\Vert _{S}\right\} .\label{eq:hinge as upper bound}
\end{equation}
The hinge loss functions (\ref{eq:hinge as margin}) and (\ref{eq:hinge as upper bound})
are rather different, and are the result of different types of relaxations.
We review these differences in light of two possible ways in which
the optimization problem of SVM for binary classification is typically
interpreted. The first interpretation is that the SVM objective function
is a convex and continuous upper bound to the non-convex and discontinuous
zero-one loss function. The regularization term can be interpreted
as a standard Tikhonov regularization term. The motivation of changing
the zero-one loss to a hinge-type loss function is that the resulting
ERM problem can be more efficiently solved compared to the ERM problem
for the zero-one loss. With this interpretation, the choice of hinge
loss function is arbitrary, and, in principle, any convex upper bound
on the zero-one loss is also appropriate. The second interpretation
of the optimization problem of SVM is in terms of margin maximization.
For inseparable datasets, the objective function of SVM is interpreted
as a balance between increasing the margin and increasing the classification
errors. As well known, for binary classification, both interpretations
lead to exactly the same optimization problem \cite[Ch. 15]{shalev2014understanding}.
This is, however, not the case for the decoder learning problem. The
hinge loss function in \cite{weinberger2021generalization} follows
the first interpretation for minimizing the error probability of the
channel decoder. This approach leads to an ERM problem with the hinge
loss over the noise samples, and an implicit regularization in the
form of maximal eigenvalue constraint. However, this hinge loss is
not directly related to the margin. In this paper, we follow the second
interpretation for maximizing the margin induced by the channel decoder.
This approach leads to an RLM problem with the hinge loss over the
transformed noise samples and a codebook related regularization in
(\ref{eq:stable RLM}). Hence, unlike SVM, the two approaches lead
to different optimization problems for the channel decoding problem,
where the main source of difference is in fact due to the lower bound
taken in step 2. We argue that for the channel decoding problem the
second approach is better since error probability is strongly related
to margin, as discussed above.

The development of the optimization problem (\ref{eq:stable RLM})
will be made in several steps, which we next describe in detail.

\paragraph*{Step 1 \textendash{} maximization of the minimum margin}

We begin with the assumption that the dataset $\boldsymbol{D}$ is
\emph{separable}, i.e., there exists a precision matrix $S$ that
achieves zero loss over $\boldsymbol{D}$. While this is a rather
strong assumption for a general dataset, it is only a staring point
which will be relaxed in the following steps. It is also analogous
to the linear separability assumption made for hard SVM, which is
also used as a starting point to soft SVM \cite[Ch. 15]{shalev2014understanding}.
In fact, in our setting, separability can always be achieved by setting
the training-SNR parameter $\Gamma$ to be large enough. The \emph{margin}
of a hyperplane w.r.t. a dataset is defined to be the minimal distance
between a point in the dataset and the hyperplane \cite[Ch. 15]{shalev2014understanding}.
The learner's goal is to find a precision matrix $S$ that maximizes
the minimum margin, over all codeword-pairs $x_{p},x_{q}\in C$, according
to the Mahalanobis distance w.r.t. $S$. This learning problem is
formulated as the following margin-maximization problem, given as
follows:
\begin{claim}
\label{maximum margin form claim}The maximum margin induced by a
Mahalanobis distance NN decoder with precision matrix $S$ is
\begin{equation}
\max_{S\in\mathbb{S}_{+}}\min_{1\le p<q\le m}\min_{i\in\boldsymbol{D}_{p}\cup\boldsymbol{D}_{q}}\frac{a_{pqi}^{T}S\delta_{pq}}{\left\Vert S\delta_{pq}\right\Vert }.\label{eq:margin objective}
\end{equation}
\end{claim}

\paragraph*{Step 2 \textendash{} a convex lower bound}

The previous problem is not necessarily convex, and therefore we proceed
to maximize the following convex lower bound on its value.
\begin{claim}
\label{lower bound claim}The problem
\begin{equation}
\begin{array}{cc}
\max_{S\in\mathbb{S}_{+}}\min_{1\le p<q\le m}\min_{i\in\boldsymbol{D}_{p}\cup\boldsymbol{D}_{q}} & a_{pqi}^{T}S\delta_{pq}\\
\st & \max_{1\le p<q\le m}\left\Vert S\delta_{pq}\right\Vert \le1
\end{array}\label{eq:lower bound margin objective}
\end{equation}
is a convex optimization problem, whose value is a lower bound on
the value of (\ref{eq:margin objective}).
\end{claim}

\paragraph*{Step 3 \textendash{} minimum norm formulation}

With the prospect removal of the separability assumption, we next
derive a minimum norm optimization problem, so that every solution
to it is a solution to (\ref{eq:lower bound margin objective}). This
step is analogous to the equivalent formulation of hard SVM as a quadratic
optimization problem \cite[Ch. 15.1]{shalev2014understanding}.
\begin{lem}
\label{equiv min norm lemma}Every solution to the following minimum
norm problem

\begin{equation}
\begin{array}{cc}
\min_{S\in\mathbb{S}_{+}}\max_{1\le p<q\le m} & \left\Vert S\delta_{pq}\right\Vert ^{2}\\
\st & \min_{1\le p<q\le m}\min_{i\in\boldsymbol{D}_{p}\cup\boldsymbol{D}_{q}}a_{pqi}^{T}S\delta_{pq}\ge1
\end{array},\label{eq:minimum norm form}
\end{equation}
is a solution to (\ref{eq:lower bound margin objective}).
\end{lem}

\paragraph*{Step 4 \textendash{} relaxation of the separability assumption}

Next, we introduce slack variables in order to relax the assumption
that the dataset $\boldsymbol{D}$ is separable. This step is analogous
to the relaxation made for soft SVM \cite[Ch. 15.2]{shalev2014understanding}.
A short derivation in Appendix \ref{subsec:Additive proofs} results
the RLM problem 
\begin{equation}
\min_{S\in\mathbb{S}_{+}}\mathring{L}_{\boldsymbol{D}}^{\text{hinge}}\left(S\right)+\lambda\cdot\max_{1\le p<q\le m}\left\Vert S\delta_{pq}\right\Vert ^{2}.\label{eq:not stable RLM}
\end{equation}

\paragraph*{Step 5 \textendash{} inducing stability by a generalization of the
regularization}

Some of the generalization bounds for SVM are based on the \emph{stability}
of its learning rule. However, the problem (\ref{eq:not stable RLM})
is, in general, not stable, due to the fact that the regularization
term $\max_{1\le p<q\le m}\|S\delta_{pq}\|^{2}$ is indifferent to
changes in directions orthogonal to the maximizer $\delta_{pq}$.
Nonetheless, we next assume, w.l.o.g., that $\Span\{\delta_{pq}\}_{1\le p<q\le m}=\mathbb{R}^{d}$,
and slightly modify the learning rule to a stable one. The final RLM
rule for finding a maximum minimum margin decoder is defined for given
positive parameters $\{\eta_{i}\}_{i\in\left[d+1\right]}$ which satisfy
$\sum_{i=1}^{d+1}\eta_{i}=1$, and a proper partition $\{P_{j}\}_{j\in\left[d+1\right]}$,
as
\begin{equation}
A\left(\boldsymbol{D}\right)=\argmin_{S\in\mathbb{S}_{+}}\mathring{L}_{\boldsymbol{D}}^{\text{hinge}}\left(S\right)+\lambda\sum_{i=1}^{d+1}\eta_{i}\max_{j\in P_{i}}\left\Vert S\delta_{j}\right\Vert ^{2}.\label{eq:stable RLM rule}
\end{equation}
The stability of this learning rule will be used in Sec. \ref{sec:Generalization-Error-Bounds}
to derive generalization bounds. 

\subsection{The Non-linear Gaussian Noise Channel RLM\label{subsec:The-Non-linear-RLM}}

The RLM problem for the non-linear Gaussian noise channel uses the
following definitions. Let $K\in\mathcal{K}\triangleq\mathbb{S}_{+}^{d_{x}}$
be an auxiliary matrix variable. We denote by
\begin{equation}
\mathring{\ell}^{\text{hinge}}(H,K,i)\triangleq\frac{1}{m-1}\sum_{j'\in\left[m\right]\backslash\left\{ j_{i}\right\} }\max\left\{ 0,1-\left[y_{i}^{T}H\delta_{j_{i}j'}-\frac{1}{2}\left(x_{j_{i}}+x_{j'}\right)^{T}K\delta_{j_{i}j'}\right]\right\} \label{eq:hinge loss as margin}
\end{equation}
a hinge-type surrogate loss function, and by 
\begin{equation}
\mathring{L}_{\boldsymbol{D}}^{\text{hinge}}\left(H,K\right)\triangleq\frac{1}{n}\sum_{i\in\left[n\right]}\mathring{\ell}^{\text{hinge}}\left(H,K,i\right)\label{eq:hinge as margin-1}
\end{equation}
the average \emph{hinge loss} over the dataset. As before, let $P=\bigcup_{i=1}^{d+1}P_{i}$
be a proper partition, let $\{\eta_{i}\}_{i\in\left[d+1\right]}$
be positive parameters which sum to $1$, and let $\lambda>0$ be
a regularization parameter. We show in Appendix \ref{subsec:Non linear Dev}
that the RLM learning rule 
\begin{equation}
\begin{array}{cc}
\min_{\left(H,K\right)\in\mathcal{H}\times\mathcal{K}} & \mathring{L}_{\boldsymbol{D}}^{\text{hinge}}\left(H,K\right)+\lambda\sum_{i=1}^{d+1}\eta_{i}\left[\max_{j\in P_{i}}\left\Vert H\delta_{j}\right\Vert ^{2}+\max_{j'\in P_{i}}\left\Vert K\delta_{j'}\right\Vert ^{2}\right]\\
\st & H^{T}H\preceq K
\end{array},\label{eq:stable RLM-1}
\end{equation}
learns a linear kernel $H$ which approximately maximizes a lower
bound on the margin among all codeword-pairs. The derivation is similar
to the one for the additive noise channel. In what follows we will
describe only the main differences, and the full derivation can be
found in Appendix \ref{subsec:Non linear Dev}.

The key difference between the two derivations is that while the decoder
class for the additive noise channel is linear in its parameter $S$,
the decoder class for the non-linear Gaussian noise channel is quadratic
in its parameter $H$. The first consequence of this difference is
that even after taking the lower bound in step $2$, the problem is
still not necessarily convex. The second consequence of this difference
is in establishing a minimum norm problem, in step 3. The technique
which we use (following SVM-type analysis) is only suitable for linear
classifiers. Therefore, we proceed by a linearization of the decoder,
replacing $H^{T}H$ with an auxiliary PSD matrix $K\in\mathcal{K}\triangleq\mathbb{S}_{+}^{d_{x}}$.
Nonetheless, we note that the learned decoder will be parameterized
only by $H$. Therefore, at the next step we add a constraint that
links the auxiliary matrix $K$ to $H$ as $H^{T}H=K$, and then further
take a convex relaxation, namely $H^{T}H\preceq K$.

\section{Generalization Error Bounds\label{sec:Generalization-Error-Bounds}}

In this section, we state average generalization error bounds on the
expected error probability, for the RLM learning rules (\ref{eq:stable RLM})
and (\ref{eq:stable RLM-1}). Additionally, we show an optimal choice
for $\lambda$, the regularization parameter. Finally, we state a
uniform high-probability generalization error bound for the decoder
class of the non-linear channel model (\ref{eq:hypothesis class}),
which complements a similar bound from \cite[Th. 3]{weinberger2021generalization}
for the additive noise channel model. We say that the random vector
$Z$ is sub-Gaussian with variance proxy $\sigma_{Z}^{2}$ if $\p[\|Z\|_{2}\geq t]\leq2\exp(-t^{2}/(2\sigma_{Z}^{2}))$
for all $t\geq0$.
\begin{thm}
\label{thm: generalization theorem}Let $A$ be the RLM rule (\ref{eq:stable RLM}),
and let $\mu$ be the distribution of the noise $Z$. If $r_{x}\triangleq\max_{x\in C}\|x\|,\eta_{\text{\emph{min}}}\triangleq\min_{i\in\left[d+1\right]}\{\eta_{i}\}$,
the noise $Z$ is sub-Gaussian with variance proxy $\sigma_{Z}^{2}$,
and $\zeta\triangleq32r_{x}^{2}\sqrt{6\left(128r_{x}^{2}\sigma_{Z}^{2}+1024\sigma_{Z}^{4}+r_{x}^{4}\right)}$,
then,
\begin{equation}
\mathbb{E}_{\boldsymbol{D}\sim\mu}\left[\boldsymbol{p}_{\mu}\left(A\left(\boldsymbol{D}\right)\right)\right]\le\left(m-1\right)\mathbb{E}_{\boldsymbol{D}\sim\mu}\left[\mathring{L}_{\boldsymbol{D}}^{\text{\emph{hinge}}}\left(A\left(\boldsymbol{D}\right)\right)\right]+\frac{\left(m-1\right)\log\left(n\right)}{\lambda n}\cdot\frac{\zeta}{\eta_{\min}\min_{1\le p<q\le m}\left\Vert \delta_{pq}\right\Vert ^{2}}.\label{eq:gen bound wlambda}
\end{equation}
Let ${\cal S}_{B}\triangleq\{S\in\mathbb{S}_{+}\colon\max_{j\in[\frac{1}{2}m(m-1)]}\Vert S\delta_{j}\Vert\le B\}$,
and let $\lambda=\sqrt{\frac{\zeta\log\left(n\right)}{B^{2}n\eta_{\min}\min_{1\le p<q\le m}\left\Vert \delta_{pq}\right\Vert ^{2}}}$.
Then, the RLM rule (\ref{eq:stable RLM rule}) with $\mathbb{S}_{+}$
replaced with ${\cal S}_{B}$ satisfies
\[
\mathbb{E}_{\boldsymbol{D}\sim\mu}\left[\boldsymbol{p}_{\mu}\left(A\left(\boldsymbol{D}\right)\right)\right]\le\left(m-1\right)\min_{S\in{\cal S}_{B}}\mathring{L}_{\mu}^{\text{\emph{hinge}}}\left(S\right)+\left(m-1\right)B\sqrt{\frac{4\zeta\log\left(n\right)}{n\eta_{\min}\min_{1\le p<q\le m}\left\Vert \delta_{pq}\right\Vert ^{2}}}.
\]
\end{thm}

\paragraph*{Proof outline}

The proof is based on an on-average-replace-one-stability argument.
We begin by proving that the regularization function in (\ref{eq:stable RLM})
is $2\eta_{\text{\emph{min}}}\min_{1\le p<q\le m}\left\Vert \delta_{pq}\right\Vert ^{2}$-strongly
convex, and the loss function in (\ref{eq:stable RLM}) is convex
and Lipschitz with seminorm 
\[
\max_{1\le p<q\le m}\max_{i\in\left[n\right]}\sqrt{4\left|\left\langle z_{i},\delta_{pq}\right\rangle \right|\left\Vert \delta_{pq}\right\Vert ^{2}+2\left\langle z_{i},\delta_{pq}\right\rangle ^{2}+2\left\Vert z_{i}\right\Vert ^{2}\left\Vert \delta_{pq}\right\Vert ^{2}+\left\Vert \delta_{pq}\right\Vert ^{4}}.
\]
Next, we apply\cite[Cor. 13.6]{shalev2014understanding}, where we
replace the $2$-strong-convexity of the Tikhonov regularization with
the appropriate constant for the regularization function of (\ref{eq:stable RLM}).
Then, we get from \cite[Cor. 13.6]{shalev2014understanding} that
the RLM problem (\ref{eq:stable RLM}) is on-average-replace-one-stable
with a rate that depends on the expected value of $\max_{i\in[n]}\|Z_{i}\|^{2}$.
We then decompose the expected rate to an expectation conditioned
on a ``good'' event, where all the samples are bounded by some constant,
and an expectation conditioned on a ``bad'' event, where not all
samples are bounded. Next, we use the sub-Gaussian assumption to bound
the expected rate conditioned on each event. Finally, we follow \cite[Cor. 13.9]{shalev2014understanding}
to derive an optimal choice for the regularization parameter $\lambda$.

\paragraph*{Comparison to \cite{weinberger2021generalization}}

In \cite{weinberger2021generalization}, a $\tilde{O}(m\sqrt{\frac{d}{n}}+\sqrt{\frac{\log\left(1/\delta\right)}{n}})$
high-probability generalization error bound for the error probability
loss function, as well as a $\tilde{O}(\sqrt{\frac{d\left(d+m\right)}{n}}+\sqrt{\frac{\log\left(1/\delta\right)}{n}})$
high probability generalization error bound for the surrogate hinge-type
upper bound (\ref{eq:hinge as upper bound}) was proved. In comparison,
here we prove a $\tilde{O}(m/n)$ generalization error bound on the
error probability. The convergence rate of this bound is much faster,
however, this is only an average error bound, and does not have a
high probability guarantee.

\paragraph*{Theoretical guidance for the choice of training SNR}

In classification problems, the generalization bound is typically
used to bound the expected error of the learned classifier, given
the empirical error. Here, the generalization bound has an additional
and important role in providing a theoretical guidance for the choice
of training SNR. For the decoder learning problem, the dataset $\boldsymbol{D}$
can be generated for any arbitrary training SNR parameter $\Gamma$
of the input codebook $C$. As discussed in various previous works
(e.g., \cite{gruber2017deep,kim2018communication,o2017introduction}),
this raises the question of how to optimize the training SNR. Intuitively,
on one hand, training with a sufficiently high SNR leads to zero empirical
error for many decoders in the class, not necessarily the one with
the lowest expected error. On the other hand, training with SNR too
low may produce a decoder which has high error probability (as most
evident from the extreme case of zero SNR), and may be too pessimistic
in assessing the error probability. In \cite{kim2018communication},
a rule-of-thumb for choosing the training SNR was proposed, based
on the capacity of the Gaussian channel. This rule, however, did not
take into account generalization error aspects. The generalization
error bound of Theorem \ref{thm: generalization theorem} hints a
different rule for choosing the training SNR. Specifically, we propose
to choose the training SNR so that the empirical error $\mathring{L}_{\boldsymbol{D}}^{\text{hinge}}(A(\boldsymbol{D}))$
roughly equals to the generalization bound on the hinge-type loss
(the right-hand side of (\ref{eq:gen bound wlambda}) divided by $m-1$).
With this training SNR, it is guaranteed that the expected error is
on the same order as the empirical error. In practice, this can be
accomplished by tuning the training SNR as a hyperparameter by cross
validation.

The following theorem states the generalization error bound for the
learning rule (\ref{eq:stable RLM-1}), with a similar proof outline.
\begin{thm}
\label{thm: generalization theorem-1}Let $A$ be the RLM rule (\ref{eq:stable RLM-1}),
and let $\mu$ be the distribution of the channel output for the channel
transformation $f$. If $r_{x}\triangleq\max_{x\in C}\|x\|,R_{x}\triangleq\max_{x\in C}\|f(x)\|,\eta_{\text{\emph{min}}}\triangleq\min_{i\in\left[d+1\right]}\{\eta_{i}\}$,
the noise $W$ is sub-Gaussian with variance proxy $\sigma_{W}^{2}$,
and $\zeta\triangleq24r_{x}^{2}\sqrt{25r_{x}^{4}+16R_{x}^{4}+64\sigma_{W}^{4}}$,
then,
\[
\mathbb{E}_{\boldsymbol{D}\sim\mu}\left[\boldsymbol{p}_{\mu}\left(A\left(\boldsymbol{D}\right)\right)\right]\le\left(m-1\right)\mathbb{E}_{\boldsymbol{D}\sim\mu}\left[\mathring{L}_{\boldsymbol{D}}^{\text{\emph{hinge}}}\left(A\left(\boldsymbol{D}\right)\right)\right]+\frac{\left(m-1\right)\log\left(n\right)}{\lambda n}\cdot\frac{\zeta}{\eta_{\min}\min_{1\le p<q\le m}\left\Vert \delta_{pq}\right\Vert ^{2}}.
\]
Let ${\cal H}_{B}\triangleq\{H\in\mathcal{H}\colon\max_{j\in[\frac{1}{2}m(m-1)]}\Vert H\delta_{j}\Vert\le B_{H}\},{\cal K}_{B}\triangleq\{K\in\mathcal{K}\colon\max_{j\in[\frac{1}{2}m(m-1)]}\Vert K\delta_{j}\Vert\le B_{K}\}$,
such that $B=B_{H}+B_{K}$, and let $\lambda=\sqrt{\frac{\zeta\log\left(n\right)}{B^{2}n\eta_{\min}\min_{1\le p<q\le m}\left\Vert \delta_{pq}\right\Vert ^{2}}}$.
Then, the RLM rule (\ref{eq:stable RLM-1}) with $\mathcal{H}\times\mathcal{K}$
replaced with ${\cal H}_{B}\times{\cal K}_{B}$ satisfies 
\[
\mathbb{E}_{\boldsymbol{D}\sim\mu}\left[\boldsymbol{p}_{\mu}\left(A\left(\boldsymbol{D}\right)\right)\right]\le\left(m-1\right)\min_{\left(H,K\right)\in\mathcal{H}_{B}\times\mathcal{K}_{B}}\mathring{L}_{\mu}^{\text{\emph{hinge}}}\left(H,K\right)+\left(m-1\right)B\sqrt{\frac{4\zeta\log\left(n\right)}{n\eta_{\min}\min_{1\le p<q\le m}\left\Vert \delta_{pq}\right\Vert ^{2}}}.
\]
\end{thm}
Note that Theorem \ref{thm: generalization theorem-1} holds for general
sub-Gaussian noise, and specifically under our Gaussian noise assumption
(used here to justify the structure of the decoder). Finally, we state
a uniform high-probability generalization error bound for the decoder
class of the non-linear Gaussian noise channel (\ref{eq:hypothesis class}).
This uniform bound can be used to bound the generalization error of
\emph{any} learning algorithm for the problem, e.g., ERM.
\begin{thm}
\label{thm:uniform hinge gen}Assume that $\mathcal{H}=\{H\in\mathbb{R}^{d_{y}\times d_{x}}:\max_{i\in[\min\{d_{x},d_{y}\}]}\sigma_{i}^{2}(H)\le r_{H}^{2}\}$
and denote $d_{m}\triangleq\min\{d_{x},d_{y}\}$. Then, with probability
$1-\delta$, for all $H\in\mathcal{H}$

\begin{align}
 & \left|\mathring{L}_{\mu}^{\text{\emph{hinge}}}\left(H\right)-\mathring{L}_{\boldsymbol{D}}^{\text{\emph{hinge}}}\left(H\right)\right|\nonumber \\
 & \le24\sqrt{\frac{d_{x}^{2}+d_{y}^{2}+d_{m}}{n\log\left(12d_{m}r_{H}\right)}}\left[2\log\left(12d_{m}r_{H}\right)+1-\exp\left(2\log\left(12d_{m}r_{H}\right)-\frac{2}{3}\sqrt{\frac{n\log\left(12d_{m}r_{H}\right)}{d_{x}^{2}+d_{y}^{2}+d_{m}}}\right)\right]\nonumber \\
 & \hphantom{==}+\sqrt{\frac{2\left[2\left(R_{x}+r_{z}\right)r_{H}r_{x}+r_{x}^{2}r_{H}^{2}\right]^{2}\log\left(2/\delta\right)}{n}.}
\end{align}
\end{thm}
The proof is based on bounding the generalization error using the
Rademacher complexity of the loss class, i.e., the hinge loss class
for a decoder from the decoder class (\ref{eq:hypothesis class}).
In turn, the Rademacher complexity is bounded via Dudley's entropy
integral. This bound complements the uniform high-probability generalization
error bound for the decoder class of the additive noise channel (\ref{eq:NN decoder family}),
which was proved in \cite[Th. 3]{weinberger2021generalization}

\section{Stochastic Sub-gradient Descent Algorithms\label{sec:SGD}}

In this section, we propose a stochastic sub-gradient descent algorithms
for solving (\ref{eq:stable RLM}) and (\ref{eq:stable RLM-1}), inspired
by an algorithm for classification called PEGASOS \cite{pegasos}.
These algorithms achieve an $\epsilon$-accurate solution in $\tilde{O}(1/\epsilon)$
iterations. The run-time is independent of the dataset size $n$,
which makes the algorithms especially suited for learning from large
datasets. This is the case, in an offline design of the decoder (i.e.,
prior to data communication), in which noise samples are readily available.
Moreover, even if the decoder is learned online, during a training
phase, and so the number of samples is relatively small, low-complexity
of each iteration is typically of importance due to limited computational
power of the communication device (as, for example, motivates learning
equalizers by the least mean squares (LMS) algorithm \cite[Ch. 9]{barry2012digital}).
In comparison, and as discussed in \cite{bottou2007support}, the
computational cost of solving a standard SVM problem grows at least
like $O(n^{2})$. Moreover, it was shown in \cite{shalev2008svm}
that even if the solver is efficient in the data-laden regime, in
which data is virtually unlimited, it has a worse dependence on $\epsilon$,
compared to the sub-gradient descent algorithm. The pseudo code of
our proposed algorithm is given in Algorithm \ref{subgradient alg},
where $\Gamma$ denotes the hypothesis (either $S$ or $(H,K)$).
\begin{algorithm}[t]
\caption{RLM Sub-gradient Descent Algorithm\label{subgradient alg}}

\begin{algorithmic}[1]

\State  \textbf{input }$\boldsymbol{D}\in\mathbb{R}^{d\times n}\times\mathbb{N}^{n},\lambda\in\mathbb{R}^{+},\{\eta\}_{i=1}^{d+1}\in\Delta^{d},T\in\mathbb{N}^{+},c\in\mathbb{N}^{+}$

\State  \textbf{begin}

\State \textbf{ ~}Set $\Gamma_{1}=0$

\State  \textbf{~for} $t=1,2,\dots,T$

\State  \textbf{~~}Choose $c$ samples from the dataset $\boldsymbol{D}$,
uniformly at random

\State  \textbf{~~}Perform a sub-gradient step $\Gamma_{t+1}\gets\Gamma_{t}-\frac{1}{\lambda t}\nabla_{t}$
\Comment{ Update }

\State  \textbf{~~}Project to the set of admissible solutions $\Gamma_{t+1}\gets\Pi(\Gamma_{t+1})$
\Comment{ Project }

\State  \textbf{~end for}

\State  \textbf{end}

\State  \textbf{output }$\Gamma_{T+1}$

\end{algorithmic}
\end{algorithm}
 Next, we describe the \emph{Update }and \emph{Project }steps for
each model.

Notice that all matrix derivatives are w.r.t. symmetric matrices.
The derivative of a matrix function $f(S)$ w.r.t. to a symmetric
matrix $S$ \cite[Cor. 1]{mcculloch1982symmetric} is
\[
\frac{\d f}{\d S}=\dsym\left(\frac{\partial f}{\partial S}\right)\triangleq\frac{\partial f}{\partial S}+\frac{\partial f}{\partial S}^{T}-\diag\frac{\partial f}{\partial S},
\]
where $\d f/\d S$ denotes the symmetric derivative and $\partial f/\partial S$
denotes the general matrix derivative.

For the additive noise channel model we denote the following. Denote
the RLM objective by $f(S)$, the set of $c$ samples from round $t$
by $\boldsymbol{A}_{t}$, and the RLM objective with $\boldsymbol{D}$
replaced by $\boldsymbol{A}_{t}$ by $f_{t}(S)$. Let $p_{a},q_{a}$
be the codeword-pair related to a transformed sample $a$, $\delta_{a}\triangleq\delta_{p_{a},q_{a}}$,
and let $j_{t,k}\triangleq\argmax_{j\in P_{k}}\Vert S_{t}\delta_{j}\Vert^{2}$.
Then, the sub-gradient of $f_{t}(S)$ is
\begin{equation}
\nabla_{t}\triangleq\lambda\sum_{k=1}^{d+1}\eta_{k}\dsym\left(2S_{t}\delta_{j_{t,k}}\delta_{j_{t,k}}^{T}\right)-\frac{1}{\left|\boldsymbol{A}_{t}\right|}\sum_{i\in\boldsymbol{A}_{t}}\mathbb{I}\left[a_{i}^{T}S_{t}\delta_{a_{i}}<1\right]\dsym\left(a_{i}\delta_{a_{i}}^{T}\right).\label{eq:objective subgradient}
\end{equation}
In general, a gradient step may result in $S_{t+1}\notin\mathbb{S}_{+}$
hence we include a projection step to circumvent this problem. We
note in passing that in \cite[Sec. 2.2]{pegasos} the projection step
was optional and used to limit the set of admissible solutions to
the ball of $1/\sqrt{\lambda}$ radius. This lead to two separate
cases in the analysis. In our problem the projection is obligatory
due to the definition of the Mahalanobis distance. According to \cite[Ch. 8]{Boyd},
the projection of a symmetric matrix to the positive semidefinite
cone w.r.t. the Frobenius norm is
\begin{equation}
\Pi_{\mathbb{S}_{+}}(S)\triangleq\sum_{i=1}^{d}\max\left\{ \lambda_{i},0\right\} v_{i}v_{i}^{T},\label{eq:PSD projection}
\end{equation}
and this is the projection used here.

For the non-linear channel model we denote the following. Denote the
RLM rule objective by $f(H,K)$, the set of $c$ samples from round
$t$ by $\boldsymbol{A}_{t}$, and the RLM objective with $\boldsymbol{D}$
replaced by $\boldsymbol{A}_{t}$ by $f_{t}(H,K)$. Let $j_{t,k}^{\left(1\right)}\triangleq\argmax_{j\in P_{k}}\Vert H_{t}\delta_{j}\Vert^{2}$
and $j_{t,k}^{\left(2\right)}\triangleq\argmax_{j\in P_{k}}\Vert K_{t}\delta_{j}\Vert^{2}$
. Then, the sub-gradients of $f_{t}(H,K)$ are
\begin{align}
\nabla_{t}^{(1)}f_{t}\left(H_{t},K_{t}\right) & \triangleq\frac{\partial f_{t}}{\partial H}\left(H_{t},K_{t}\right)\nonumber \\
 & =2\lambda\sum_{k=1}^{d+1}\eta_{k}H_{t}\delta_{j_{t,k}^{\left(1\right)}}\delta_{j_{t,k}^{\left(1\right)}}^{T}\nonumber \\
 & \phantom{=}-\frac{1}{\left|\boldsymbol{A}_{t}\right|}\sum_{i\in\boldsymbol{A}_{t}}\frac{1}{m-1}\sum_{j'\in\left[m\right]\backslash\left\{ j_{i}\right\} }\mathbb{I}\left[y_{i}^{T}H_{t}\delta_{j_{i}j'}-\frac{1}{2}\left(x_{j_{i}}+x_{j'}\right)^{T}K_{t}\delta_{j_{i}j'}<1\right]y_{i}\delta_{j_{i}j'}^{T}\label{eq:subgrad H}
\end{align}
and
\begin{align}
\nabla_{t}^{(2)} & f_{t}\left(H_{t},K_{t}\right)\triangleq\frac{\partial f_{t}}{\partial K}\left(H_{t},K_{t}\right)\nonumber \\
 & =\lambda\sum_{k=1}^{d+1}\eta_{k}\dsym\left(2K_{t}\delta_{j_{t,k}^{\left(2\right)}}\delta_{j_{t,k}^{\left(2\right)}}^{T}\right)+\frac{1}{\left|\boldsymbol{A}_{t}\right|}\sum_{i\in\boldsymbol{A}_{t}}\frac{1}{m-1}\nonumber \\
 & \phantom{==}\times\sum_{j'\in\left[m\right]\backslash\left\{ j_{i}\right\} }\mathbb{I}\left[y_{i}^{T}H_{t}\delta_{j_{i}j'}-\frac{1}{2}\left(x_{j_{i}}+x_{j'}\right)^{T}K_{t}\delta_{j_{i}j'}<1\right]\dsym\left(\frac{1}{2}\left(x_{j_{i}}+x_{j'}\right)\delta_{j_{i}j'}^{T}\right).\label{eq:subgrad K}
\end{align}
In general, a gradient step may result in $H_{t+1},K_{t+1}:H^{T}H\npreceq K$,
or even $K\prec0$, hence we include a projection step, w.r.t. the
Frobenius norm, to circumvent this problem. We formulate this projection
as a convex optimization problem, which does not depend on the sample
size.
\begin{claim}
\label{claim:projection claim}The projection of $(H_{t},K_{t})\in\mathcal{H}\times\mathcal{K}$
to the set $\{(H,K)\in\mathcal{H}\times\mathcal{K}:H^{T}H\preceq K\}$,
w.r.t. the Frobenius norm, is a convex optimization problem
\begin{equation}
\begin{array}{cc}
\min_{\left(H,K\right)\in\mathcal{H}\times\mathcal{K}} & \left\Vert H-H_{t}\right\Vert _{F}^{2}+\left\Vert K-K_{t}\right\Vert _{F}^{2}\\
\st\,\, & H^{T}H\preceq K
\end{array}.
\end{equation}
\end{claim}
We prove the following optimization error bound for the two algorithms.
\begin{thm}
\label{thm: optimization bound}Let $\Gamma^{*}$ be the solution
of the RLM ((\ref{eq:stable RLM}) or (\ref{eq:stable RLM-1})), and
$\Gamma_{t}$ be the hypothesis generated by Algorithm \ref{subgradient alg}
at a random round $t\in[T]$. Denote the objective of the RLM by $f(\Gamma)$.
Assume that for all $t$, each element in $\boldsymbol{A}_{t}$ is
sampled uniformly at random from the dataset (with or without replacement).
Then, 
\[
f\left(\Gamma_{t}\right)-f\left(\Gamma^{*}\right)=O\left(\frac{\ln^{3}\left(T\right)\cdot\ln\left(1/\delta\right)}{\lambda T}\right)
\]
with probability larger than $\frac{1-4\delta\ln\left(T\right)}{2}$.
\end{thm}
Thus, as discussed in \cite{pegasos}, roughly \emph{two} validation
attempts are required to obtain a good solution. Using this result
combined with the expected generalization error bounds, stated in
Sec. \ref{sec:Generalization-Error-Bounds}, we can describe the total
error bound, which is comprised of three terms. First, the choice
of the class of decoders, which do not necessarily contain the optimal,
maximum likelihood decoder, inflicts an \emph{approximation error}.
Second, learning the decoder based on samples instead of the unknown
channel's distribution, inflicts a \emph{generalization error}, which
was bounded in Theorems \ref{thm: generalization theorem} and \ref{thm: generalization theorem-1},
that established an expected generalization error of rate $\tilde{O}(m/(n)).$
Third, solving the optimization problem only approximately, using
a finite number $T$ of iterations of Algorithm \ref{subgradient alg},
inflicts an \emph{optimization error}, which is bounded in Theorem
\ref{thm: optimization bound}.

\section{Experiments for the Sub-gradient Descent Algorithms\label{sec:Examples}}

In this section, we exemplify the empirical performance of the proposed
algorithm for different codebooks and noise distributions. In each
experiment, the learner was provided with the codebook of $m$ codewords
in $\mathbb{R}^{d}$ and $n_{\text{train}}$ i.i.d. training samples
in $\text{SNR}_{\text{train}}$. Then, the proposed Algorithm \ref{subgradient alg}
has run for $T$ iterations with a batch size of $n_{\text{batch}}$
and a regularization parameter $\lambda$. Finally, test sets for
various SNR values were used, each one with $n_{\text{test}}$ noise
samples.

\subsection{Stochastic Sub-gradient Descent Algorithm for the Additive Noise
Channel\label{subsec:Examples-additive}}

The algorithm's performance for the additive noise channel, is compared
with the real precision matrix $\Sigma^{-1}$, the estimated precision
matrix $\hat{\Sigma}^{-1}$ and the identity matrix $I$. We denote
by $S_{T}$ the hypothesis chosen during the training phase, whether
it is the hypothesis generated at iteration $T$ or at another iteration,
guided by cross validation over the training samples.
\begin{enumerate}
\item Two-dimensional noise with a Gaussian component and a uniform component:
Assume $d=2$, a codebook of $m=2$ codewords $C=\{(1,1),(-1,-1)\}$,
and let $Z\in\mathbb{R}^{2}$ be a zero-mean noise with two independent
components of equal variance. Specifically, $Z_{1}$ is a uniform
random variable and $Z_{2}$ is a Gaussian random variable. In this
simple example, in high SNR the optimal matrix is obviously $S=\left(\begin{array}{cc}
1 & 0\\
0 & 0
\end{array}\right)$ which induces the linear separator $S\delta_{12}=(2,0)^{T}$, a vertical
separator. The proposed algorithm indeed learns such a decoder, and
its classification of the test samples is shown in the left panel
of Fig. \ref{fig:uniform-gaussian-1}. The classification of the test
samples by the decoder using $\Sigma^{-1}$ is shown in the right
panel of Fig. \ref{fig:uniform-gaussian-1}. The performance of $\Sigma^{-1}$
is worse, even though this decoder reflects some knowledge on the
noise distribution. Nonetheless, in low SNR there is a change of trends
due to the uniform noise being bounded. In low SNR values the optimal
decoder is no longer a vertical separator, as is evident in Fig. \ref{fig:uniform-gaussian-1-1}.
Fig. \ref{fig:uniform-gaussian-2} displays the error probability
of the final hypothesis decoder, under various SNR values. The experiment
parameters are listed in Table \ref{tab:additive-exp-1}.
\begin{table}

\begin{centering}
\begin{tabular}{|c|c|c|c|c|c|c|c|}
\hline 
$d$ & $m$ & $n_{\text{train}}$ & $n_{\text{batch}}$ & $T$ & $n_{\text{test}}$ & $\lambda$ & $\text{SNR}_{\text{train}}$\tabularnewline
\hline 
\hline 
$2$ & $2$ & $10^{4}$ & $50$ & $600$ & $10^{4}$ & $4.8\cdot10^{-4}$ & $4.8$\tabularnewline
\hline 
\end{tabular}
\par\end{centering}
\centering{}\caption{Additive noise channel with a two-dimensional noise comprised of a
Gaussian component and a uniform component. Experiment parameters.\label{tab:additive-exp-1}}
\end{table}
\begin{figure}
\begin{centering}
\includegraphics[scale=0.5]{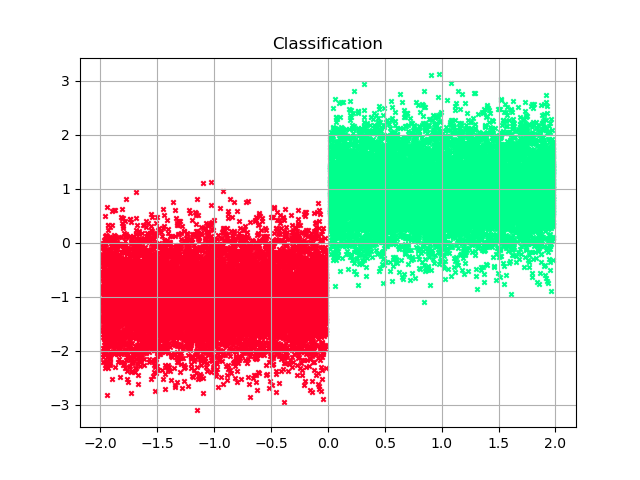}\includegraphics[scale=0.5]{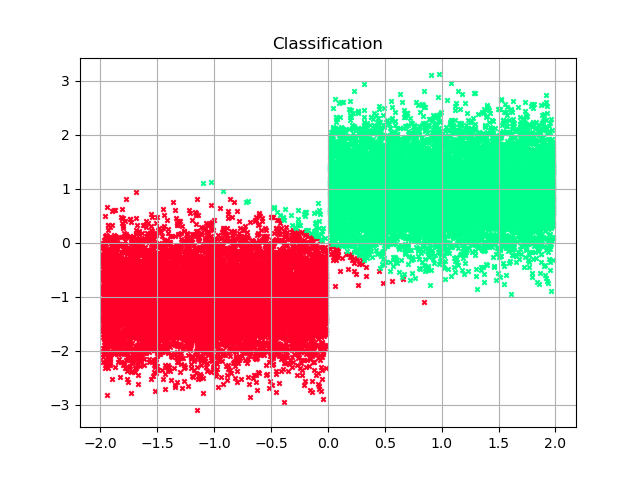}
\par\end{centering}
\caption{Additive noise channel with a two-dimensional noise comprised of a
Gaussian component and a uniform component. Classification in high
SNR. Left: $S_{T}$ classification of test samples. Right: $\Sigma^{-1}$
classification of test samples.\label{fig:uniform-gaussian-1}}
\end{figure}
\begin{figure}
\begin{centering}
\includegraphics[scale=0.5]{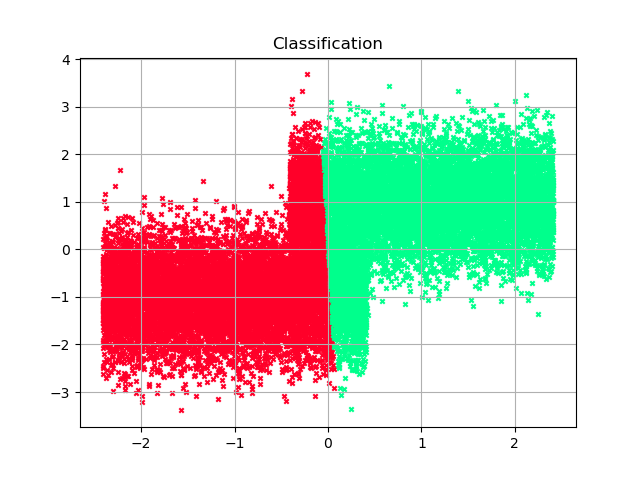}\includegraphics[scale=0.5]{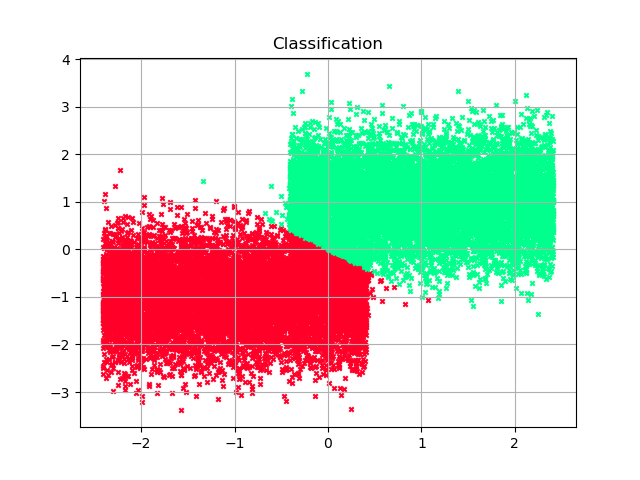}
\par\end{centering}
\caption{Additive noise channel with a two-dimensional noise comprised of a
Gaussian component and a uniform component. Classification in low
SNR. Left: $S_{T}$ classification of test samples. Right: $\Sigma^{-1}$
classification of test samples.\label{fig:uniform-gaussian-1-1}}
\end{figure}
\begin{figure}
\centering{}\includegraphics[scale=0.5]{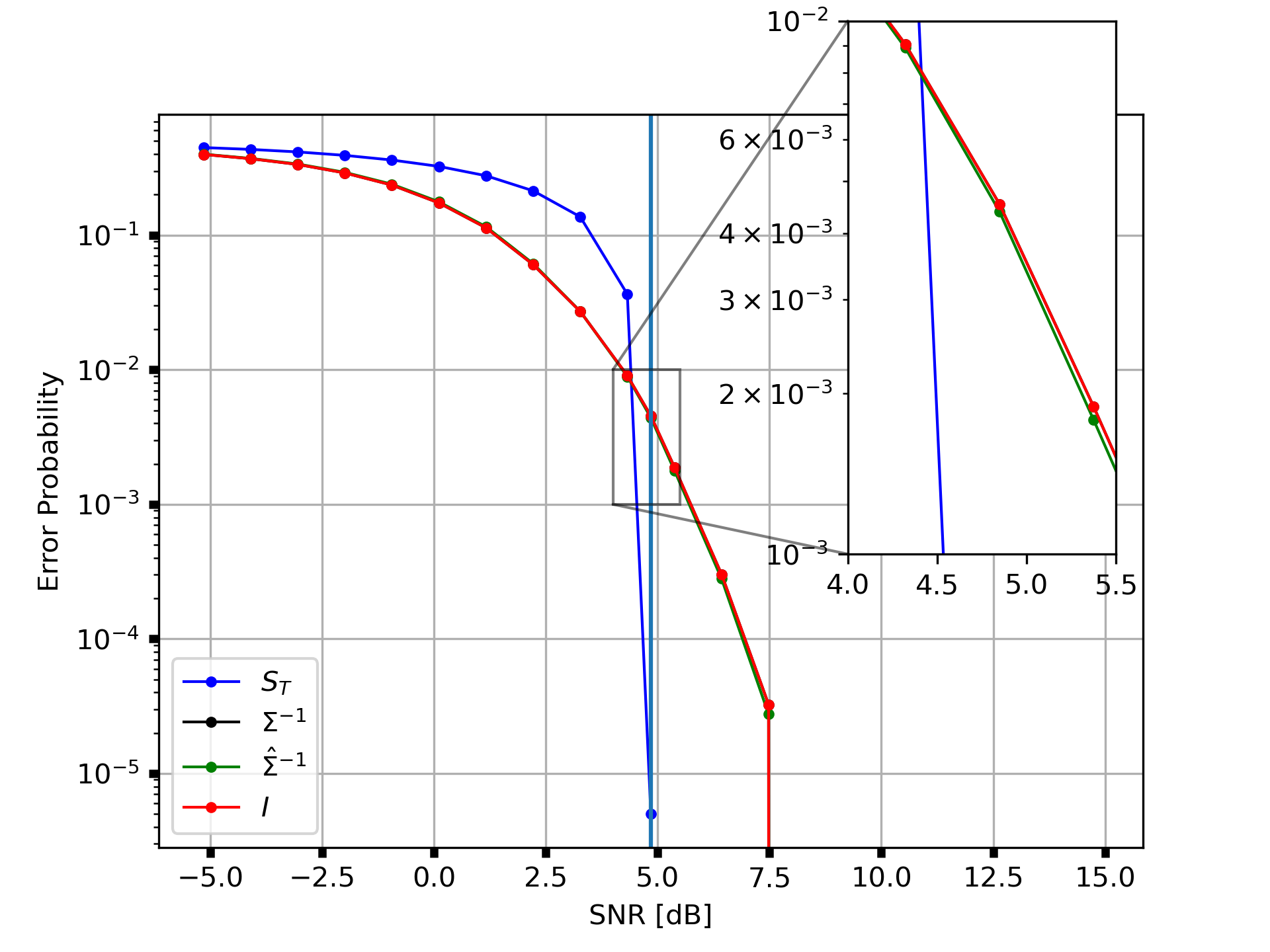}\caption{Additive noise channel with a two-dimensional noise comprised of a
Gaussian component and a uniform component. Error probability vs.
various SNRs {[}dB{]}.\label{fig:uniform-gaussian-2}}
\end{figure}
\item Two-dimensional Gaussian mixture noise: We perform a simulation similar
to the previous example, where now the codebook is a $64$-QAM constellation
in $2$D \cite[Fig. 2, (c)]{batshon2008proposal}, and $Z\in\mathbb{R}^{2}$
is a zero-mean Gaussian Mixture noise of $l=2$ components $Z_{i}\sim\mathcal{N}(0,K_{i})$,
and mixture weights $\{\omega_{i}\}_{i=1}^{l}$. The right panel of
Fig. \ref{fig:gaussian mixture sim2} displays the error probability
of the final hypothesis decoder, under various SNR values. In Fig.
\ref{fig:gaussian mixture sim2}, we observe that the error probability
scales similarly for both $S_{T}$ and $\Sigma^{-1}=\left(\frac{1}{l}\sum_{i=1}^{l}\omega_{i}K_{i}\right)^{-1}$,
even for SNR values that are significantly smaller than the training
SNR. This shows the generalization of the learned decoder, which is
a desired trait for systems where the SNR at operation time is unknown
durning training.
\begin{table}
\begin{centering}
\begin{tabular}{|c|c|c|c|c|c|c|c|}
\hline 
$d$ & $m$ & $n_{\text{train}}$ & $n_{\text{batch}}$ & $T$ & $n_{\text{test}}$ & $\lambda$ & $\text{SNR}_{\text{train}}$\tabularnewline
\hline 
\hline 
$2$ & $64$ & $220$ & $5$ & $10^{3}$ & $10^{4}$ & $10^{-7}$ & $22.34$\tabularnewline
\hline 
\end{tabular}
\par\end{centering}
\centering{}\caption{Additive noise channel with a two-dimensional Gaussian mixture noise.
Experiment parameters.\label{tab:additive-exp-2}}
\end{table}
\begin{figure}
\centering{}\includegraphics[scale=0.5]{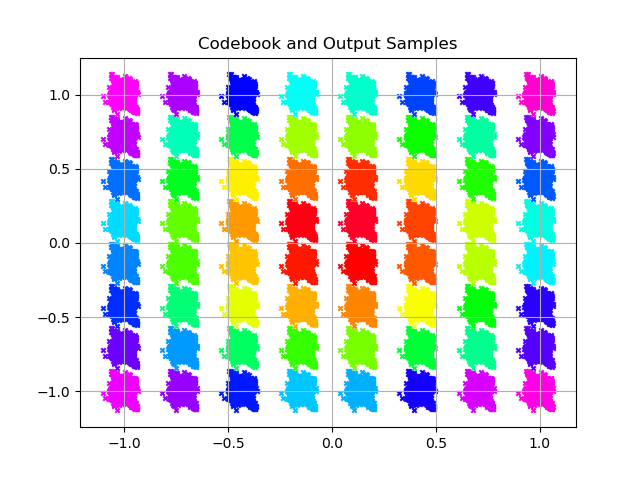}\includegraphics[scale=0.5]{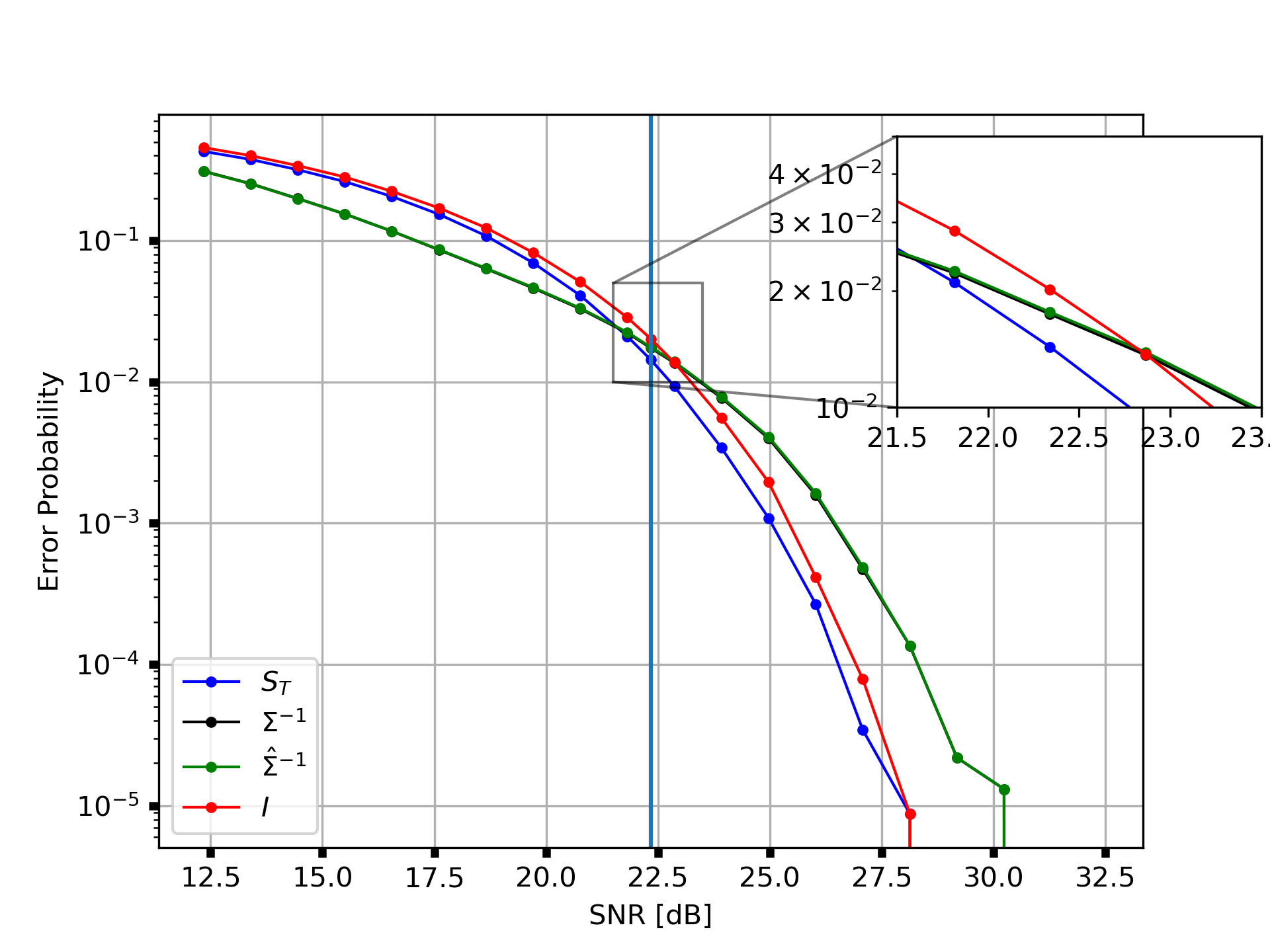}\caption{Additive noise channel with a two-dimensional Gaussian mixture noise.
Left: the test samples in SNR 27 {[}dB{]}. Right: error probability
vs. various SNRs {[}dB{]}. \label{fig:gaussian mixture sim2}}
\end{figure}
\end{enumerate}

\subsection{Stochastic Sub-gradient Descent Algorithm for the Non-Linear White
Gaussian Noise Channel\label{subsec:Examples-non-linear}}

The algorithm's performance, for the non-linear channel, is compared
with a NN decoder using the real transformation $f(\cdot)$, which
is the maximum likelihood decoder, and a NN decoder using the mean
of the transformed samples of each codeword as the transformed value
of the codeword, denoted $\mu_{X}$. We denote by $H_{T}$ the hypothesis
chosen during the training phase, whether it is the hypothesis generated
at iteration $T$ or at another iteration, guided by cross validation
over the training samples.
\begin{enumerate}
\item Two-dimensional linear channel: Assume $d=2$, a codebook of $m=8$
codewords as proposed in \cite[Fig. 2, (a)]{batshon2008proposal},
and a linear and invertible channel transformation
\[
f\begin{pmatrix}x_{1}\\
x_{2}
\end{pmatrix}=\left(\begin{array}{cc}
1.04 & 0.19\\
0.19 & 1.96
\end{array}\right)\begin{pmatrix}x_{1}\\
x_{2}
\end{pmatrix}.
\]
 In this simple example, the optimal matrix is obviously $H=\left(\begin{array}{cc}
1.04 & 0.19\\
0.19 & 1.96
\end{array}\right)$. The training samples are shown in the left panel of Fig. \ref{fig:sim},
and the right panel displays the error probability of the final hypothesis
decoder, under various SNR values. The right panel shows that the
learned decoder's performance is practically the same as that of the
optimal decoder. The experiment parameters are listed in Table \ref{tab:non-linear-exp-1}.
\begin{table}
\begin{centering}
\begin{tabular}{|c|c|c|c|c|c|c|c|}
\hline 
$d$ & $m$ & $n_{\text{train}}$ & $n_{\text{batch}}$ & $T$ & $n_{\text{test}}$ & $\lambda$ & $\text{SNR}_{\text{train}}$\tabularnewline
\hline 
\hline 
$2$ & $8$ & $1.6\cdot10^{3}$ & $10$ & $1.4\cdot10^{3}$ & $10^{4}$ & $(10^{-3},10^{-4})$ & $3.74$\tabularnewline
\hline 
\end{tabular}
\par\end{centering}
\centering{}\caption{Two-dimensional linear white Gaussian noise channel. Experiment parameters.\label{tab:non-linear-exp-1}}
\end{table}
\begin{figure}
\begin{centering}
\includegraphics[scale=0.5]{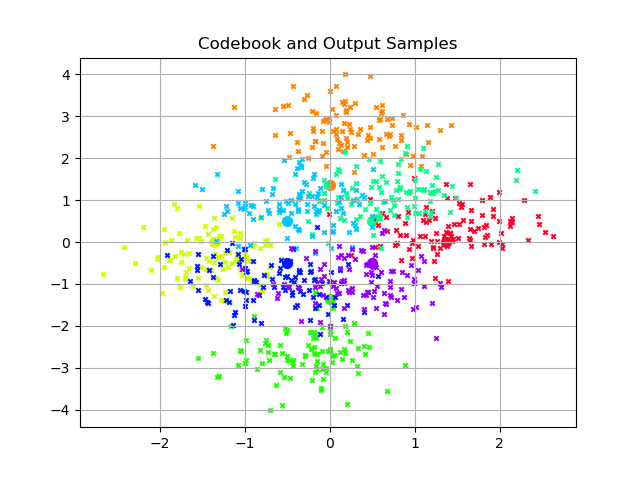}\includegraphics[scale=0.5]{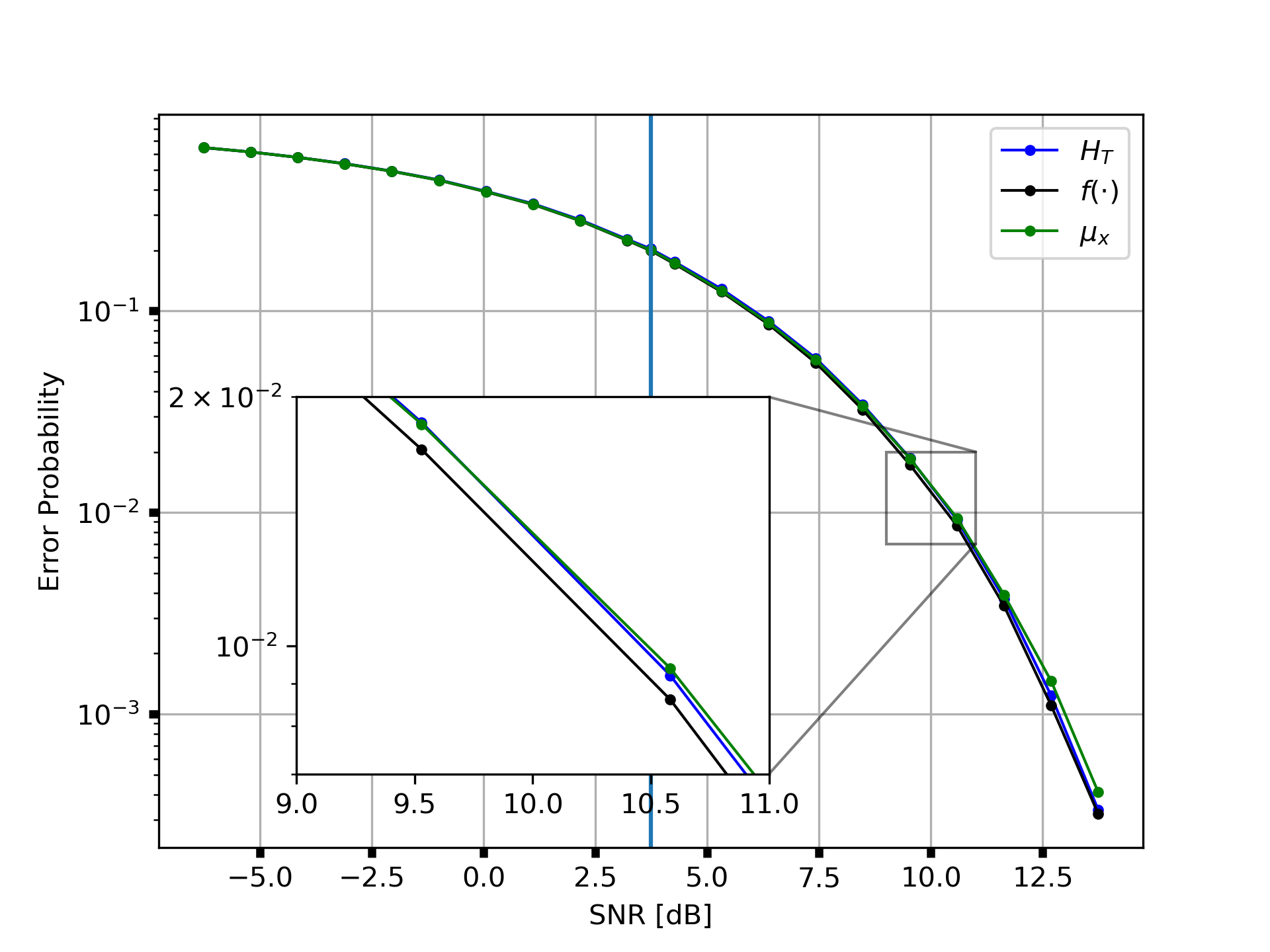}
\par\end{centering}
\caption{Two-dimensional linear white Gaussian noise channel. Left: codewords
(circles) and output samples (X's). Right: error probability vs. various
SNRs {[}dB{]}. \label{fig:sim}}
\end{figure}
\item Two-dimensional non-linear channel: We perform a simulation similar
to the previous example, where now the channel transformation is non-linear
\[
f\begin{pmatrix}x_{1}\\
x_{2}
\end{pmatrix}=\left(\begin{array}{ccc}
0.94 & 0.07 & 0.15\\
0.46 & 0.72 & 0.4
\end{array}\right)\begin{pmatrix}x_{1}\\
x_{2}\\
x_{1}x_{2}
\end{pmatrix}.
\]
 In this example, the learner is required to learn a linear transformation
which approximates the non-linear channel transformation. The training
samples are shown in the left panel of Fig. \ref{fig:sim-1}, and
the right panel displays the error probability of the final hypothesis
decoder, under various SNR values. The right panel shows that the
learned decoder's performance is close to that of the optimal decoder,
however the performance gap increases with the SNR. This shows the
difficulty of learning a decoder for a non-linear channel, compared
to the linear channel from the previous experiment. The experiment
parameters are listed in Table \ref{tab:non-linear-exp-2}.
\begin{table}
\begin{centering}
\begin{tabular}{|c|c|c|c|c|c|c|c|}
\hline 
$d$ & $m$ & $n_{\text{train}}$ & $n_{\text{batch}}$ & $T$ & $n_{\text{test}}$ & $\lambda$ & $\text{SNR}_{\text{train}}$\tabularnewline
\hline 
\hline 
$2$ & $8$ & $1.6\cdot10^{3}$ & $10$ & $3\cdot10^{3}$ & $10^{4}$ & $(10^{-2},10^{-3})$ & $11.7$\tabularnewline
\hline 
\end{tabular}
\par\end{centering}
\centering{}\caption{Two-dimensional non-linear white Gaussian noise channel. Experiment
parameters.\label{tab:non-linear-exp-2}}
\end{table}
\begin{figure}
\begin{centering}
\includegraphics[scale=0.5]{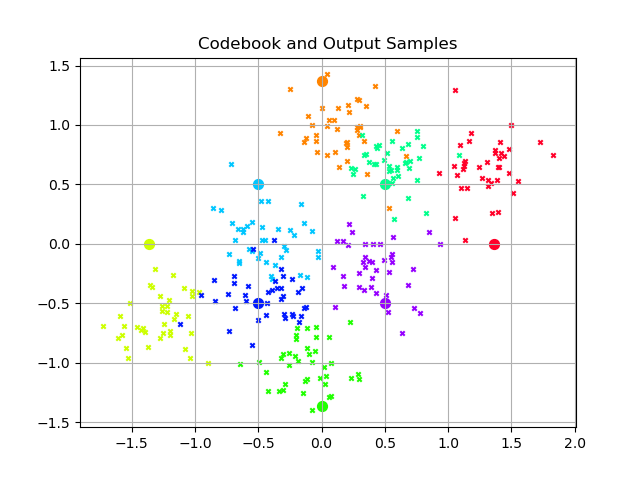}\includegraphics[scale=0.5]{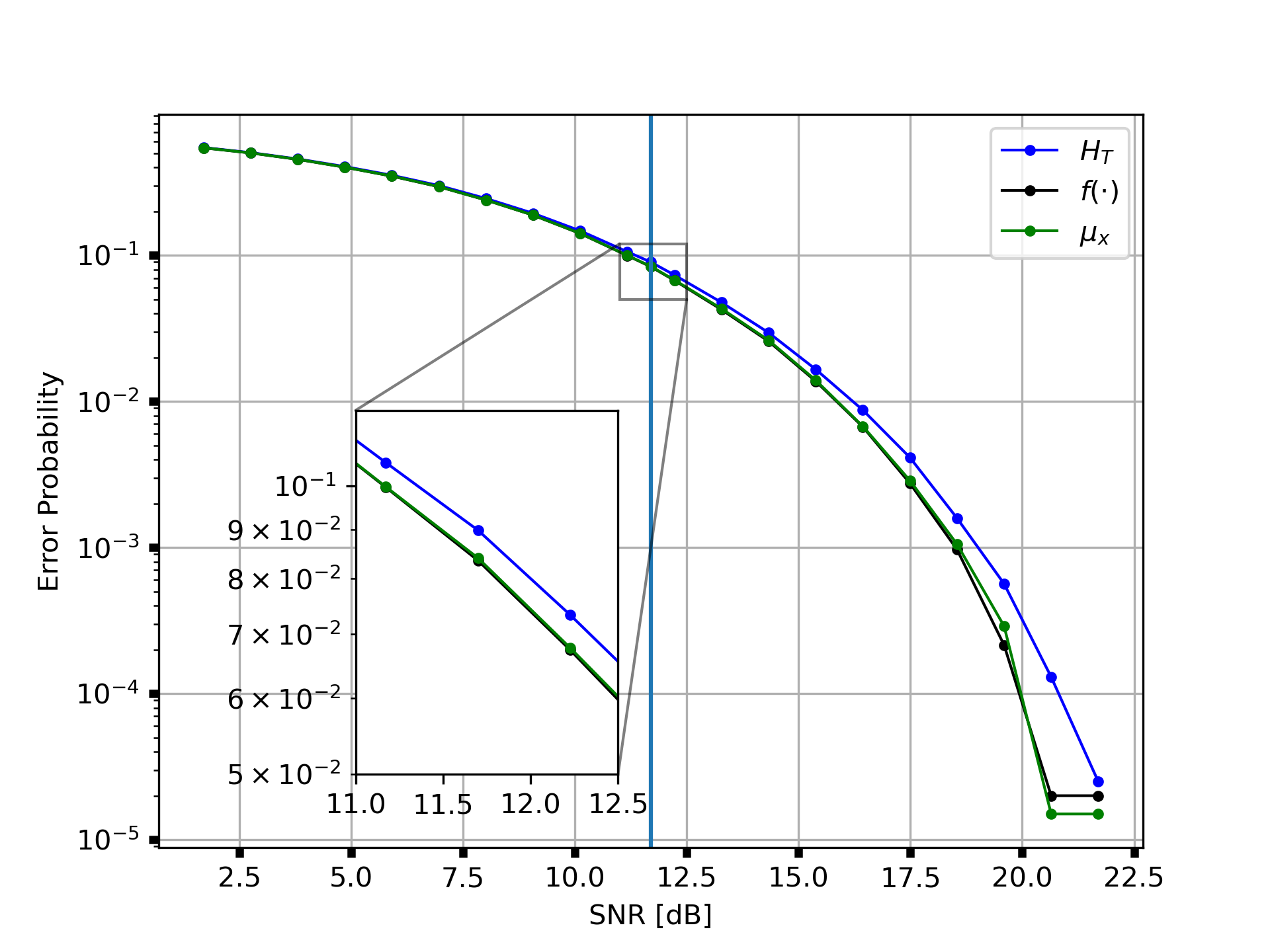}
\par\end{centering}
\caption{Two-dimensional non-linear white Gaussian noise channel. Left: codewords
(circles) and output samples (X's). Right: error probability vs. various
SNRs {[}dB{]}. \label{fig:sim-1}}
\end{figure}
\end{enumerate}

\section{Summary and Future Research\label{sec:Future-Research}}

In this section, we summarize the main results of the paper, point
out some open problems and propose ideas for future research.

In Sec. \ref{sec:Maximum-Margin-by} we have derived an RLM problem
for learning a maximum margin NN decoder. We have developed a convex
lower bound to the maximum margin problem which can be efficiently
solved. Specifically, in step 2 we used a lower bound in order to
simplify the difficult, non-convex in general, problem. A goal for
future research is to consider methods to tighten this lower bound.
For example, such a method can be based on elimination of irrelevant
codeword-pairs, e.g., far apart pairs connected with a line in the
same direction as the line connecting a closer pair. In the derivation
for the non-linear Gaussian noise channel, a linearization of the
decoder was proposed as a part of the derivation of the minimum norm
problem. Future research could study other derivations of minimum
norm formulations, which are suitable for non-linear classifiers.

Next, in Sec. \ref{sec:Generalization-Error-Bounds}, we have proved
an expected generalization error bound for the error probability loss.
The stated generalization error bound holds only on average, and a
goal for future research is to prove a generalization bound which
holds with high probability. Such generalization bounds were recently
proved for SVM in \cite{gronlund2020near}, possibly the same methods
could be used for this problem. Another possibility for improving
the generalization bound is by making assumptions on the noise distribution,
similar to what was established for vector quantizers in \cite{linder2000training,levrard2013fast,antos2005improved,antos2005individual}.

For channels with complicated non-linear transformations, a linear
kernel might perform poorly. Nonetheless, it is possible to enrich
the expressive power of the NN decoder, by first mapping the codewords
into a high-dimensional feature space \cite[Ch. 16]{shalev2014understanding}.
If such a mapping is applied to the codebook, then the matrices dimensions
will increase accordingly, $\mathcal{H}=\mathbb{R}^{d_{y}\times D_{x}}$
and $\mathcal{K}=\mathbb{S}^{D_{x}\times D_{x}}$. Therefore, a hypothesis
will be determined by $O\left(d_{y}D_{x}+D_{x}^{2}\right)$ parameters.
Furthermore, the complexity of every operation involving codewords
would increase significantly. These computations become infeasible
when the mapping is to an infinite dimension feature space. As well
known, in order to alleviate this dimensionality problem, one can
use the \emph{kernel trick }\cite[Ch. 16.2]{shalev2014understanding}
for the non-linear channel model. In this regard, it can be shown
that the RLM and its solution, the decoder, can be expressed solely
by the kernel function and the samples. However, the constraint $H^{T}H\preccurlyeq K$
cannot be easily expressed in terms of the kernel. It is a goal for
future research to solve this problem and enable the use of a kernel
trick in this problem.

In Sec. \ref{sec:SGD}, we proposed a stochastic sub-gradient descent
algorithm for solving the RLM problem. This algorithm involves a projection
step in each iteration. While the complexity of the projection is
independent of the sample size, it still poses a hard computational
requirement for high-dimensional problems. The projection step in
the non-linear channel model stems from the linearization of the decoder,
and therefore it might be unnecessary for alternative variants of
minimum norm problems.

Finally, a different direction for future research is exploring the
margin properties of DNNs in the context of the decoder learning problem.
For classification problems, it was recently identified that the large
margin principle is beneficial to the generalization ability of DNNs
(see, e.g., the survey \cite{guo2021recent} and references therein).
It is of interest to identify similar properties for the decoder learning
problem.

\appendices{}

\section{Derivations for Section \ref{sec:Maximum-Margin-by}\label{sec:Formulation Proofs}}

In this appendix, we provide the proofs for the derivation of the
RLM for the additive noise channel (Appendix \ref{subsec:Additive proofs}),
and provide the development steps and proofs for the non-linear channel
(Appendix \ref{subsec:Non linear Dev}).

\subsection{Proofs for the Additive Noise Channel RLM problem \label{subsec:Additive proofs}}

We start by justifying the margin based objective (step 1) by proving
claim \ref{maximum margin form claim}.
\begin{IEEEproof}[Proof of Claim \ref{maximum margin form claim}]
Consider the decision regions of a given codeword-pair $x_{p},x_{q}\in C$
\[
\left\Vert x_{p}-y\right\Vert _{S}\stackrel[p]{q}{\gtrless}\left\Vert x_{q}-y\right\Vert _{S}\iff\left(y-\frac{1}{2}\left(x_{p}+x_{q}\right)\right)^{T}S\delta_{pq}\stackrel[q]{p}{\gtrless}0.
\]
Note that the factor $-\frac{1}{2}(x_{p}+x_{q})$ centers the codebook
and accordingly the subset $\boldsymbol{D}_{p}\bigcup\boldsymbol{D}_{q}$
of that dataset. We remain with a linear classifier $\bar{y}S\delta_{pq}\stackrel[q]{p}{\gtrless}0$
for the centered samples $\bar{y}\triangleq y-\frac{1}{2}(x_{p}+x_{q})$.
By using \cite[Claim 15.1]{shalev2014understanding} we get that the
distance between a point $v$ and the hyperplane with normal vector
$S\delta_{pq}$ is
\[
\frac{\left|v^{T}S\delta_{pq}\right|}{\left\Vert S\delta_{pq}\right\Vert }.
\]
Therefore, the margin induced by $S$ for the codeword-pair $p,q$
is
\[
\min_{i\in\boldsymbol{D}_{p}\cup\boldsymbol{D}_{q}}\frac{\left|\left(y_{i}-\frac{1}{2}\left(x_{p}+x_{q}\right)\right)^{T}S\delta_{pq}\right|}{\left\Vert S\delta_{pq}\right\Vert }.
\]
Taking the minimum over all codeword-pairs and maximizing over $S$
gives
\[
\max_{S\in\mathbb{S}_{+}}\min_{1\le p<q\le m}\min_{i\in\boldsymbol{D}_{p}\cup\boldsymbol{D}_{q}}\frac{\left|\left(y_{i}-\frac{1}{2}\left(x_{p}+x_{q}\right)\right)^{T}S\delta_{pq}\right|}{\left\Vert S\delta_{pq}\right\Vert }.
\]
Under the assumption of separability we can equivalently write 
\[
\max_{S\in\mathbb{S}_{+}}\min_{1\le p<q\le m}\min_{i\in\boldsymbol{D}_{p}\cup\boldsymbol{D}_{q}}\left(-1\right)^{\mathbb{I}\left(i\in\boldsymbol{D}_{q}\right)}\frac{\left(y_{i}-\frac{1}{2}\left(x_{p}+x_{q}\right)\right)^{T}S\delta_{pq}}{\left\Vert S\delta_{pq}\right\Vert }.
\]
\end{IEEEproof}
We proceed to show the lower bound validity (step 2) by proving claim
\ref{lower bound claim}.
\begin{IEEEproof}[Proof of Claim \ref{lower bound claim}]
Due to invariance to scaling of $S$ in (\ref{eq:margin objective}),
we can add the constraint $\max_{1\le p<q\le m}\left\Vert S\delta_{pq}\right\Vert \le1$
(from (\ref{eq:lower bound margin objective})), and get an equivalent
problem. Under this constraint
\begin{multline}
\max_{S\in\mathbb{S}_{+}}\min_{1\le p<q\le m}\min_{i\in\boldsymbol{D}_{p}\cup\boldsymbol{D}_{q}}\left(-1\right)^{\mathbb{I}\left(i\in\boldsymbol{D}_{q}\right)}\left(y_{i}-\frac{1}{2}\left(x_{p}+x_{q}\right)\right)^{T}S\delta_{pq}\\
\le\max_{S\in\mathbb{S}_{+}}\min_{1\le p<q\le m}\min_{i\in\boldsymbol{D}_{p}\cup\boldsymbol{D}_{q}}\left(-1\right)^{\mathbb{I}\left(i\in\boldsymbol{D}_{q}\right)}\frac{\left(y_{i}-\frac{1}{2}\left(x_{p}+x_{q}\right)\right)^{T}S\delta_{pq}}{\left\Vert S\delta_{pq}\right\Vert },
\end{multline}
which directly leads to the claim.
\end{IEEEproof}
Now we turn to prove the equivalent minimum norm formulation (step
3). This is done by proving Lemma \ref{equiv min norm lemma}.
\begin{IEEEproof}[Proof of Lemma \ref{equiv min norm lemma}]
Let $S^{*}$ be a solution of (\ref{eq:lower bound margin objective})
and $\gamma^{*}$ be its margin value, i.e., 
\[
\gamma^{*}=\min_{1\le p<q\le m}\min_{i\in\boldsymbol{D}_{p}\cup\boldsymbol{D}_{q}}a_{pqi}^{T}S^{*}\delta_{pq}.
\]
Let $S^{\#}$ be the solution to (\ref{eq:minimum norm form}) and
denote 
\[
\hat{S}\triangleq\frac{S^{\#}}{\max_{1\le p<q<m}\left\Vert S^{\#}\delta_{pq}\right\Vert }.
\]
We will show that $\hat{S}$ is a solution to (\ref{eq:lower bound margin objective}).

First, we show that $\hat{S}$ achieves a margin value of at least
$\gamma^{*}$. By definition of $\gamma^{*}$ we have that for all
$1\le p<q\le m,i\in\boldsymbol{D}_{p}\cup\boldsymbol{D}_{q}$ it holds
that $a_{pqi}^{T}\frac{S^{*}}{\gamma^{*}}\delta_{pq}\ge1$. Hence,
$\frac{S^{*}}{\gamma^{*}}$ satisfies the constraint in (\ref{eq:minimum norm form}),
combined with $S^{\#}$ being a minimizer for (\ref{eq:minimum norm form})
it yields
\begin{equation}
\max_{1\le p<q\le m}\left\Vert S^{\#}\delta_{pq}\right\Vert ^{2}\le\max_{1\le p<q\le m}\left\Vert \frac{S^{*}}{\gamma^{*}}\delta_{pq}\right\Vert ^{2}.\label{eq:norm compare}
\end{equation}
Using this, we can bound 
\begin{equation}
\max_{1\le p<q\le m}\left\Vert S^{\#}\delta_{pq}\right\Vert \overset{\eqref{eq:norm compare}}{\le}\max_{1\le p<q\le m}\left\Vert \frac{S^{*}}{\gamma^{*}}\delta_{pq}\right\Vert =\frac{1}{\gamma^{*}}\max_{1\le p<q\le m}\left\Vert S^{*}\delta_{pq}\right\Vert \le\frac{1}{\gamma^{*}},\label{eq:gamma_star inequality-1}
\end{equation}
where the last inequality is due to $S^{*}$ satisfying the constraint
$\left\Vert S^{*}\delta_{pq}\right\Vert \le1$ for all $1\le p<q\le m,i\in\boldsymbol{D}_{p}\cup\boldsymbol{D}_{q}$.
It follows that for all $1\le p<q\le m,i\in\boldsymbol{D}_{p}\cup\boldsymbol{D}_{q}$
\[
a_{pqi}^{T}\hat{S}\delta_{pq}=\frac{1}{\max_{1\le p<q\le m}\left\Vert S^{\#}\delta_{pq}\right\Vert }a_{pqi}^{T}S^{\#}\delta_{pq}\overset{(a)}{\ge}\frac{1}{\max_{1\le p<q\le m}\left\Vert S^{\#}\delta_{pq}\right\Vert }\overset{(b)}{\ge}\gamma^{*},
\]
where $(a)$ is due to $S^{\#}$ satisfying the constraint $a_{pqi}^{T}S\delta_{pq}\ge1$
for all $1\le p<q\le m,i\in\boldsymbol{D}_{p}\cup\boldsymbol{D}_{q}$,
and $(b)$ follows from (\ref{eq:gamma_star inequality-1}).

Second, $\hat{S}$ complies with the constraints of (\ref{eq:lower bound margin objective}),
since, by definition
\[
\max_{1\le p<q\le m}\left\Vert \hat{S}\delta_{pq}\right\Vert =\max_{1\le p<q<m}\left\Vert \frac{S^{\#}}{\max_{0\le p<q\le m}\left\Vert S^{\#}\delta_{pq}\right\Vert }\delta_{pq}\right\Vert =1.
\]
Notice that due to invariance to scaling, in a NN decoder, using $S^{\#}$
is equivalent to using $\hat{S}$. Maximizing $\left\Vert S\delta_{pq}\right\Vert $
is equivalent to maximizing $\left\Vert S\delta_{pq}\right\Vert ^{2}$.
This argument completes the proof.
\end{IEEEproof}
We next consider step 4, and show that relaxing the separability assumption
of a minimum norm problem leads to an RLM problem. To this end, we
allow the constraints in the minimum norm problem (\ref{eq:minimum norm form})
to be violated by adding non-negative slack variables $\{\xi_{pqi}\}$,
one for each sample. We penalize for violations by minimizing over
the slack variables and add $\lambda$ as a regularization parameter
that controls the tradeoff between the two terms. This leads to the
following optimization problem
\begin{equation}
\begin{array}{ccc}
\min_{S\in\mathbb{S}_{+}}\min_{\{\xi_{pqi}\}} & \lambda\max_{0\le p<q<m}\left\Vert S\delta_{pq}\right\Vert ^{2}+\sum_{\substack{1\le p<q\le m\\
i\in\boldsymbol{D}_{p}\cup\boldsymbol{D}_{q}
}
}\xi_{pqi}\\
\st & a_{pqi}^{T}S\delta_{pq}\ge1-\xi_{pqi} & \forall\,1\le p<q\le m,\,i\in\boldsymbol{D}_{p}\cup\boldsymbol{D}_{q}.\\
 & \xi_{pqi}\ge0 & \forall\,1\le p<q\le m,\,i\in\boldsymbol{D}_{p}\cup\boldsymbol{D}_{q}
\end{array}\label{eq:with literaly slack vars}
\end{equation}
For any $1\le p<q\le m,i\in\boldsymbol{D}_{p}\cup\boldsymbol{D}_{q}$
either $a_{pqi}^{T}S\delta_{pq}\ge1$ and then $\xi_{pqi}=0$, or
$a_{pqi}^{T}S\delta_{pq}<1$ and then $\xi_{pqi}=1-a_{pqi}^{T}S\delta_{pq}$.
Substituting $\xi_{pqi}=\max\{0,1-a_{pqi}^{T}S\delta_{pq}\}$ in (\ref{eq:with literaly slack vars})
gives (\ref{eq:not stable RLM}).

\subsection{Development Steps and Proofs for The Non-linear White Gaussian Channel\label{subsec:Non linear Dev}}

We next develop an RLM problem from a maximum margin approach for
the non-linear white Gaussian noise channel. The derivation follows
the same general steps as the ones for the additive noise channel
in Sec. \ref{subsec:The-Additive-RLM}, and so we omit details whenever
they are similar to that of Sec. \ref{subsec:The-Additive-RLM}. Nonetheless,
we highlight the step of developing a minimum norm formulation (Step
3), since it is delicate due to the quadratic form of the decoder,
and so requires an additional approximation step.

\paragraph*{Step 1 \textendash{} maximization of the minimum margin}

As before, we begin with the assumption that the dataset $\boldsymbol{D}$
is \emph{separable}, which means here that there exists a decoder
$H$ that achieves zero loss over $\boldsymbol{D}$, and this assumption
will be relaxed in the following steps. Now, the learner's goal is
to find a decoder $H$ that maximizes the minimum margin, over all
transformed codeword-pairs $Hx_{p},Hx_{q}\in C$. This margin-maximization
problem is given as follows:
\begin{claim}
\label{maximum margin form claim-1}The maximum margin induced by
a decoder $H$ is
\begin{equation}
\max_{H\in\mathcal{H}}\min_{i\in[n]}\min_{j'\in[m]\backslash\{j_{i}\}}\frac{y_{i}^{T}H\delta_{j_{i}j'}-\frac{1}{2}\left(x_{j_{i}}+x_{j'}\right)^{T}H^{T}H\delta_{j_{i}j'}}{\left\Vert H\delta_{j_{i}j'}\right\Vert }.\label{eq:margin objective-1}
\end{equation}
\end{claim}
\begin{IEEEproof}
For every codeword-pair $x_{p},x_{q}\in C$ the decision regions are
\[
\left\Vert y-Hx_{p}\right\Vert \stackrel[p]{q}{\gtrless}\left\Vert y-Hx_{q}\right\Vert \iff y^{T}H\delta_{pq}-\frac{1}{2}\left(x_{p}+x_{q}\right)^{T}H^{T}H\delta_{pq}\stackrel[q]{p}{\gtrless}0.
\]
Using analogous arguments to those used in the proof of Claim \ref{maximum margin form claim}
completes the proof. 
\end{IEEEproof}

\paragraph*{Step 2 \textendash{} a lower bound}

The problem (\ref{eq:margin objective-1}) is not necessarily convex
and is hard to solve directly. Therefore, we proceed to maximize the
following lower bound on its value. Unlike the analogous lower bound
in Sec. \ref{subsec:The-Additive-RLM}, finding this lower bound is
not a convex optimization problem, but it still serves as a useful
step for next derivations. We denote $r_{x}=\max_{j\in[m]}\|x_{j}\|$
and $R_{x}=\max_{j\in[m]}\|f(x_{j})\|$.
\begin{claim}
\label{lower bound claim-1}The value of the problem
\begin{equation}
\begin{array}{cc}
\max_{H\in\mathcal{H}}\min_{i\in\left[n\right]}\min_{j'\in\left[m\right]\backslash\left\{ j_{i}\right\} } & y_{i}^{T}H\delta_{j_{i}j'}-\frac{1}{2}\left(x_{j_{i}}+x_{j'}\right)^{T}H^{T}H\delta_{j_{i}j'}\\
\st\,\, & \max_{1\le p<q\le m}\left\Vert H\delta_{pq}\right\Vert \le\sqrt{\frac{r_{x}^{2}}{r_{x}^{2}+R_{x}^{2}}}
\end{array}\label{eq:lower bound margin objective-1}
\end{equation}
is a lower bound on the value of (\ref{eq:margin objective-1}).
\end{claim}
The proof is similar to the proof for claim \ref{lower bound claim}.
We note, however, that since the current considered class of decoders
is not scale-invariant, we suspect that this lower bound might be
less tight in general. So tightening of this bound is an interesting
open problem.
\begin{IEEEproof}
By adding the constraint 
\[
\max_{1\le p<q\le m}\left\Vert H\delta_{pq}\right\Vert \le\sqrt{\frac{r_{x}^{2}}{r_{x}^{2}+R_{x}^{2}}},
\]
 we effectively restrict the hypothesis class without increasing the
value of the problem. This is since $\sqrt{\frac{r_{x}^{2}}{r_{x}^{2}+R_{x}^{2}}}<1$
and so under this additional constraint
\begin{multline}
\max_{\left(H,K\right)\in\mathcal{H}\times\mathcal{K}}\min_{i\in\left[n\right]}\min_{j'\in\left[m\right]\backslash\left\{ j_{i}\right\} }y_{i}^{T}H\delta_{j_{i}j'}-\frac{1}{2}\left(x_{j_{i}}+x_{j'}\right)^{T}H^{T}H\delta_{j_{i}j'}\\
\le\max_{\left(H,K\right)\in\mathcal{H}\times\mathcal{K}}\min_{i\in\left[n\right]}\min_{j'\in\left[m\right]\backslash\left\{ j_{i}\right\} }\frac{y_{i}^{T}H\delta_{j_{i}j'}-\frac{1}{2}\left(x_{j_{i}}+x_{j'}\right)^{T}H^{T}H\delta_{j_{i}j'}}{\left\Vert H\delta_{j_{i}j'}\right\Vert }.
\end{multline}
This directly leads to the claim.
\end{IEEEproof}

\paragraph*{Step 3 \textendash{} minimum norm formulation}

As we have seen, obtaining an RLM problem and the removal of the separability
assumption is done via a minimum norm formulation. In order to obtain
a minimum norm formulation, the objective function should be linear
in the parameter determining the decoder, yet (\ref{eq:lower bound margin objective-1})
violates this requirement. We therefore resort to the the following
approximation strategy. First, we write $K=H^{T}H\in\mathcal{K}\triangleq\mathbb{S}_{+}^{d_{x}}$
and thus linearize the decoder's parametrization. We then drop this
constraint, yet add a judicious constraint on $\max_{1\le p<q\le m}\left\Vert H\delta_{pq}\right\Vert ^{2}+\max_{1\le p'<q'\le m}\left\Vert K\delta_{p'q'}\right\Vert ^{2}$,
which replaces the constraint $\max_{1\le p<q\le m}\|H\delta_{pq}\|\le\sqrt{\frac{r_{x}^{2}}{r_{x}^{2}+R_{x}^{2}}}$
in (\ref{eq:lower bound margin objective-1}). With this approximation,
we then convert the problem (\ref{eq:lower bound margin objective-1})
to a minimum norm problem. Then, since the learned decoder will eventually
use a learned linear kernel $H$ and will (implicitly) set $K=H^{T}H$,
we re-introduce the constraint $K=H^{T}H$, but in a convex-relaxed
form $H^{T}H\preceq K$.

Specifically, we choose the constraint on $\max_{1\le p<q\le m}\left\Vert H\delta_{pq}\right\Vert ^{2}+\max_{1\le p'<q'\le m}\left\Vert K\delta_{p'q'}\right\Vert ^{2}$
as follows: Since the channel transformation $f$ increases the norm
of $x$ by a factor of at most $R_{x}/r_{x}$ it is reasonable to
bound $\lambda_{\max}(H)\le\frac{R_{x}}{r_{x}}$.\footnote{We however do not explicitly add this constraint on $H$, but rather
use it to develop a constraint on $K$.} Now, the constraint on $H$ in (\ref{eq:lower bound margin objective-1})
implies that
\[
\max_{1\le p<q\le m}\delta_{pq}^{T}K\delta_{pq}=\max_{1\le p<q\le m}\left\Vert H\delta_{pq}\right\Vert ^{2}\le\frac{r_{x}^{2}}{r_{x}^{2}+R_{x}^{2}}.
\]
Then, 
\[
\max_{1\le p<q\le m}\left\Vert K\delta_{pq}\right\Vert ^{2}\le\lambda_{\max}(K)\cdot\max_{1\le p<q\le m}\delta_{pq}^{T}K\delta_{pq}\le\frac{R_{x}^{2}}{r_{x}^{2}+R_{x}^{2}},
\]
which holds since $K\in{\cal K}$ is symmetric. By replacing the constraint
in (\ref{eq:lower bound margin objective-1}), we get the following
problem
\begin{equation}
\begin{array}{cc}
\max_{(H,K)\in\mathcal{H}\times\mathcal{K}}\min_{i\in\left[n\right]}\min_{j'\in\left[m\right]\backslash\left\{ j_{i}\right\} } & y_{i}^{T}H\delta_{j_{i}j'}-\frac{1}{2}\left(x_{j_{i}}+x_{j'}\right)^{T}K\delta_{j_{i}j'}\\
\st\,\, & \max_{1\le p<q\le m}\left\Vert H\delta_{pq}\right\Vert ^{2}+\max_{1\le p'<q'\le m}\left\Vert K\delta_{p'q'}\right\Vert ^{2}\le1.
\end{array}\label{eq:lower bound with S const}
\end{equation}
The formulation (\ref{eq:lower bound with S const}) then leads to
a minimum norm problem, as desired.
\begin{lem}
\label{equiv min norm lemma-1}Every solution to the following minimum
norm problem
\begin{equation}
\begin{array}{cc}
\min_{(H,K)\in\mathcal{H}\times\mathcal{K}} & \max_{\substack{1\le p<q\le m\\
1\le p'<q'\le m
}
}\left\Vert H\delta_{pq}\right\Vert ^{2}+\left\Vert K\delta_{p'q'}\right\Vert ^{2}\\
\st\,\, & \min_{i\in[n]}\min_{j'\in[m]\backslash\{j_{i}\}}y_{i}^{T}H\delta_{j_{i}j'}-\frac{1}{2}\left(x_{j_{i}}+x_{j'}\right)^{T}K\delta_{j_{i}j'}\ge1,
\end{array}\label{eq:minimum norm form-1}
\end{equation}
is a solution to (\ref{eq:lower bound with S const}).
\end{lem}
The proof of Lemma \ref{equiv min norm lemma-1} shares similar structure
to the proof of Lemma \ref{equiv min norm lemma} but it is more delicate,
due to the more complicated parametrization of the linearized decoder.
\begin{IEEEproof}
Let $(H^{*},K^{*})$ be a solution of (\ref{eq:lower bound with S const})
and $\gamma^{*}$ its margin value, i.e., 
\[
\gamma^{*}=\min_{i\in[n]}\min_{j'\in[m]\backslash\{j_{i}\}}y_{i}^{T}H\delta_{j_{i}j'}-\frac{1}{2}\left(x_{j_{i}}+x_{j'}\right)^{T}K\delta_{j_{i}j'}.
\]
Let $(H^{\#},K^{\#})$ be a solution to (\ref{eq:minimum norm form-1})
and denote the corresponding normalized solution
\[
\hat{H}\triangleq\frac{H^{\#}}{\sqrt{\max_{1\le p<q\le m}\left\Vert H^{\#}\delta_{pq}\right\Vert ^{2}+\max_{1\le p'<q'\le m}\left\Vert K^{\#}\delta_{p'q'}\right\Vert ^{2}}},
\]
and 
\[
\hat{K}\triangleq\frac{K^{\#}}{\sqrt{\max_{1\le p<q\le m}\left\Vert H^{\#}\delta_{pq}\right\Vert ^{2}+\max_{1\le p'<q'\le m}\left\Vert K^{\#}\delta_{p'q'}\right\Vert ^{2}}}.
\]
We will show that $(\hat{H},\hat{K})$ is a solution to (\ref{eq:lower bound with S const}).

First, we show that $(\hat{H},\hat{K})$ achieves a margin value of
at least $\gamma^{*}$. By definition of $\gamma^{*}$ we have that
for all $i\in[n],j'\in[m]\backslash\{j_{i}\}$:
\[
y_{i}^{T}\frac{H^{*}}{\gamma^{*}}\delta_{j_{i}j'}-\frac{1}{2}\left(x_{j_{i}}+x_{j'}\right)^{T}\frac{K^{*}}{\gamma^{*}}\delta_{j_{i}j'}\ge1.
\]
Hence, $(\frac{H^{*}}{\gamma^{*}},\frac{K^{*}}{\gamma^{*}})$ satisfies
the constraint in (\ref{eq:minimum norm form-1}). Combined with $(H^{\#},K^{\#})$
being a minimizer for (\ref{eq:minimum norm form-1}), it yields
\begin{equation}
\max_{1\le p<q\le m}\left\Vert H^{\#}\delta_{pq}\right\Vert ^{2}+\max_{1\le p'<q'\le m}\left\Vert K^{\#}\delta_{p'q'}\right\Vert ^{2}\le\max_{1\le p<q\le m}\left\Vert \frac{H^{*}}{\gamma^{*}}\delta_{pq}\right\Vert ^{2}+\max_{1\le p'<q'\le m}\left\Vert \frac{K^{*}}{\gamma^{*}}\delta_{p'q'}\right\Vert ^{2}.\label{eq:norm compare-1}
\end{equation}
Using this we can bound 
\begin{align}
\max_{1\le p<q\le m}\left\Vert H^{\#}\delta_{pq}\right\Vert ^{2}+\max_{1\le p'<q'\le m}\left\Vert K^{\#}\delta_{p'q'}\right\Vert ^{2} & \overset{\eqref{eq:norm compare-1}}{\le}\max_{1\le p<q\le m}\left\Vert \frac{H^{*}}{\gamma^{*}}\delta_{pq}\right\Vert ^{2}+\max_{1\le p'<q'\le m}\left\Vert \frac{K^{*}}{\gamma^{*}}\delta_{p'q'}\right\Vert ^{2}\label{eq:gamma_star inequality-1-1}\\
 & =\left(\frac{1}{\gamma^{*}}\right)^{2}\max_{1\le p<q\le m}\left\Vert H^{*}\delta_{pq}\right\Vert ^{2}+\left(\frac{1}{\gamma^{*}}\right)^{2}\max_{1\le p'<q'\le m}\left\Vert K^{*}\delta_{p'q'}\right\Vert ^{2}\\
 & \le\left(\frac{1}{\gamma^{*}}\right)^{2},
\end{align}
where the last inequality is due to $(H^{*},K^{*})$ satisfying the
constraints of (\ref{eq:lower bound with S const}). It follows that
for all $i\in[n],j'\in[m]\backslash\{j_{i}\}$:
\begin{align}
y_{i}^{T}\hat{H}\delta_{j_{i}j'}-\frac{1}{2}\left(x_{j_{i}}+x_{j'}\right)^{T}\hat{K}\delta_{j_{i}j'} & =\frac{y_{i}^{T}H^{\#}\delta_{j_{i}j'}-\frac{1}{2}\left(x_{j_{i}}+x_{j'}\right)^{T}K^{\#}\delta_{j_{i}j'}}{\sqrt{\max_{1\le p<q\le m}\left\Vert H^{\#}\delta_{pq}\right\Vert ^{2}+\max_{1\le p'<q'\le m}\left\Vert K^{\#}\delta_{p'q'}\right\Vert ^{2}}}\\
 & \overset{(a)}{\ge}\frac{1}{\sqrt{\max_{1\le p<q\le m}\left\Vert H^{\#}\delta_{pq}\right\Vert ^{2}+\max_{1\le p'<q'\le m}\left\Vert K^{\#}\delta_{p'q'}\right\Vert ^{2}}}\\
 & \overset{(b)}{\ge}\gamma^{*},
\end{align}
where $(a)$ is due to $(H^{\#},K^{\#})$ satisfying the constraint
\[
y_{i}^{T}H^{\#}\delta_{j_{i}j'}-\frac{1}{2}\left(x_{j_{i}}+x_{j'}\right)^{T}K^{\#}\delta_{j_{i}j'}\ge1
\]
for all $i\in[n],j'\in[m]\backslash\{j_{i}\}$, and $(b)$ follows
from (\ref{eq:gamma_star inequality-1-1}).

Second, we show that $(\hat{H},\hat{K})$ complies with the constraints
of (\ref{eq:lower bound with S const}). By definition
\[
\max_{1\le p<q\le m}\left\Vert \hat{H}\delta_{pq}\right\Vert ^{2}+\max_{1\le p'<q'\le m}\left\Vert \hat{K}\delta_{p'q'}\right\Vert ^{2}=\frac{\max_{1\le p<q\le m}\left\Vert H^{\#}\delta_{pq}\right\Vert ^{2}+\max_{1\le p'<q'\le m}\left\Vert K^{\#}\delta_{p'q'}\right\Vert ^{2}}{\max_{1\le p<q\le m}\left\Vert H^{\#}\delta_{pq}\right\Vert ^{2}+\max_{1\le p'<q'\le m}\left\Vert K^{\#}\delta_{p'q'}\right\Vert ^{2}}=1.
\]
As evident from (\ref{eq:margin objective-1}) with $H^{T}H$ replaced
by $K$, the problem is scale invariant and so using $(H^{\#},K^{\#})$
is equivalent to using $(\hat{H},\hat{K})$.
\end{IEEEproof}
As said, after obtaining a minimum norm formulation (\ref{eq:minimum norm form-1})
we add the convex constraint $H^{T}H\preceq K$.

\paragraph*{Step 4 \textendash{} relaxation of the separability assumption}

Next, we introduce slack variables in order to relax the assumption
that the dataset $\boldsymbol{D}$ is separable. To this end, we allow
the constraints in the minimum norm problem to be violated by adding
non-negative slack variables $\{\xi_{ij'}\}$, one for each sample.
We penalize for violations by minimizing over the slack variables
and add $\lambda$ as a regularization parameter. This leads to the
following optimization problem

\begin{equation}
\begin{array}{cc}
\min_{\left(H,K\right)\in\mathcal{H}\times\mathcal{K}}\min_{\left\{ \xi\right\} } & \lambda\left[\max_{1\le p<q\le m}\left\Vert H\delta_{pq}\right\Vert ^{2}+\max_{1\le p'<q'\le m}\left\Vert K\delta_{p'q'}\right\Vert ^{2}\right]+\sum_{i\in\left[n\right]}\sum_{j'\in\left[m\right]\backslash\left\{ j_{i}\right\} }\xi_{ij'}\\
\st & y_{i}^{T}H\delta_{j_{i}j'}-\frac{1}{2}\left(x_{j_{i}}+x_{j'}\right)^{T}K\delta_{j_{i}j'}\ge1-\xi_{ij'}\\
 & \xi_{ij'}\ge0
\end{array}.\label{eq:with literaly slack vars-1}
\end{equation}
For any $i\in[n],j'\in[m]\backslash\{j_{i}\}$ either $y_{i}^{T}H\delta_{j_{i}j'}-\frac{1}{2}(x_{j_{i}}+x_{j'})^{T}K\delta_{j_{i}j'}\ge1$
and then $\xi_{ij'}=0$, or $y_{i}^{T}H\delta_{j_{i}j'}-\frac{1}{2}(x_{j_{i}}+x_{j'})^{T}K\delta_{j_{i}j'}<1$
and then $\xi_{ij'}=1-[y_{i}^{T}H\delta_{j_{i}j'}-\frac{1}{2}(x_{j_{i}}+x_{j'})^{T}K\delta_{j_{i}j'}]$.
By substituting 
\[
\xi_{ij'}=\max\left\{ 0,1-\left[y_{i}^{T}H\delta_{j_{i}j'}-\frac{1}{2}\left(x_{j_{i}}+x_{j'}\right)^{T}K\delta_{j_{i}j'}\right]\right\} 
\]
in (\ref{eq:with literaly slack vars-1}), the problem (\ref{eq:with literaly slack vars-1})
is equivalent to the RLM problem for the hinge loss function $\mathring{\ell}^{\text{hinge}}(H,K,i)$
in (\ref{eq:hinge loss as margin}). The RLM problem is then given
by
\begin{equation}
\begin{array}{cc}
\min_{\left(H,K\right)\in\mathcal{H}\times\mathcal{K}} & \mathring{L}_{\boldsymbol{D}}^{\text{hinge}}\left(H,K\right)+\lambda\left[\max_{1\le p<q\le m}\left\Vert H\delta_{pq}\right\Vert ^{2}+\max_{1\le p'<q'\le m}\left\Vert K\delta_{p'q'}\right\Vert ^{2}\right]\\
\st\,\, & H^{T}H\preceq K,
\end{array}\label{eq:not stable RLM-1}
\end{equation}
where $\mathring{L}_{\boldsymbol{D}}^{\text{hinge}}(H,K)$ is as in
(\ref{eq:hinge as margin-1}), and $\lambda>0$ is a regularization
parameter.

\paragraph*{Step 5 \textendash{} inducing stability by a generalization of the
regularization}

The regularization function is similar to the one from Sec. \ref{subsec:The-Additive-RLM}.
In the same manner we modify the learning rule to a stable one by
considering a proper partition. The final RLM rule for finding a maximum
minimum margin decoder is defined for a given set of positive parameters
$\{\eta_{i}\}_{i\in\left[d+1\right]}$ which satisfy $\sum_{i=1}^{d+1}\eta_{i}=1$,
and a proper partition $\{P_{j}\}_{j\in\left[d+1\right]}$, as
\begin{equation}
\begin{array}{cc}
A\left(\boldsymbol{D}\right)= & \argmin_{\left(H,K\right)\in\mathcal{H}\times\mathcal{K}}\mathring{L}_{\boldsymbol{D}}^{\text{hinge}}\left(H,K\right)+\lambda\sum_{i=1}^{d+1}\eta_{i}\left[\max_{j\in P_{i}}\left\Vert H\delta_{j}\right\Vert ^{2}+\max_{j'\in P_{i}}\left\Vert K\delta_{j'}\right\Vert ^{2}\right]\\
\st\,\, & H^{T}H\preceq K
\end{array}.\label{eq:stable RLM rule-1}
\end{equation}

\section{Proofs for Section \ref{sec:Generalization-Error-Bounds}\label{sec:Generalization proofs}}

Throughout this section we will use only Euclidean norms. For two
matrices $H$ and $K$ we will let
\[
\left\Vert \left(H,K\right)\right\Vert _{F}\triangleq\sqrt{\left\Vert H\right\Vert _{F}^{2}+\left\Vert K\right\Vert _{F}^{2}}.
\]

\subsection{Proof of Theorem \ref{thm: generalization theorem}\label{subsec:Expected Gen Proofs}}

The proof will use the strong convexity of the regularization function
and the convexity and Lipschitzness of the loss function. The following
lemma establishes the strong convexity constant of the regularization
function for the additive noise channel model.
\begin{lem}
\label{strong convexity lemma}Assume that $\Span\{\delta_{pq}\}_{1\le p\le q\le m}=\mathbb{R}^{d}$,
and let $\{P_{i}\}_{i\in\left[d+1\right]}$ be a proper partition.
Then, $g(S)\triangleq\sum_{i=1}^{d+1}\eta_{i}\max_{j\in P_{i}}\|S\delta_{j}\|^{2}$
is $2\gamma$-strongly convex w.r.t. the Frobenius norm, where
\begin{equation}
\gamma\triangleq\min_{S\in\mathbb{S}_{+}}\sum_{i=1}^{d+1}\eta_{i}\cdot\min_{j\in P_{i}}\frac{\left\Vert S\delta_{j}\right\Vert ^{2}}{\left\Vert S\right\Vert _{F}^{2}},\label{eq: definition of gamma strong convexity}
\end{equation}
and the lower bound
\[
\gamma=\min_{\left\{ j_{i}\in P_{i}\right\} _{i=1}^{d+1}}\lambda_{\text{\emph{min}}}\left(\sum_{i=1}^{d+1}\eta_{i}\delta_{j_{i}}\delta_{j_{i}}^{T}\right)\ge\eta_{\text{\emph{min}}}\min_{1\le p<q\le m}\left\Vert \delta_{pq}\right\Vert ^{2}
\]
holds with $\eta_{\text{\emph{min}}}\triangleq\min_{i\in\left[d+1\right]}\{\eta_{i}\}$.
\end{lem}
We prove Lemma \ref{strong convexity lemma} for an arbitrary matrix
$H\in\mathbb{R}^{d_{y}\times d_{x}}$, and specifically, the result
is applicable to any PSD matrix $S$.
\begin{IEEEproof}
We prove the claim by verifying strong convexity directly from its
definition (e.g., \cite[Definition 13.4]{shalev2014understanding}).
Let $\alpha\in[0,1]$ be given. Then,
\begin{align}
 & g\left(\alpha H_{1}+\left(1-\alpha\right)H_{2}\right)-\alpha g\left(H_{1}\right)-\left(1-\alpha\right)g\left(H_{2}\right)\nonumber \\
 & =\sum_{i=1}^{d+1}\eta_{i}\cdot\max_{j\in P_{i}}\left\Vert \left(\alpha H_{1}+\left(1-\alpha\right)H_{2}\right)\delta_{j}\right\Vert ^{2}\nonumber \\
 & \hphantom{=}-\alpha\sum_{i=1}^{d+1}\eta_{i}\cdot\max_{j'\in P_{i}}\left\Vert H_{1}\delta_{j'}\right\Vert ^{2}-\left(1-\alpha\right)\sum_{i=1}^{d+1}\eta_{i}\cdot\max_{j''\in P_{i}}\left\Vert H_{2}\delta_{j''}\right\Vert ^{2}\\
 & =\sum_{i=1}^{d+1}\eta_{i}\cdot\max_{j\in P_{i}}\left[\alpha^{2}\left\Vert H_{1}\delta_{j}\right\Vert ^{2}+2\alpha\left(1-\alpha\right)\left\langle H_{1}\delta_{j},H_{2}\delta_{j}\right\rangle +\left(1-\alpha\right)^{2}\left\Vert H_{2}\delta_{j}\right\Vert ^{2}\right]\nonumber \\
 & \hphantom{=}-\alpha\sum_{i=1}^{d+1}\eta_{i}\cdot\max_{j'\in P_{i}}\left\Vert H_{1}\delta_{j'}\right\Vert ^{2}-\left(1-\alpha\right)\sum_{i=1}^{d+1}\eta_{i}\cdot\max_{j''\in P_{i}}\left\Vert H_{2}\delta_{j''}\right\Vert ^{2}\\
 & \le\sum_{i=1}^{d+1}\eta_{i}\cdot\max_{j\in P_{i}}\left[\alpha^{2}\left\Vert H_{1}\delta_{j}\right\Vert ^{2}+2\alpha\left(1-\alpha\right)\left\langle H_{1}\delta_{j},H_{2}\delta_{j}\right\rangle +\left(1-\alpha\right)^{2}\left\Vert H_{2}\delta_{j}\right\Vert ^{2}\right.\nonumber \\
 & \hphantom{=}\left.-\alpha\left\Vert H_{1}\delta_{j}\right\Vert ^{2}-\left(1-\alpha\right)\left\Vert H_{2}\delta_{j}\right\Vert ^{2}\right]\\
 & =\sum_{i=1}^{d+1}\eta_{i}\cdot\max_{j\in P_{i}}\left[\alpha\left(\alpha-1\right)\left\Vert H_{1}\delta_{j}\right\Vert ^{2}+\alpha\left(\alpha-1\right)\left\Vert H_{2}\delta_{j}\right\Vert ^{2}-2\alpha\left(\alpha-1\right)\left\langle H_{1}\delta_{j},H_{2}\delta_{j}\right\rangle \right]\\
 & =\sum_{i=1}^{d+1}\eta_{i}\cdot\max_{j\in P_{i}}\alpha\left(\alpha-1\right)\left[\left\Vert H_{1}\delta_{j}\right\Vert ^{2}-2\left\langle H_{1}\delta_{j},H_{2}\delta_{j}\right\rangle +\left\Vert H_{2}\delta_{j}\right\Vert ^{2}\right]\\
 & =-\alpha\left(1-\alpha\right)\sum_{i=1}^{d+1}\eta_{i}\cdot\min_{j\in P_{i}}\left\Vert \left(H_{1}-H_{2}\right)\delta_{j}\right\Vert ^{2}\nonumber \\
 & \le-\alpha\left(1-\alpha\right)\gamma\left\Vert H_{1}-H_{2}\right\Vert _{F}^{2},
\end{align}
where the first inequality is by decreasing the absolute value of
the negative element, and where $\gamma$ is as defined in (\ref{eq: definition of gamma strong convexity})
(Lemma \ref{strong convexity lemma}).

Denote the singular value decomposition (SVD) of $H$ by $H\triangleq U\Lambda V^{T}$,
where $U\in\mathbb{R}^{d_{y}\times d_{y}},V\in\mathbb{R}^{d_{x}\times d_{x}}$
are unitary matrices and their columns are the left-singular vectors
and right-singular vectors respectively, and $\Lambda\in\mathbb{R}^{d_{y}\times d_{x}}$
is a diagonal matrix whose elements are the corresponding singular
values. Denote the main diagonal of $\Lambda^{2}$ by $\alpha$. Further
denote $\delta_{j}^{\left(V\right)}\triangleq V^{T}\delta_{j}$ and
its element-wise squaring by $d_{j}^{\left(V\right)}$. With these
definitions we can establish the following relation:
\begin{equation}
\left\Vert H\delta\right\Vert ^{2}=\delta^{T}V\Lambda U^{T}U\Lambda V^{T}\delta=\left\Vert \Lambda\delta^{\left(V\right)}\right\Vert ^{2}=d^{\left(V\right)^{T}}\alpha.\label{eq:squared norm of decom}
\end{equation}
Using the above, we bound $\gamma$ as follows
\begin{align}
\gamma=\min_{H\in\mathcal{H}}\sum_{i=1}^{d+1}\eta_{i}\min_{j\in P_{i}}\frac{\left\Vert H\delta_{j}\right\Vert ^{2}}{\left\Vert H\right\Vert _{F}^{2}} & =\min_{H\in\mathcal{H}:\left\Vert H\right\Vert _{F}=1}\sum_{i=1}^{d+1}\eta_{i}\min_{j\in P_{i}}\left\Vert H\delta_{j}\right\Vert ^{2}\\
 & =\min_{\left\{ j_{i}\in P_{i}\right\} _{i=1}^{d+1}}\min_{H\in\mathcal{H}:\left\Vert H\right\Vert _{F}=1}\sum_{i=1}^{d+1}\eta_{i}\left\Vert H\delta_{j_{i}}\right\Vert ^{2}\\
 & \overset{(a)}{=}\min_{\left\{ j_{i}\in P_{i}\right\} _{i=1}^{d+1}}\min_{\Lambda\in\text{diag}\left(\Delta^{d}\right),V:V^{T}=V^{-1}}\sum_{i=1}^{d+1}\eta_{i}\left\Vert \Lambda\delta_{j_{i}}^{\left(V\right)}\right\Vert ^{2}\\
 & \overset{\left(b\right)}{=}\min_{\left\{ j_{i}\in P_{i}\right\} _{i=1}^{d+1}}\min_{\alpha\in\Delta^{d},V:V^{T}=V^{-1}}\sum_{i=1}^{d+1}\eta_{i}d_{j_{i}}^{\left(V\right)^{T}}\alpha\\
 & =\min_{\left\{ j_{i}\in P_{i}\right\} _{i=1}^{d+1}}\min_{\alpha\in\Delta^{d},V:V^{T}=V^{-1}}\left(\sum_{i=1}^{d+1}\eta_{i}d_{j_{i}}^{\left(V\right)}\right)^{T}\alpha\\
 & \overset{\left(c\right)}{=}\min_{\left\{ j_{i}\in P_{i}\right\} _{i=1}^{d+1}}\min_{\alpha\in\Delta^{d},V:V^{T}=V^{-1}}\begin{pmatrix}\sum_{i=1}^{d+1}\eta_{i}\left(v_{1}^{T}\delta_{j_{i}}\right)^{2}\\
\vdots\\
\sum_{i=1}^{d+1}\eta_{i}\left(v_{d}^{T}\delta_{j_{i}}\right)^{2}
\end{pmatrix}^{T}\alpha\\
 & =\min_{\left\{ j_{i}\in P_{i}\right\} _{i=1}^{d+1}}\min_{V:V^{T}=V^{-1}}\min\left\{ \sum_{i=1}^{d+1}\eta_{i}\left(v_{k}^{T}\delta_{j_{i}}\right)^{2}\right\} _{k=1}^{d}\\
 & \overset{\left(d\right)}{=}\min_{\left\{ j_{i}\in P_{i}\right\} _{i=1}^{d+1}}\min_{v_{1}\in\mathbb{R}^{d}:\left\Vert v_{1}\right\Vert =1}\left(\sum_{i=1}^{d+1}\eta_{i}\left(v_{1}^{T}\delta_{j_{i}}\right)^{2}\right)\\
 & =\min_{\left\{ j_{i}\in P_{i}\right\} _{i=1}^{d+1}}\min_{v_{1}\in\mathbb{R}^{d}:\left\Vert v_{1}\right\Vert =1}\left(\sum_{i=1}^{d+1}\eta_{i}v_{1}^{T}\delta_{j_{i}}\delta_{j_{i}}^{T}v_{1}\right)\\
 & =\min_{\left\{ j_{i}\in P_{i}\right\} _{i=1}^{d+1}}\min_{v_{1}\in\mathbb{R}^{d}:\left\Vert v_{1}\right\Vert =1}\left(v_{1}^{T}\left(\sum_{i=1}^{d+1}\eta_{i}\delta_{j_{i}}\delta_{j_{i}}^{T}\right)v_{1}\right)\\
 & \overset{\left(e\right)}{=}\min_{\left\{ j_{i}\in P_{i}\right\} _{i=1}^{d+1}}\lambda_{\text{min}}\left(\sum_{i=1}^{d+1}\eta_{i}\delta_{j_{i}}\delta_{j_{i}}^{T}\right)\\
 & \ge\min_{\left\{ j_{i}\in P_{i}\right\} _{i=1}^{d+1}}\eta_{\text{min}}\lambda_{\text{min}}\left(\sum_{i=1}^{d+1}\delta_{j_{i}}\delta_{j_{i}}^{T}\right)\\
 & \ge\eta_{\text{min}}\cdot\min_{1\le p<q\le m}\left\Vert \delta_{pq}\right\Vert ^{2},
\end{align}
where $(a)$ follows from \ref{eq:squared norm of decom}, $(b)$
follows since $\sum_{k=1}^{d}\alpha(k)=\left\Vert H\right\Vert _{F}^{2}=1$
implies that $\alpha\in\Delta^{d}$, in $(c)$ $\{v_{k}\}_{k\in[d]}$
are the columns of $V$, and $(d)$ is due to $V^{T}$ and $\Lambda$
being decided up to row permutations, so we can always choose them
such that the maximal index $k\in[d]$ will be $k=1$. At $(d)$,
we note that due to the partition being proper, there is no $v_{1}$
for which $v_{1}^{T}\delta_{j_{i}}=0$ for all $i\in[d+1]$, so the
bound is positive. Otherwise, one could choose representatives $\{j_{i}\in P_{i}\}_{i=1}^{d+1}$
such that they are all perpendicular to some singular vector and the
bound will be zero. Finally, $(e)$ is due to the Rayleigh quotient
bounds $R(M,x)=x^{T}Mx/\|x\|^{2}\in[\lambda_{\text{min}}(M),\lambda_{\text{max}}(M)]$
for all $x\in\mathbb{R}^{d}$.
\end{IEEEproof}
The next lemma establishes the convexity and Lipschitz constant of
the loss function for the additive noise channel model. The proof
of this lemma will use the following claim.
\begin{claim}
\label{claim:dsym norm bound}Let $a,b\in\mathbb{R}^{d}$ then
\[
\left\Vert \dsym\left(ab^{T}\right)\right\Vert _{F}^{2}\le2\|a\|^{2}\|b\|^{2}+2\left\langle a,b\right\rangle ^{2}.
\]
\end{claim}
\begin{IEEEproof}
Denote $A\triangleq ab^{T}$. Then, using the properties of the trace
operation,
\begin{align}
\left\Vert \dsym\left(A\right)\right\Vert _{F}^{2} & =\left\Vert A+A^{T}-\diag\left(A\right)\right\Vert _{F}^{2}\\
 & =\left\langle A,A\right\rangle +\left\langle A,A^{T}\right\rangle +\left\langle A^{T},A\right\rangle +\left\langle A^{T},A^{T}\right\rangle \nonumber \\
 & \phantom{=}-\left\langle A,\diag\left(A\right)\right\rangle -\left\langle A^{T},\diag\left(A\right)\right\rangle +\left\langle \diag\left(A\right),\diag\left(A\right)\right\rangle \\
 & =2\left\Vert A\right\Vert _{F}^{2}+2\left\langle A,A^{T}\right\rangle +\left\Vert \diag\left(A\right)\right\Vert _{F}^{2}-2\left\langle A,\diag\left(A\right)\right\rangle \\
 & =2\left\Vert A\right\Vert _{F}^{2}+2\left\langle A,A^{T}\right\rangle -\left\Vert \diag\left(A\right)\right\Vert _{F}^{2}\\
 & \le2\left\Vert A\right\Vert _{F}^{2}+2\left\langle A,A^{T}\right\rangle ,
\end{align}
and substituting back we get
\[
\left\Vert \dsym\left(ab^{T}\right)\right\Vert _{F}^{2}\le2\|a\|^{2}\|b\|^{2}+2\left\langle a,b\right\rangle ^{2}.
\]
\end{IEEEproof}
\begin{lem}
\label{lipschitz lemma}$S\rightarrow\max\{0,1-a_{pqi}^{T}S\delta_{pq}\}$
is convex and\textup{ $L$-}Lipschitz, w.r.t. the Frobenius norm,
with
\[
L\triangleq\max_{1\le p<q\le m}\max_{i\in\left[n\right]}\sqrt{4\left|\left\langle z_{i},\delta_{pq}\right\rangle \right|\left\Vert \delta_{pq}\right\Vert ^{2}+2\left\langle z_{i},\delta_{pq}\right\rangle ^{2}+2\left\Vert z_{i}\right\Vert ^{2}\left\Vert \delta_{pq}\right\Vert ^{2}+\left\Vert \delta_{pq}\right\Vert ^{4}},
\]
where $z_{i}\in\mathbb{R}^{d}$ is the noise sample that was transformed
to $a_{pqi}\triangleq\left(-1\right)^{\mathbb{I}(i\in\boldsymbol{D}_{q})}(y_{i}-\frac{1}{2}(x_{p}+x_{q}))$
.
\end{lem}
\begin{IEEEproof}
$\mathring{\ell}^{\text{hinge}}(S,p,q,i)$ is a pointwise maximization
of convex (linear) functions of $S$ therefore it is convex. We prove
that it is $L$-Lipschitz by bounding its sub-gradient \cite[Lemma 14.7]{shalev2014understanding}.
The Frobenius norm of any sub-gradient $V$ is bounded by
\begin{equation}
\left\Vert V\right\Vert _{F}^{2}\le\left\Vert \dsym\left(a_{pqi}\delta_{pq}^{T}\right)\right\Vert _{F}^{2}\le2\|a_{pqi}\|^{2}\|\delta_{pq}\|^{2}+2\left\langle a_{pqi},\delta_{pq}\right\rangle ^{2},\label{eq: upper bound on the sub-gradient}
\end{equation}
where the second inequality is due to Claim \ref{claim:dsym norm bound}.
We thus get that $L=\sqrt{2(\langle a_{pqi},\delta_{pq}\rangle^{2}+\|\delta_{pq}\|^{2}\|a_{pqi}\|^{2})}$.
We next further simplify this expression. Note that,
\begin{align}
\left\langle a_{pqi},\delta_{pq}\right\rangle ^{2} & =\left[\left(-1\right)^{\mathbb{I}\left(i\in\boldsymbol{D}_{q}\right)}\left(x_{p}-x_{q}\right)^{T}\left(y_{i}-\frac{1}{2}\left(x_{p}+x_{q}\right)\right)\right]^{2}\\
 & =\left[\delta_{pq}y_{i}-\frac{1}{2}\left(\left\Vert x_{p}\right\Vert ^{2}-\left\Vert x_{q}\right\Vert ^{2}\right)\right]^{2}\\
 & =\left(\delta_{pq}y_{i}\right)^{2}-\delta_{pq}y_{i}\left(\left\Vert x_{p}\right\Vert ^{2}-\left\Vert x_{q}\right\Vert ^{2}\right)+\frac{1}{4}\left(\left\Vert x_{p}\right\Vert ^{4}-2\left\Vert x_{p}\right\Vert ^{2}\left\Vert x_{q}\right\Vert ^{2}+\left\Vert x_{q}\right\Vert ^{4}\right)\\
 & =\left(\left\langle y_{i},x_{p}\right\rangle -\left\langle y_{i},x_{q}\right\rangle \right)^{2}-\left(\left\langle y_{i},x_{p}\right\rangle -\left\langle y_{i},x_{q}\right\rangle \right)\left(\left\Vert x_{p}\right\Vert ^{2}-\left\Vert x_{q}\right\Vert ^{2}\right)\nonumber \\
 & \hphantom{=}+\frac{1}{4}\left(\left\Vert x_{p}\right\Vert ^{4}-2\left\Vert x_{p}\right\Vert ^{2}\left\Vert x_{q}\right\Vert ^{2}+\left\Vert x_{q}\right\Vert ^{4}\right),\label{eq: identity 1}
\end{align}
and,
\begin{equation}
\left\Vert \delta_{pq}\right\Vert ^{2}=\left(x_{p}-x_{q}\right)^{T}\left(x_{p}-x_{q}\right)=\left\Vert x_{p}\right\Vert ^{2}-2\left\langle x_{p},x_{q}\right\rangle +\left\Vert x_{q}\right\Vert ^{2},\label{eq: identity 2}
\end{equation}
as well as
\begin{align}
\left\Vert a_{pqi}\right\Vert ^{2} & =\left(y_{i}-\frac{1}{2}\left(x_{p}+x_{q}\right)\right)^{T}\left(y_{i}-\frac{1}{2}\left(x_{p}+x_{q}\right)\right)\\
 & =\left\Vert y_{i}\right\Vert ^{2}-\left\langle y_{i},x_{p}\right\rangle -\left\langle y_{i},x_{q}\right\rangle +\frac{1}{4}\left\Vert x_{p}\right\Vert ^{2}+\frac{1}{2}\left\langle x_{p},x_{q}\right\rangle +\frac{1}{4}\left\Vert x_{q}\right\Vert ^{2}.\label{eq: identity 3}
\end{align}
Using the above three identities (\ref{eq: identity 1}), (\ref{eq: identity 2})
and (\ref{eq: identity 3}), we get after some tedious algebra that
the upper bound on $\|V\|_{F}^{2}$ in (\ref{eq: upper bound on the sub-gradient})
is
\begin{align}
\left\langle a_{pqi},\delta_{pq}\right\rangle ^{2}+\left\Vert \delta_{pq}\right\Vert ^{2}\left\Vert a_{pqi}\right\Vert ^{2} & =\left(\left\langle y_{i},x_{p}\right\rangle -\left\langle y_{i},x_{q}\right\rangle \right)^{2}+\left\Vert y_{i}\right\Vert ^{2}\left\langle x_{p}-x_{q},x_{p}-x_{q}\right\rangle \nonumber \\
 & \phantom{=}+2\left\langle y_{i},x_{p}\right\rangle \left(\left\langle x_{p},x_{q}\right\rangle -\left\Vert x_{p}\right\Vert ^{2}\right)+2\left\langle y_{i},x_{q}\right\rangle \left(\left\langle x_{p},x_{q}\right\rangle -\left\Vert x_{q}\right\Vert ^{2}\right)\nonumber \\
 & \phantom{=}+\frac{1}{2}\left\Vert x_{p}\right\Vert ^{4}-\left\langle x_{p},x_{q}\right\rangle ^{2}+\frac{1}{2}\left\Vert x_{q}\right\Vert ^{4}.\label{eq:bound on a delta norm}
\end{align}
Notice that the above expression is symmetric in switching the roles
of $p$ and $q$. Therefore, we can continue with $y_{i}=x_{p}+z_{i}$
and the result for $y_{i}=x_{q}+z_{i}$ will be immediate by switching
$p\longleftrightarrow q$. After some more algebra, we get that for
$y_{i}=x_{p}+z_{i}$
\[
\left\langle a_{pqi},\delta_{pq}\right\rangle ^{2}+\left\Vert \delta_{pq}\right\Vert ^{2}\left\Vert a_{pqi}\right\Vert ^{2}=2\left\langle z_{i},\delta_{pq}\right\rangle \left\Vert \delta_{pq}\right\Vert ^{2}+\left\langle z_{i},\delta_{pq}\right\rangle ^{2}+\left\Vert z_{i}\right\Vert ^{2}\left\Vert \delta_{pq}\right\Vert ^{2}+\frac{1}{2}\left\Vert \delta_{pq}\right\Vert ^{4}.
\]
Therefore,
\[
y_{i}=x_{p}+z_{i}\Rightarrow\left\Vert V\right\Vert \le\sqrt{4\left\langle z_{i},\delta_{pq}\right\rangle \left\Vert \delta_{pq}\right\Vert ^{2}+2\left\langle z_{i},\delta_{pq}\right\rangle ^{2}+2\left\Vert z_{i}\right\Vert ^{2}\left\Vert \delta_{pq}\right\Vert ^{2}+\left\Vert \delta_{pq}\right\Vert ^{4}},
\]
and 
\[
y_{i}=x_{q}+z_{i}\Rightarrow\left\Vert V\right\Vert \le\sqrt{-4\left\langle z_{i},\delta_{pq}\right\rangle \left\Vert \delta_{pq}\right\Vert ^{2}+2\left\langle z_{i},\delta_{pq}\right\rangle ^{2}+2\left\Vert z_{i}\right\Vert ^{2}\left\Vert \delta_{pq}\right\Vert ^{2}+\left\Vert \delta_{pq}\right\Vert ^{4}}.
\]
We thus conclude that
\[
L\le\sqrt{4\left|\left\langle z_{i},\delta_{pq}\right\rangle \right|\left\Vert \delta_{pq}\right\Vert ^{2}+2\left\langle z_{i},\delta_{pq}\right\rangle ^{2}+2\left\Vert z_{i}\right\Vert ^{2}\left\Vert \delta_{pq}\right\Vert ^{2}+\left\Vert \delta_{pq}\right\Vert ^{4}},
\]
and this bound completes the proof.
\end{IEEEproof}
Now we are ready to prove Theorem \ref{thm: generalization theorem}
using the above Lemmas.
\begin{IEEEproof}[Proof of Th. \ref{thm: generalization theorem}]
We begin by applying \cite[Cor. 13.6]{shalev2014understanding},
where we replace the $2\lambda$-strong-convexity of the Tikhonov
regularization with the $2\lambda\gamma$-strong-convexity of the
regularization function from Lemma \ref{strong convexity lemma}.
Additionally, we take the Lipschitzness constant from Lemma \ref{lipschitz lemma}.
Then, we get from \cite[Cor. 13.6]{shalev2014understanding} that
the RLM problem (\ref{eq:stable RLM-1}) is on-average-replace-one-stable
with rate $2L^{2}/(\lambda\gamma n)$ and this implies the expected
generalization bound
\begin{align}
 & \mathbb{E}_{\boldsymbol{D}\sim\mu}\left[\mathring{L}_{\mu}^{\text{\emph{hinge}}}\left(A\left(\boldsymbol{D}\right)\right)-\mathring{L}_{\boldsymbol{D}}^{\text{\emph{hinge}}}\left(A\left(\boldsymbol{D}\right)\right)\right]\nonumber \\
 & \le\mathbb{E}_{\boldsymbol{D}\sim\mu}\left[\frac{\max_{1\le p<q\le m}\max_{i\in\left[n\right]}8\left|\left\langle Z_{i},\delta_{pq}\right\rangle \right|\left\Vert \delta_{pq}\right\Vert ^{2}+4\left\langle Z_{i},\delta_{pq}\right\rangle ^{2}+4\left\Vert Z_{i}\right\Vert ^{2}\left\Vert \delta_{pq}\right\Vert ^{2}+2\left\Vert \delta_{pq}\right\Vert ^{4}}{\lambda n\eta_{\text{\emph{min}}}\min_{1\le p<q\le m}\left\Vert \delta_{pq}\right\Vert ^{2}}\right]\\
 & \le\mathbb{E}_{\boldsymbol{D}\sim\mu}\left[\frac{32r_{x}^{3}\max_{i\in\left[n\right]}\left\Vert Z_{i}\right\Vert +16r_{x}^{2}\max_{i\in\left[n\right]}\left\Vert Z_{i}\right\Vert ^{2}+8r_{x}^{4}}{\lambda n\eta_{\text{\emph{min}}}\min_{1\le p<q\le m}\left\Vert \delta_{pq}\right\Vert ^{2}}\right],
\end{align}
where the last inequality is by Cauchy-Schwartz. Now, we use the sub-Gaussian
assumption in order to bound this expected value. Denote the event
where all $\{Z_{i}\}_{i\in[n]}$ are bounded by $\sqrt{2\sigma_{Z}^{2}\log\left(\frac{2n}{\delta}\right)}$,
for $\delta>0$, i.e.,
\[
\mathcal{E}=\bigcap_{i=1}^{n}\left\{ \left\Vert Z_{i}\right\Vert \le\sqrt{2\sigma_{Z}^{2}\log\left(\frac{2n}{\delta}\right)}\right\} .
\]
Now, since $\{Z_{i}\}_{i\in[n]}$ are i.i.d. sub-Gaussian random variables
with variance proxy $\sigma_{Z}^{2}$, then under the sub-Gaussian
assumption and using the union bound,
\[
\p\left(\mathcal{E}^{c}\right)=\p\left(\bigcup_{i=1}^{n}\left\{ \left\Vert Z_{i}\right\Vert \ge\sqrt{2\sigma_{Z}^{2}\log\left(\frac{2n}{\delta}\right)}\right\} \right)\le2n\exp\left(-\frac{2\sigma_{Z}^{2}\log\left(\frac{2n}{\delta}\right)}{2\sigma_{Z}^{2}}\right)=\delta.
\]
By conditioning the expectation on this event, we get
\begin{align}
 & \mathbb{E}_{\boldsymbol{D}\sim\mu}\left[\frac{32r_{x}^{3}\max_{i\in\left[n\right]}\left\Vert Z_{i}\right\Vert +16r_{x}^{2}\max_{i\in\left[n\right]}\left\Vert Z_{i}\right\Vert ^{2}+8r_{x}^{4}}{\lambda n\eta_{\text{\emph{min}}}\min_{1\le p<q\le m}\left\Vert \delta_{pq}\right\Vert ^{2}}\right]\nonumber \\
 & =\mathbb{E}_{\boldsymbol{D}\sim\mu}\left[\frac{32r_{x}^{3}\max_{i\in\left[n\right]}\left\Vert Z_{i}\right\Vert +16r_{x}^{2}\max_{i\in\left[n\right]}\left\Vert Z_{i}\right\Vert ^{2}+8r_{x}^{4}}{\lambda n\eta_{\text{\emph{min}}}\min_{1\le p<q\le m}\left\Vert \delta_{pq}\right\Vert ^{2}}\left(\mathbb{I}\left(\mathcal{E}\right)+\mathbb{I}\left(\mathcal{E}^{c}\right)\right)\right]\\
 & =\mathbb{E}_{\boldsymbol{D}\sim\mu}\left[\frac{32r_{x}^{3}\max_{i\in\left[n\right]}\left\Vert Z_{i}\right\Vert +16r_{x}^{2}\max_{i\in\left[n\right]}\left\Vert Z_{i}\right\Vert ^{2}+8r_{x}^{4}}{\lambda n\eta_{\text{\emph{min}}}\min_{1\le p<q\le m}\left\Vert \delta_{pq}\right\Vert ^{2}}\mathbb{I}\left(\mathcal{E}\right)\right]\nonumber \\
 & \phantom{=}+\mathbb{E}_{\boldsymbol{D}\sim\mu}\left[\frac{32r_{x}^{3}\max_{i\in\left[n\right]}\left\Vert Z_{i}\right\Vert +16r_{x}^{2}\max_{i\in\left[n\right]}\left\Vert Z_{i}\right\Vert ^{2}+8r_{x}^{4}}{\lambda n\eta_{\text{\emph{min}}}\min_{1\le p<q\le m}\left\Vert \delta_{pq}\right\Vert ^{2}}\mathbb{I}\left(\mathcal{E}^{c}\right)\right].
\end{align}
Next we will bound each term.

The first term is bounded as follows
\begin{align}
 & \mathbb{E}_{\boldsymbol{D}\sim\mu}\left[\frac{32r_{x}^{3}\max_{i\in\left[n\right]}\left\Vert Z_{i}\right\Vert +16r_{x}^{2}\max_{i\in\left[n\right]}\left\Vert Z_{i}\right\Vert ^{2}+8r_{x}^{4}}{\lambda n\eta_{\text{\emph{min}}}\min_{1\le p<q\le m}\left\Vert \delta_{pq}\right\Vert ^{2}}\mathbb{I}\left(\mathcal{E}\right)\right]\nonumber \\
 & =\p\left(\mathcal{E}\right)\mathbb{E}_{\boldsymbol{D}\sim\mu}\left[\frac{32r_{x}^{3}\max_{i\in\left[n\right]}\left\Vert Z_{i}\right\Vert +16r_{x}^{2}\max_{i\in\left[n\right]}\left\Vert Z_{i}\right\Vert ^{2}+8r_{x}^{4}}{\lambda n\eta_{\text{\emph{min}}}\min_{1\le p<q\le m}\left\Vert \delta_{pq}\right\Vert ^{2}}\;\middle|\;\mathcal{E}\right]\\
 & \le\mathbb{E}_{\boldsymbol{D}\sim\mu}\left[\frac{32r_{x}^{3}\max_{i\in\left[n\right]}\left\Vert Z_{i}\right\Vert +16r_{x}^{2}\max_{i\in\left[n\right]}\left\Vert Z_{i}\right\Vert ^{2}+8r_{x}^{4}}{\lambda n\eta_{\text{\emph{min}}}\min_{1\le p<q\le m}\left\Vert \delta_{pq}\right\Vert ^{2}}\;\middle|\;\mathcal{E}\right]\\
 & \le\frac{32r_{x}^{3}\sqrt{2\sigma_{Z}^{2}\log\left(\frac{2n}{\delta}\right)}+32r_{x}^{2}\sigma_{Z}^{2}\log\left(\frac{2n}{\delta}\right)+8r_{x}^{4}}{\lambda n\eta_{\text{\emph{min}}}\min_{1\le p<q\le m}\left\Vert \delta_{pq}\right\Vert ^{2}}\\
 & \le\frac{\log\left(\frac{2n}{\delta}\right)\sqrt{\left(32r_{x}^{3}\sqrt{2\sigma_{Z}^{2}}+32r_{x}^{2}\sigma_{Z}^{2}+8r_{x}^{4}\right)^{2}}}{\lambda n\eta_{\text{\emph{min}}}\min_{1\le p<q\le m}\left\Vert \delta_{pq}\right\Vert ^{2}}\\
 & \overset{(a)}{\le}\frac{\log\left(\frac{2n}{\delta}\right)\sqrt{3\left(2048r_{x}^{6}\sigma_{Z}^{2}+1024r_{x}^{4}\sigma_{Z}^{4}+64r_{x}^{8}\right)}}{\lambda n\eta_{\text{\emph{min}}}\min_{1\le p<q\le m}\left\Vert \delta_{pq}\right\Vert ^{2}},
\end{align}
where $(a)$ is due to $\left(a+b+c\right)^{q}=3^{q}\left(\frac{1}{3}a+\frac{1}{3}b+\frac{1}{3}c\right)^{q}\le3^{q}\left(\frac{1}{3}a^{q}+\frac{1}{3}b^{q}+\frac{1}{3}c^{q}\right)=3^{q-1}\left(a^{q}+b^{q}+c^{q}\right).$

The second term is bounded by
\begin{align}
 & \mathbb{E}_{\boldsymbol{D}\sim\mu}\left[\frac{32r_{x}^{3}\max_{i\in\left[n\right]}\left\Vert Z_{i}\right\Vert +16r_{x}^{2}\max_{i\in\left[n\right]}\left\Vert Z_{i}\right\Vert ^{2}+8r_{x}^{4}}{\lambda n\eta_{\text{\emph{min}}}\min_{1\le p<q\le m}\left\Vert \delta_{pq}\right\Vert ^{2}}\mathbb{I}\left(\mathcal{E}^{c}\right)\right]\nonumber \\
 & \overset{(a)}{\le}\sqrt{\mathbb{E}_{\boldsymbol{D}\sim\mu}\left[\frac{32r_{x}^{3}\max_{i\in\left[n\right]}\left\Vert Z_{i}\right\Vert +16r_{x}^{2}\max_{i\in\left[n\right]}\left\Vert Z_{i}\right\Vert ^{2}+8r_{x}^{4}}{\lambda n\eta_{\text{\emph{min}}}\min_{1\le p<q\le m}\left\Vert \delta_{pq}\right\Vert ^{2}}\right]^{2}\mathbb{E}_{\boldsymbol{D}\sim\mu}\left[\mathbb{I}^{2}\left(\mathcal{E}^{c}\right)\right]}\\
 & =\frac{\sqrt{\p\left(\mathcal{E}^{c}\right)}}{\lambda n\eta_{\min}\min_{1\le p<q\le m}\left\Vert \delta_{pq}\right\Vert ^{2}}\sqrt{\mathbb{E}_{\boldsymbol{D}\sim\mu}\left[32r_{x}^{3}\max_{i\in\left[n\right]}\left\Vert Z_{i}\right\Vert +16r_{x}^{2}\max_{i\in\left[n\right]}\left\Vert Z_{i}\right\Vert ^{2}+8r_{x}^{4}\right]^{2}}\\
 & \overset{(b)}{\le}\frac{1}{\lambda n\eta_{\min}\min_{1\le p<q\le m}\left\Vert \delta_{pq}\right\Vert ^{2}}\sqrt{3\delta\mathbb{E}_{\boldsymbol{D}\sim\mu}\left[1024r_{x}^{6}\max_{i\in\left[n\right]}\left\Vert Z_{i}\right\Vert ^{2}+256r_{x}^{4}\max_{i\in\left[n\right]}\left\Vert Z_{i}\right\Vert ^{4}+64r_{x}^{8}\right]}\\
 & \le\frac{1}{\lambda n\eta_{\min}\min_{1\le p<q\le m}\left\Vert \delta_{pq}\right\Vert ^{2}}\sqrt{3\delta\mathbb{E}_{\boldsymbol{D}\sim\mu}\left[1024r_{x}^{6}\sum_{i\in\left[n\right]}\left\Vert Z_{i}\right\Vert ^{2}+256r_{x}^{4}\sum_{i\in\left[n\right]}\left\Vert Z_{i}\right\Vert ^{4}+64r_{x}^{8}\right]}\\
 & \overset{(c)}{\le}\frac{1}{\lambda n\eta_{\min}\min_{1\le p<q\le m}\left\Vert \delta_{pq}\right\Vert ^{2}}\sqrt{3\delta\left(8192r_{x}^{6}n\sigma_{Z}^{2}+65536r_{x}^{4}n\sigma_{Z}^{4}+64r_{x}^{8}\right)}\\
 & \le\frac{1}{\lambda n\eta_{\min}\min_{1\le p<q\le m}\left\Vert \delta_{pq}\right\Vert ^{2}}\sqrt{3\delta n\left(8192r_{x}^{6}\sigma_{Z}^{2}+65536r_{x}^{4}\sigma_{Z}^{4}+64r_{x}^{8}\right)},
\end{align}
where $(a)$ follows from the Cauchy-Schwartz inequality, $(b)$ follows
again from $(a+b+c)^{q}\le3^{q-1}(a^{q}+b^{q}+c^{q})$, and $(c)$
follows from standard bounds on the absolute moments of sub-Gaussian
random variables (e.g., \cite[Lemma 1.4]{rigollet2015high}). Taking
$\delta=2/n$, we get
\[
\mathbb{E}_{\boldsymbol{D}\sim\mu}\left[\mathring{L}_{\mu}^{\text{\emph{hinge}}}\left(A\left(\boldsymbol{D}\right)\right)-\mathring{L}_{\boldsymbol{D}}^{\text{\emph{hinge}}}\left(A\left(\boldsymbol{D}\right)\right)\right]\le\frac{\log\left(n\right)}{n}\cdot\frac{32r_{x}^{2}\sqrt{6\left(128r_{x}^{2}\sigma_{Z}^{2}+1024\sigma_{Z}^{4}+r_{x}^{4}\right)}}{\lambda\eta_{\min}\min_{1\le p<q\le m}\left\Vert \delta_{pq}\right\Vert ^{2}}.
\]
The above bound is on the expected hinge-type loss, next, we bound
the expected error probability by this loss
\begin{align}
\boldsymbol{p}_{\mu}\left(A\left(\boldsymbol{D}\right)\right) & =\frac{1}{m}\sum_{j=1}^{m}\mathbb{E}_{\mu}\left[\mathbb{I}\left\{ \min_{j^{'}\in\left[m\right]\backslash\left\{ j\right\} }\left\Vert x_{j'}-Y\right\Vert _{S}^{2}<\left\Vert x_{j}-Y\right\Vert _{S}^{2}\right\} \;\middle|\;Y=x_{j}+Z\right]\\
 & =\frac{1}{m}\sum_{j=1}^{m}\mathbb{E}_{\mu}\left[\mathbb{I}\left\{ \min_{j^{'}\in\left[m\right]\backslash\left\{ j\right\} }2Y^{T}Sx_{j}-2Y^{T}Sx_{j'}+x_{j'}^{T}Sx_{j'}-x_{j}^{T}Sx_{j}<0\right\} \;\middle|\;Y=x_{j}+Z\right]\\
 & =\frac{1}{m}\sum_{j=1}^{m}\mathbb{E}_{\mu}\left[\mathbb{I}\left\{ \min_{j^{'}\in\left[m\right]\backslash\left\{ j\right\} }\left(Y-\frac{1}{2}\left(x_{j}+x_{j'}\right)\right)^{T}S\delta_{jj'}<0\right\} \;\middle|\;Y=x_{j}+Z\right]\\
 & \le\frac{1}{m}\sum_{j=1}^{m}\mathbb{E}_{\mu}\left[\sum_{j^{'}\in\left[m\right]\backslash\left\{ j\right\} }\max\left\{ 0,1-\left(Y-\frac{1}{2}\left(x_{j}+x_{j'}\right)\right)^{T}S\delta_{jj'}\right\} \;\middle|\;Y=x_{j}+Z\right]\\
 & \le\frac{m-1}{m}\sum_{j=1}^{m}\mathbb{E}_{\mu}\left[\frac{1}{m-1}\sum_{j^{'}\in\left[m\right]\backslash\left\{ j\right\} }\max\left\{ 0,1-\left(Y-\frac{1}{2}\left(x_{j}+x_{j'}\right)\right)^{T}S\delta_{jj'}\right\} \;\middle|\;Y=x_{j}+Z\right]\\
 & =\left(m-1\right)\mathring{L}_{\mu}^{\text{\emph{hinge}}}\left(A\left(\boldsymbol{D}\right)\right).\label{eq:error prob upperbound hinge}
\end{align}
Therefore,
\begin{align}
\mathbb{E}_{\boldsymbol{D}\sim\mu}\left[\boldsymbol{p}_{\mu}\left(A\left(\boldsymbol{D}\right)\right)\right] & \le\left(m-1\right)\mathbb{E}_{\boldsymbol{D}\sim\mu}\left[\mathring{L}_{\boldsymbol{D}}^{\text{\emph{hinge}}}\left(A\left(\boldsymbol{D}\right)\right)\right]\nonumber \\
 & \phantom{\le}+\frac{\left(m-1\right)\log\left(n\right)}{n}\cdot\frac{32r_{x}^{2}\sqrt{6\left(128r_{x}^{2}\sigma_{Z}^{2}+1024\sigma_{Z}^{4}+r_{x}^{4}\right)}}{\lambda\eta_{\min}\min_{1\le p<q\le m}\left\Vert \delta_{pq}\right\Vert ^{2}}.
\end{align}
Now, following the proof of \cite[Cor. 13.9]{shalev2014understanding},
for any $S\in\mathcal{S}_{B}$,
\begin{align}
\mathbb{E}_{\boldsymbol{D}\sim\mu}\left[\mathring{L}_{\boldsymbol{D}}^{\text{\emph{hinge}}}\left(A\left(\boldsymbol{D}\right)\right)\right] & \le\mathbb{E}_{\boldsymbol{D}\sim\mu}\left[\mathring{L}_{\boldsymbol{D}}^{\text{\emph{hinge}}}\left(A\left(\boldsymbol{D}\right)\right)+\lambda\sum_{i=1}^{d+1}\eta_{i}\max_{j\in P_{i}}\left\Vert A\left(\boldsymbol{D}\right)\delta_{j}\right\Vert ^{2}\right]\\
 & \le\mathbb{E}_{\boldsymbol{D}\sim\mu}\left[\mathring{L}_{\boldsymbol{D}}^{\text{\emph{hinge}}}\left(S\right)+\lambda\sum_{i=1}^{d+1}\eta_{i}\max_{j\in P_{i}}\left\Vert S\delta_{j}\right\Vert ^{2}\right]\\
 & \le\mathring{L}_{\mu}^{\text{\emph{hinge}}}\left(S\right)+\lambda B^{2}.
\end{align}
With that and the upper bound on the expected error probability by
hinge (\ref{eq:error prob upperbound hinge}), we conclude that
\[
\mathbb{E}_{\boldsymbol{D}\sim\mu}\left[\boldsymbol{p}_{\mu}\left(A\left(\boldsymbol{D}\right)\right)\right]\le\left(m-1\right)\min_{S\in\mathcal{S}_{B}}\mathring{L}_{\mu}^{\text{\emph{hinge}}}\left(S\right)+\left(m-1\right)\lambda B^{2}+\frac{\left(m-1\right)\log\left(n\right)}{n}\cdot\frac{32r_{x}^{2}\sqrt{6\left(128r_{x}^{2}\sigma_{Z}^{2}+1024\sigma_{Z}^{4}+r_{x}^{4}\right)}}{\lambda\eta_{\min}\min_{1\le p<q\le m}\left\Vert \delta_{pq}\right\Vert ^{2}}.
\]
Optimizing this bound w.r.t. the choice of $\lambda$ yields 
\[
\lambda=\sqrt{\frac{32r_{x}^{2}\log\left(n\right)\sqrt{6\left(128r_{x}^{2}\sigma_{Z}^{2}+1024\sigma_{Z}^{4}+r_{x}^{4}\right)}}{B^{2}n\eta_{\min}\min_{1\le p<q\le m}\left\Vert \delta_{pq}\right\Vert ^{2}}},
\]
and the bound
\[
\mathbb{E}_{\boldsymbol{D}\sim\mu}\left[\boldsymbol{p}_{\mu}\left(A\left(\boldsymbol{D}\right)\right)\right]\le\left(m-1\right)\min_{S\in\mathcal{S}_{B}}\mathring{L}_{\mu}^{\text{\emph{hinge}}}\left(S\right)+\left(m-1\right)B\sqrt{\frac{128r_{x}^{2}\log\left(n\right)\sqrt{6\left(128r_{x}^{2}\sigma_{Z}^{2}+1024\sigma_{Z}^{4}+r_{x}^{4}\right)}}{n\eta_{\min}\min_{1\le p<q\le m}\left\Vert \delta_{pq}\right\Vert ^{2}}}.
\]
\end{IEEEproof}

\subsection{Proof of Theorem \ref{thm: generalization theorem-1}\label{subsec:Expected Gen Proofs-1}}

Similarly to the proof of Theorem \ref{thm: generalization theorem},
the proof will use the strong convexity of the regularization function
and the convexity and Lipschitzness of the loss function. The following
lemma establishes the strong convexity constant of the regularization
function for the non-linear channel model.
\begin{lem}
\label{strong convexity lemma-1}Assume that $\Span\{\delta_{pq}\}_{1\le p\le q\le m}=\mathbb{R}^{d}$,
and let $\{P_{i}\}_{i\in\left[d+1\right]}$ be a proper partition.
Then, $g(H)\triangleq\sum_{i=1}^{d+1}\eta_{i}[\max_{j\in P_{i}}\|H\delta_{j}\|^{2}+\max_{j'\in P_{i}}\|K\delta_{j'}\|^{2}]$
is $2\gamma$-strongly convex w.r.t. the Frobenius norm, where
\[
\gamma\triangleq\min\left[\min_{H\in\mathcal{H}}\sum_{i=1}^{d+1}\eta_{i}\cdot\min_{j\in P_{i}}\frac{\left\Vert H\delta_{j}\right\Vert ^{2}}{\left\Vert H\right\Vert _{F}^{2}},\min_{K\in\mathcal{K}}\sum_{i=1}^{d+1}\eta_{i}\cdot\min_{j\in P_{i}}\frac{\left\Vert K\delta_{j}\right\Vert ^{2}}{\left\Vert K\right\Vert _{F}^{2}}\right]
\]
and the lower bound
\[
\gamma=\min_{\left\{ j_{i}\in P_{i}\right\} _{i=1}^{d+1}}\lambda_{\text{\emph{min}}}\left(\sum_{i=1}^{d+1}\eta_{i}\delta_{j_{i}}\delta_{j_{i}}^{T}\right)\ge\eta_{\text{\emph{min}}}\min_{1\le p<q\le m}\left\Vert \delta_{pq}\right\Vert ^{2}
\]
holds with $\eta_{\text{\emph{min}}}\triangleq\min_{i\in\left[d+1\right]}\{\eta_{i}\}$.
\end{lem}
Lemma \ref{strong convexity lemma-1} follows directly from Lemma
\ref{strong convexity lemma} and the following claim:
\begin{claim}
\label{strongly convex sum}If $f(H)$ is $\beta_{f}$-strongly convex
and $g(K)$ is $\beta_{g}$-strongly convex. Then, $h(H,K)\triangleq f(H)+g(K)$
is $\min\{\beta_{f},\beta_{g}\}$-strongly convex.
\end{claim}
\begin{IEEEproof}
We validate this property directly from the definition of strong convexity
(e.g., \cite[Definition 13.4]{shalev2014understanding}), 
\begin{align}
h\left(\alpha\left(H_{1},K_{1}\right)+\left(1-\alpha\right)\left(H_{2},K_{2}\right)\right) & =f\left(\alpha H_{1}+\left(1-\alpha\right)H_{2}\right)+g\left(\alpha K_{1}+\left(1-\alpha\right)K_{2}\right)\\
 & \le\alpha f\left(H_{1}\right)+\left(1-\alpha\right)f\left(H_{2}\right)-\frac{\beta_{f}}{2}\alpha\left(1-\alpha\right)\left\Vert H_{1}-H_{2}\right\Vert ^{2}\nonumber \\
 & \phantom{\le}+\alpha g\left(K_{1}\right)+\left(1-\alpha\right)g\left(K_{2}\right)-\frac{\beta_{g}}{2}\alpha\left(1-\alpha\right)\left\Vert K_{1}-K_{2}\right\Vert ^{2}\\
 & =\alpha h\left(H_{1},K_{1}\right)+\left(1-\alpha\right)h\left(H_{2},K_{2}\right)\nonumber \\
 & \phantom{=}-\frac{\beta_{f}}{2}\alpha\left(1-\alpha\right)\left\Vert H_{1}-H_{2}\right\Vert ^{2}-\frac{\beta_{g}}{2}\alpha\left(1-\alpha\right)\left\Vert K_{1}-K_{2}\right\Vert ^{2}\\
 & \le\alpha h\left(H_{1},K_{1}\right)+\left(1-\alpha\right)h\left(H_{2},K_{2}\right)\nonumber \\
 & \phantom{\le}-\frac{1}{2}\alpha\left(1-\alpha\right)\min\left\{ \beta_{f},\beta_{g}\right\} \left(\left\Vert H_{1}-H_{2}\right\Vert ^{2}+\left\Vert K_{1}-K_{2}\right\Vert ^{2}\right)\\
 & =\alpha h\left(H_{1},K_{1}\right)+\left(1-\alpha\right)h\left(H_{2},K_{2}\right)\nonumber \\
 & \phantom{=}-\frac{1}{2}\alpha\left(1-\alpha\right)\min\left\{ \beta_{f},\beta_{g}\right\} \left\Vert \left(H_{1},K_{1}\right)-\left(H_{2},K_{2}\right)\right\Vert ^{2}.
\end{align}
\end{IEEEproof}
The next lemma establishes the convexity and Lipschitz constant of
the loss function for the non-linear Gaussian noise channel model.
\begin{lem}
\label{lipschitz lemma-1}$(H,K)\to\max\{0,1-[y_{i}^{T}H\delta_{j_{i}j'}-\frac{1}{2}(x_{j_{i}}+x_{j'})^{T}K\delta_{j_{i}j'}]\}$
is convex and\textup{ $L$-}Lipschitz, w.r.t. the Frobenius norm,
with
\[
L\triangleq\max_{i\in\left[n\right]}\sqrt{5}r_{x}^{2}+2r_{x}\left(R_{x}+\left\Vert w_{i}\right\Vert \right),
\]
where $w_{i}\in\mathbb{R}^{d}$ is the noise sample that was added
to $f(x_{j_{i}})$, to create $y_{i}$, $r_{x}\triangleq\max_{x\in C}\|x\|$
and $R_{x}\triangleq\max_{x\in C}\|f(x)\|$.
\end{lem}
The proof requires the following claim:
\begin{claim}
\label{sum of lipschitz}If $g(H,K)$ is $L_{H}$-Lipschitz w.r.t
$H$ and $L_{K}$-Lipschitz w.r.t $K$ then it is $L$-Lipschitz w.r.t.
$(H,K)$ for $L\triangleq L_{H}+L_{K}.$
\end{claim}
\begin{IEEEproof}
We prove by definition
\begin{align}
\left|g\left(H_{1},K_{1}\right)-g\left(H_{2},K_{2}\right)\right| & =\left|g\left(H_{1},K_{1}\right)-g\left(H_{2},K_{1}\right)+g\left(H_{2},K_{1}\right)-g\left(H_{2},K_{2}\right)\right|\\
 & \le\left|g\left(H_{1},K_{1}\right)-g\left(H_{2},K_{1}\right)\right|+\left|g\left(H_{2},K_{1}\right)-g\left(H_{2},K_{2}\right)\right|\\
 & \le L_{H}\left\Vert H_{1}-H_{2}\right\Vert +L_{K}\left\Vert K_{1}-K_{2}\right\Vert \\
 & \le L_{H}\left\Vert \left(H_{1},K_{1}\right)-\left(H_{2},K_{2}\right)\right\Vert +L_{K}\left\Vert \left(H_{1},K_{1}\right)-\left(H_{2},K_{2}\right)\right\Vert \\
 & \le L\left\Vert \left(H_{1},K_{1}\right)-\left(H_{2},K_{2}\right)\right\Vert .
\end{align}
\end{IEEEproof}
The proof of Lemma \ref{lipschitz lemma-1} is then as follows:
\begin{IEEEproof}
We prove Lipschitzness by bounding the Frobenius norm of the sub-gradient
(\cite[Lemma 14.7]{shalev2014understanding}). If $V\in\partial_{H}g$
then
\begin{align}
\left\Vert V\right\Vert _{F} & \le\left\Vert \frac{\partial}{\partial H}1-\left[y_{i}^{T}H\delta_{j_{i}j'}-\frac{1}{2}\left(x_{j_{i}}+x_{j'}\right)^{T}K\delta_{j_{i}j'}\right]\right\Vert _{F}\\
 & =\left\Vert \frac{\partial}{\partial H}y_{i}^{T}H\delta_{j_{i}j'}\right\Vert _{F}\\
 & =\left\Vert y_{i}\delta_{j_{i}j'}^{T}\right\Vert _{F}\\
 & =\sqrt{\Tr\left(\delta_{j_{i}j'}y_{i}^{T}y_{i}\delta_{j_{i}j'}^{T}\right)}\\
 & =\left\Vert y_{i}\right\Vert \left\Vert \delta_{j_{i}j'}\right\Vert \\
 & =\left\Vert f(x_{j_{i}})+w_{i}\right\Vert \left\Vert \delta_{j_{i}j'}\right\Vert \\
 & \le2r_{x}\left(\left\Vert f(x_{j_{i}})\right\Vert +\left\Vert w_{i}\right\Vert \right)\\
 & \le2r_{x}\left(R_{x}+\left\Vert w_{i}\right\Vert \right).
\end{align}
If $V\in\partial_{K}g$ then, using Claim \ref{claim:dsym norm bound},
\begin{align}
\left\Vert V\right\Vert _{F}^{2} & \le\left\Vert \dsym\left(\frac{1}{2}\left(x_{j_{i}}+x_{j'}\right)\delta_{j_{i}j'}^{T}\right)\right\Vert _{F}^{2}\\
 & \le\frac{1}{4}\left(2\left\Vert x_{j_{i}}+x_{j'}\right\Vert ^{2}\left\Vert \delta_{j_{i}j'}\right\Vert ^{2}+2\left\langle x_{j_{i}}+x_{j'},\delta_{j_{i}j'}\right\rangle ^{2}\right)\\
 & =\frac{1}{2}\left\Vert x_{j_{i}}+x_{j'}\right\Vert ^{2}\left\Vert \delta_{j_{i}j'}\right\Vert ^{2}+\frac{1}{2}\left(\left\Vert x_{j_{i}}\right\Vert ^{2}-\left\Vert x_{j'}\right\Vert ^{2}\right)^{2}\\
 & \le\frac{1}{2}\cdot4r_{x}^{2}\cdot2r_{x}^{2}+\frac{1}{2}\cdot2r_{x}^{4}\\
 & =5r_{x}^{4}.
\end{align}
\end{IEEEproof}
Now we are ready to prove Theorem \ref{thm: generalization theorem-1}
using the above Lemmas.
\begin{IEEEproof}[Proof of Th. \ref{thm: generalization theorem-1}]
The proof is similar to the proof of Theorem \ref{thm: generalization theorem},
with only a few technical differences. We begin by applying \cite[Cor. 13.6]{shalev2014understanding},
where we replace the $2\lambda$-strong-convexity of the Tikhonov
regularization with the $2\lambda\gamma$-strong-convexity of the
regularization function from Lemma \ref{strong convexity lemma-1}.
Additionally, we take the Lipschitzness constant from Lemma \ref{lipschitz lemma-1}.
Then, we get from \cite[Cor. 13.6]{shalev2014understanding} that
the RLM problem (\ref{eq:stable RLM-1}) is on-average-replace-one-stable
with rate $2L^{2}/(\lambda\gamma n)$, and the following generalization
bound holds
\[
\mathbb{E}_{\boldsymbol{D}\sim\mu}\left[\mathring{L}_{\mu}^{\text{\emph{hinge}}}\left(A\left(\boldsymbol{D}\right)\right)-\mathring{L}_{\boldsymbol{D}}^{\text{\emph{hinge}}}\left(A\left(\boldsymbol{D}\right)\right)\right]\le\mathbb{E}_{\boldsymbol{D}\sim\mu}\left[\frac{2\left[\sqrt{5}r_{x}^{2}+2r_{x}\left(R_{x}+\max_{i\in\left[n\right]}\left\Vert W_{i}\right\Vert \right)\right]^{2}}{\lambda n\eta_{\text{\emph{min}}}\min_{1\le p<q\le m}\left\Vert \delta_{pq}\right\Vert ^{2}}\right].
\]
Now, we use the sub-Gaussian assumption in the same way as in the
proof Theorem \ref{thm: generalization theorem} to get the following
bound:
\[
\mathbb{E}_{\boldsymbol{D}\sim\mu}\left[\mathring{L}_{\mu}^{\text{\emph{hinge}}}\left(A\left(\boldsymbol{D}\right)\right)-\mathring{L}_{\boldsymbol{D}}^{\text{\emph{hinge}}}\left(A\left(\boldsymbol{D}\right)\right)\right]\le\frac{\log\left(n\right)}{n}\cdot\frac{24\sqrt{25r_{x}^{8}+16r_{x}^{4}R_{x}^{4}+64r_{x}^{4}\sigma_{W}^{4}}}{\lambda\eta_{\min}\min_{1\le p<q\le m}\left\Vert \delta_{pq}\right\Vert ^{2}}.
\]
The above bound is on the expected hinge-type loss, next, we bound
the expected error probability by this loss
\begin{align}
 & \boldsymbol{p}_{\mu}\left(A\left(\boldsymbol{D}\right)\right)\nonumber \\
 & =\frac{1}{m}\sum_{j=1}^{m}\mathbb{E}_{\mu}\left[\mathbb{I}\left\{ \min_{j^{'}\in\left[m\right]\backslash\left\{ j\right\} }\left\Vert Y\right\Vert ^{2}-2Y^{T}Hx_{j'}+x_{j'}^{T}Kx_{j'}<\left\Vert Y\right\Vert ^{2}-2Y^{T}Hx_{j}+x_{j}^{T}Kx_{j}\right\} \;\middle|\;Y=f\left(x_{j}\right)+W\right]\\
 & =\frac{1}{m}\sum_{j=1}^{m}\mathbb{E}_{\mu}\left[\mathbb{I}\left\{ \min_{j^{'}\in\left[m\right]\backslash\left\{ j\right\} }Y^{T}H\delta_{jj'}-\frac{1}{2}\left(x_{j}+x_{j'}\right)^{T}K\delta_{jj'}<0\right\} \;\middle|\;Y=f\left(x_{j}\right)+W\right]\\
 & \le\frac{1}{m}\sum_{j=1}^{m}\mathbb{E}_{\mu}\left[\sum_{j^{'}\in\left[m\right]\backslash\left\{ j\right\} }\max\left\{ 0,1-\left[Y^{T}H\delta_{jj'}-\frac{1}{2}\left(x_{j}+x_{j'}\right)^{T}K\delta_{jj'}\right]\right\} \;\middle|\;Y=f\left(x_{j}\right)+W\right]\\
 & =\left(m-1\right)\frac{1}{m}\sum_{j=1}^{m}\mathbb{E}_{\mu}\left[\frac{1}{m-1}\sum_{j^{'}\in\left[m\right]\backslash\left\{ j\right\} }\max\left\{ 0,1-\left[Y^{T}H\delta_{jj'}-\frac{1}{2}\left(x_{j}+x_{j'}\right)^{T}K\delta_{jj'}\right]\right\} \;\middle|\;Y=f\left(x_{j}\right)+W\right]\\
 & =\left(m-1\right)\mathring{L}_{\mu}^{\text{\emph{hinge}}}\left(A\left(\boldsymbol{D}\right)\right).
\end{align}
Therefore,
\[
\mathbb{E}_{\boldsymbol{D}\sim\mu}\left[\boldsymbol{p}_{\mu}\left(A\left(\boldsymbol{D}\right)\right)\right]\le\left(m-1\right)\mathbb{E}_{\boldsymbol{D}\sim\mu}\left[\mathring{L}_{\boldsymbol{D}}^{\text{\emph{hinge}}}\left(A\left(\boldsymbol{D}\right)\right)\right]+\frac{\left(m-1\right)\log\left(n\right)}{n}\cdot\frac{24\sqrt{25r_{x}^{8}+16r_{x}^{4}R_{x}^{4}+64r_{x}^{4}\sigma_{W}^{4}}}{\lambda\eta_{\min}\min_{1\le p<q\le m}\left\Vert \delta_{pq}\right\Vert ^{2}}.
\]
The optimal choice of $\lambda$ then follows as in the proof of Theorem
\ref{thm: generalization theorem}.
\end{IEEEproof}

\subsection{Proof of Theorem \ref{thm:uniform hinge gen}\label{subsec:Uniform Gen Proof}}

We will need several lemmas. The first lemma characterizes the continuity
of the surrogate loss function w.r.t. $(H,K)$.
\begin{lem}
\label{lem:loss bound for cover}Suppose that $H,\tilde{H}\in\mathcal{H}$
such that $\|H-\tilde{H}\|_{\text{\emph{op}}}\le\gamma_{H}$. Then,
\[
\left|\mathring{\ell}^{\text{\emph{hinge}}}\left(H,i\right)-\mathring{\ell}^{\text{\emph{hinge}}}\left(\tilde{H},i\right)\right|\le2r_{x}\left(R_{x}+r_{z}\right)\gamma_{H}+r_{x}^{2}\gamma_{H}^{2}.
\]
\end{lem}
\begin{IEEEproof}
Denote $f(t)=\max\{0,1-t\}$, which is a 1-Lipschitz function. Then,
\begin{align}
 & \left|\mathring{\ell}^{\text{hinge}}\left(H,i,j'\right)-\mathring{\ell}^{\text{hinge}}\left(\tilde{H},i,j'\right)\right|\nonumber \\
 & \le\left|\left(y_{i}^{T}H\delta_{j_{i}j'}-\frac{1}{2}\left(x_{j_{i}}+x_{j'}\right)^{T}H^{T}H\delta_{j_{i}j'}\right)-\left(y_{i}^{T}\tilde{H}\delta_{j_{i}j'}-\frac{1}{2}\left(x_{j_{i}}+x_{j'}\right)^{T}\tilde{H}^{T}\tilde{H}\delta_{j_{i}j'}\right)\right|\\
 & =\left|\left(y_{i}^{T}H\delta_{j_{i}j'}-\frac{1}{2}x_{j}^{T}H^{T}Hx_{j}+\frac{1}{2}x_{j'}^{T}H^{T}Hx_{j'}\right)-\left(y_{i}^{T}\tilde{H}\delta_{j_{i}j'}-\frac{1}{2}x_{j}^{T}\tilde{H}^{T}\tilde{H}x_{j}+\frac{1}{2}x_{j'}^{T}\tilde{H}^{T}\tilde{H}x_{j'}\right)\right|\\
 & =\left|y_{i}^{T}\left(H-\tilde{H}\right)\delta_{j_{i}j'}-\frac{1}{2}x_{j}^{T}\left(H^{T}H-\tilde{H}^{T}\tilde{H}\right)x_{j}+\frac{1}{2}x_{j'}^{T}\left(H^{T}H-\tilde{H}^{T}\tilde{H}\right)x_{j'}\right|\\
 & \le\left(R_{x}+r_{z}\right)\gamma_{H}\cdot2r_{x}+r_{x}\gamma_{H}^{2}r_{x}\\
 & =2r_{x}\left(R_{x}+r_{z}\right)\gamma_{H}+r_{x}^{2}\gamma_{H}^{2},
\end{align}
and by the triangle inequality
\begin{align}
\left|\mathring{\ell}^{\text{hinge}}\left(H,i\right)-\mathring{\ell}^{\text{hinge}}\left(\tilde{H},i\right)\right| & =\left|\frac{1}{m-1}\sum_{j'\in\left[m\right]\backslash\left\{ j_{i}\right\} }\mathring{\ell}^{\text{hinge}}\left(H,i,j'\right)-\mathring{\ell}^{\text{hinge}}\left(\tilde{H},i,j'\right)\right|\\
 & \le\max_{j'\in\left[m\right]\backslash\left\{ j_{i}\right\} }\left|\mathring{\ell}^{\text{hinge}}\left(H,i,j'\right)-\mathring{\ell}^{\text{hinge}}\left(\tilde{H},i,j'\right)\right|.
\end{align}
\end{IEEEproof}
We denote by $N(\mathcal{H},\|\cdot\|_{\text{op}},\gamma_{H})$ the
covering number (e.g. \cite[Definition 4.2.2]{vershynin2018high})
of $\mathcal{H}$, for the operator norm and covering radius $\gamma_{H}$.
\begin{lem}
\label{lem:covering H}It holds that
\[
N\left(\mathcal{H},\left\Vert \cdot\right\Vert _{\text{\emph{op}}},\gamma_{H}\right)\le\left(\frac{12d_{m}r_{H}}{\gamma_{H}}\right)^{d_{y}^{2}}\left(\frac{12d_{m}r_{H}}{\gamma_{H}}\right)^{d_{x}^{2}}\left(\frac{4r_{H}}{\gamma_{H}}\right)^{d_{m}}\le\left(\frac{12d_{m}r_{H}}{\gamma_{H}}\right)^{d_{x}^{2}+d_{y}^{2}+d_{m}}.
\]
\end{lem}
\begin{IEEEproof}
Denote the SVD of $H=U\Sigma V^{T}$ and $d_{m}=\min\{d_{y},d_{x}\}$.
Then,
\begin{align}
\left\Vert H-\tilde{H}\right\Vert _{\text{op}} & =\left\Vert U\Sigma V^{T}-\tilde{U}\tilde{\Sigma}\tilde{V}^{T}\right\Vert _{\text{op}}\\
 & \le\left\Vert U\Sigma V^{T}-U\tilde{\Sigma}V^{T}\right\Vert _{\text{op}}+\left\Vert U\tilde{\Sigma}V^{T}-\tilde{U}\tilde{\Sigma}\tilde{V}^{T}\right\Vert _{\text{op}},\label{eq:two terms op}
\end{align}
where the inequality is the triangle inequality. Let us bound the
two terms. The first is bounded as follows:
\[
\left\Vert U\Sigma V^{T}-U\tilde{\Sigma}V^{T}\right\Vert _{\text{op}}=\left\Vert U\left(\Sigma-\tilde{\Sigma}\right)V^{T}\right\Vert _{\text{op}}\le\left\Vert U\right\Vert _{\text{op}}\left\Vert \Sigma-\tilde{\Sigma}\right\Vert _{\text{op}}\left\Vert V^{T}\right\Vert _{\text{op}}\overset{(a)}{=}\left\Vert \Sigma-\tilde{\Sigma}\right\Vert _{\text{op}}=\max_{i\in\left[d\right]}\left|\sigma_{i}-\tilde{\sigma_{i}}\right|,
\]
where $(a)$ is due to $U$ and $V$ being orthonormal matrices. The
second is bounded by
\[
\left\Vert U\tilde{\Sigma}V^{T}-\tilde{U}\tilde{\Sigma}\tilde{V}^{T}\right\Vert _{\text{op}}=\left\Vert \sum_{i=1}^{d_{m}}\tilde{\sigma_{i}}\left(u_{i}v_{i}^{T}-\tilde{u}_{i}\tilde{v}_{i}^{T}\right)\right\Vert _{\text{op}}\le\sum_{i=1}^{d_{m}}\tilde{\sigma_{i}}\left\Vert u_{i}v_{i}^{T}-\tilde{u}_{i}\tilde{v}_{i}^{T}\right\Vert _{\text{op}}\le\sum_{i=1}^{d_{m}}\tilde{\sigma_{i}}\left(\left\Vert v_{i}-\tilde{v}_{i}\right\Vert +\left\Vert u_{i}-\tilde{u}_{i}\right\Vert \right),
\]
since for any $w\in\mathbb{S}^{d-1}$
\begin{align}
\left|w^{T}\left(u_{i}v_{i}^{T}-\tilde{u}_{i}\tilde{v}_{i}^{T}\right)w\right| & =\left|v_{i}^{T}ww^{T}u_{i}-\tilde{v}_{i}^{T}ww^{T}\tilde{u}_{i}\right|\\
 & =\left|v_{i}^{T}ww^{T}u_{i}-\tilde{v}_{i}^{T}ww^{T}u_{i}+\tilde{v}_{i}^{T}ww^{T}u_{i}-\tilde{v}_{i}^{T}ww^{T}\tilde{u}_{i}\right|\\
 & \le\left|\left(v_{i}^{T}-\tilde{v}_{i}^{T}\right)ww^{T}u_{i}\right|+\left|\tilde{v}_{i}^{T}ww^{T}\left(u_{i}-\tilde{u}_{i}\right)\right|\\
 & \le\left\Vert v_{i}-\tilde{v}_{i}\right\Vert \left\Vert ww^{T}\right\Vert \left\Vert u_{i}\right\Vert +\left\Vert \tilde{v}_{i}\right\Vert \left\Vert ww^{T}\right\Vert \left\Vert u_{i}-\tilde{u}_{i}\right\Vert \\
 & \le\left\Vert v_{i}-\tilde{v}_{i}\right\Vert +\left\Vert u_{i}-\tilde{u}_{i}\right\Vert .
\end{align}
Plugging this back into (\ref{eq:two terms op}), we get that
\[
\left\Vert H-\tilde{H}\right\Vert _{\text{op}}\le\max_{i\in\left[d_{m}\right]}\left|\sigma_{i}-\tilde{\sigma_{i}}\right|+d_{m}\max_{i\in\left[d_{m}\right]}\tilde{\sigma_{i}}\left(\left\Vert v_{i}-\tilde{v}_{i}\right\Vert +\left\Vert u_{i}-\tilde{u}_{i}\right\Vert \right).
\]
Let $\epsilon_{u}=\frac{\gamma_{H}}{4d_{m}r_{H}}$ and let $\mathcal{U}$
be an $\epsilon_{u}$-net in the Euclidean distance for the unit sphere
$\mathbb{S}^{d_{y}-1}$ whose size is less than $\left|\mathcal{U}\right|\le(\frac{3}{\epsilon_{u}})^{d_{y}}$
(the existence of such net is assured from \cite[Cor. 4.2.13]{vershynin2018high}).
Let $\epsilon_{v}=\frac{\gamma_{H}}{4d_{m}r_{H}}$ and let $\mathcal{V}$
be an $\epsilon_{v}$-net in the Euclidean distance for the unit sphere
$\mathbb{S}^{d_{x}-1}$ whose size is less than $\left|\mathcal{V}\right|\le(\frac{3}{\epsilon_{v}})^{d_{x}}$,
let $\epsilon_{0}=\frac{\gamma_{H}}{2}$, and let $\mathcal{L}$ be
a proper $\epsilon_{0}$-net in the $\ell_{1}$ norm for $[-r_{H},r_{H}]$
whose size is $|\mathcal{L}|\le\frac{2r_{H}}{\epsilon_{0}}$. Then,
the set
\[
\left\{ U\Sigma V^{T}:U^{T}U=I_{d_{y}},V^{T}V=I_{d_{x}},U=\left[u_{1},\dots,u_{d_{y}}\right],\Sigma,V=\left[v_{1},\dots,v_{d_{x}}\right],u_{i}\in\mathcal{U},v_{i}\in\mathcal{V},\sigma_{i}\in\mathcal{L}\right\} 
\]
is a $\gamma_{H}$-cover of $\mathcal{H}$ whose size is $(|\mathcal{U}|\cdot|\mathcal{V}|\cdot|\mathcal{L}|)^{d}$.
\end{IEEEproof}
We denote the \emph{empirical Rademacher complexity} of a set ${\cal L}_{n}\subset\mathbb{R}^{n}$
by
\begin{equation}
\text{Rad}\left(\mathcal{L}_{n}\right)\triangleq\frac{1}{n}\mathbb{E}\left[\sup_{l^{n}\in\mathcal{L}_{n}}\sum_{i=1}^{n}R_{i}l_{i}\right],\label{eq: empircal Rademacher complexity}
\end{equation}
where $l^{n}\triangleq(l_{1},\ldots l_{n})\in\mathbb{R}^{n}$ and
$R^{n}\triangleq(R_{1},\ldots R_{n})\in\{\pm1\}^{n}$ are Rademacher
random variables (i.e., $R_{i}\sim\text{Uniform}\{-1,1\}$, i.i.d.).
We are now ready to prove Theorem \ref{thm:uniform hinge gen}.
\begin{IEEEproof}
Let $\boldsymbol{D}$ be given, and consider the loss class
\[
\boldsymbol{\mathcal{L}_{D}}\triangleq\left\{ \mathring{\ell}^{\text{hinge}}\left(H,\boldsymbol{D}\right)\in\mathbb{R}_{+}^{n}:H\in\mathcal{H}\right\} 
\]
where
\[
\mathring{\ell}^{\text{hinge}}\left(H,\boldsymbol{D}\right)=\left(\mathring{\ell}^{\text{hinge}}\left(H,y_{1}\right),\dots,\mathring{\ell}^{\text{hinge}}\left(H,y_{n}\right)\right).
\]
Let $\mathcal{\tilde{H}}$ be a $\gamma_{H}$-net of $\mathcal{H}$
in the operator norm whose size is less than $\left(\frac{12d_{m}r_{H}}{\gamma_{H}}\right)^{d_{x}^{2}+d_{y}^{2}+d_{m}}$
according to Lemma \ref{lem:covering H}. Then, by Lemma \ref{lem:loss bound for cover},
the set
\[
\boldsymbol{\tilde{\mathcal{L}}_{D}}\triangleq\left\{ \mathring{\ell}^{\text{hinge}}\left(H,\boldsymbol{D}\right)\in\mathbb{R}_{+}^{n}:H\in\tilde{\mathcal{H}}\right\} 
\]
is a $\gamma$-cover of $\boldsymbol{\mathcal{L}_{D}}$ with
\[
\gamma=2r_{x}\left(R_{x}+r_{z}\right)\gamma_{H}+r_{x}^{2}\gamma_{H}^{2}.
\]
The logarithm of the cover's size is bounded by
\[
\left(d_{x}^{2}+d_{y}^{2}+d_{m}\right)\log\left(\frac{12d_{m}r_{H}}{\gamma_{H}}\right)\triangleq a\left(\log b-\log\gamma_{H}\right).
\]
By Dudley's entropy integral (e.g., \cite[Th. 12.4]{rakhlin2012statistical}),
\begin{align}
\text{Rad}\left(\boldsymbol{\mathcal{L}_{D}}\right) & \le\inf_{\alpha\ge0}\left\{ 4\alpha+\frac{12}{\sqrt{n}}\int_{\alpha}^{1}\sqrt{\log N_{2}\left(\gamma_{H},\boldsymbol{\mathcal{L}_{D}}\right)}\d\gamma_{H}\right\} \\
 & =\inf_{\alpha\ge0}\left\{ 4\alpha+\frac{12}{\sqrt{n}}\int_{\alpha}^{1}\sqrt{a\left(\log b-\log\gamma_{H}\right)}\d\gamma_{H}\right\} \\
 & =\inf_{\alpha\ge0}\left\{ 4\alpha+12\sqrt{\frac{a\log b}{n}}\int_{\alpha}^{1}\sqrt{1-\frac{\log\gamma_{H}}{\log b}}\d\gamma_{H}\right\} \\
 & \overset{(a)}{\le}\inf_{\alpha\ge0}\left\{ 4\alpha+12\sqrt{\frac{a\log b}{n}}\int_{\alpha}^{1}1-\frac{1}{2}\frac{\log\gamma_{H}}{\log b}\d\gamma_{H}\right\} \\
 & =\inf_{\alpha\ge0}\left\{ 4\alpha+12\sqrt{\frac{a\log b}{n}}\left[\left(1-\alpha\right)+\frac{1}{2\log b}\left(\alpha\left(\log\alpha-1\right)-1\left(\log1-1\right)\right)\right]\right\} \\
 & =\inf_{\alpha\ge0}\left\{ 4\alpha+12\sqrt{\frac{a\log b}{n}}\left[1-\alpha+\frac{1}{2\log b}\left(\alpha\log\alpha-\alpha+1\right)\right]\right\} \\
 & =12\sqrt{\frac{a\log b}{n}}+6\sqrt{\frac{a}{n\log b}}+\inf_{\alpha\ge0}\left\{ \alpha\left(4-12\sqrt{\frac{a\log b}{n}}+6\sqrt{\frac{a}{n\log b}}\log a-6\sqrt{\frac{a}{n\log b}}\right)\right\} \\
 & \overset{(b)}{=}12\sqrt{\frac{a\log b}{n}}+6\sqrt{\frac{a}{n\log b}}\nonumber \\
 & \phantom{=}+\exp\left(2\log b-\frac{2}{3}\sqrt{\frac{n\log b}{a}}\right)\left(4-12\sqrt{\frac{a\log b}{n}}+6\sqrt{\frac{a}{n\log b}}\left(2\log b-\frac{2}{3}\sqrt{\frac{n\log b}{a}}\right)-6\sqrt{\frac{a}{n\log b}}\right)\\
 & =12\sqrt{\frac{a\log b}{n}}+6\sqrt{\frac{a}{n\log b}}-6\sqrt{\frac{a}{n\log b}}\exp\left(2\log b-\frac{2}{3}\sqrt{\frac{n\log b}{a}}\right)\\
 & =6\sqrt{\frac{a}{n\log b}}\left[2\log b+1-\exp\left(2\log b-\frac{2}{3}\sqrt{\frac{n\log b}{a}}\right)\right]\\
 & =6\sqrt{\frac{d_{x}^{2}+d_{y}^{2}+d_{m}}{n\log\left(12d_{m}r_{H}\right)}}\left[2\log\left(12d_{m}r_{H}\right)+1-\exp\left(2\log\left(12d_{m}r_{H}\right)-\frac{2}{3}\sqrt{\frac{n\log\left(12d_{m}r_{H}\right)}{d_{x}^{2}+d_{y}^{2}+d_{m}}}\right)\right],
\end{align}
where $(a)$ is due to $\sqrt{1-t}\le1-t/2$ and $(b)$ is due to
\[
\frac{\d}{\d x}\left(x\left(k_{1}+k_{2}\log x\right)\right)=k_{1}+k_{2}\log x+k_{2}=0\iff\log x=-\frac{k_{1}+k_{2}}{k_{2}}\iff x=\exp\left(-\frac{k_{1}+k_{2}}{k_{2}}\right).
\]
It is well-established that Rademacher complexity uniformly bounds
the deviation of empirical averages from the statistical average \cite{bartlett2002rademacher}.
We use the version from \cite[Prop. 8]{weinberger2021generalization}
with the above empirical Rademacher complexity bound, and note that
\begin{align}
\left|\mathring{\ell}^{\text{hinge}}\left(H,i,j'\right)\right| & \le\left|y_{i}^{T}H\delta_{j_{i}j'}-\frac{1}{2}\left(x_{j_{i}}+x_{j'}\right)^{T}H^{T}H\delta_{j_{i}j'}\right|\\
 & \le\left(R_{x}+r_{z}\right)r_{H}\cdot2r_{x}+r_{x}^{2}r_{H}^{2},
\end{align}
to complete the proof.
\end{IEEEproof}

\section{Proofs for Section \ref{sec:SGD}\label{sec:SGD proofs}}

In order to prove Theorem \ref{thm: optimization bound} we will need
several lemmas. We begin by bounding the Frobenius norm of each of
the sub-gradients of the approximate objectives. To this end, we begin
with the following Lemma.
\begin{lem}
\label{lem:general grad bound}Consider the update rule $S_{t+1}=\Pi(S_{t}-\frac{1}{\lambda t}\nabla_{t})$,
and a sub-gradient of the form
\[
\nabla_{t}=\lambda\sum_{k=1}^{d+1}\eta_{k}\cdot\dsym\left(2S_{t}\delta_{j_{k}^{\left(t\right)}}\delta_{j_{k}^{\left(t\right)}}^{T}\right)-V_{t},
\]
where $\|V_{t}\|_{F}\le B$ for all $t\in\mathbb{N}^{+}$ and $\Pi$
is such that $\|\Pi(S)\|_{F}\le\|S\|_{F}$ for all $S\in\mathcal{S}$.
Then, 
\[
\|\nabla_{t}\|_{F}\leq32Br_{x}^{2}\left(\ln\left(t\right)+e^{224r_{x}^{4}}\right).
\]
\end{lem}
\begin{IEEEproof}
First note that
\[
\left\Vert S_{t+1}\right\Vert _{F}=\left\Vert \Pi_{\mathbb{S}^{+}}\left(S_{t}-\frac{1}{\lambda t}\nabla_{t}\right)\right\Vert _{F}\le\left\Vert S_{t}-\frac{1}{\lambda t}\nabla_{t}\right\Vert _{F}.
\]
Now, denoting for readability

\[
D_{t}\triangleq\sum_{k=1}^{d+1}\eta_{k}\delta_{j_{k}^{\left(t\right)}}\delta_{j_{k}^{\left(t\right)}}^{T},
\]
it holds that
\[
S_{t}-\frac{1}{\lambda t}\nabla_{t}=S_{t}-\frac{1}{t}\dsym\left(2S_{t}D_{t}\right)+\frac{1}{\lambda t}V_{t}.
\]
Consequently, using the triangle inequality, 
\[
\left\Vert S_{t+1}\right\Vert _{F}\le\left\Vert S_{t}-\frac{2}{t}S_{t}D_{t}-\frac{2}{t}D_{t}S_{t}-\frac{2}{t}\diag\left(S_{t}D_{t}\right)\right\Vert _{F}+\frac{1}{\lambda t}\left\Vert V_{t}\right\Vert _{F}.
\]
We bound the first term by
\begin{align}
 & \left\Vert S_{t}-\frac{1}{t}\dsym\left(2S_{t}D_{t}\right)\right\Vert _{F}\nonumber \\
 & =\sqrt{\Tr\left[\left(S_{t}-\frac{2}{t}S_{t}D_{t}-\frac{2}{t}D_{t}S_{t}-\frac{2}{t}\diag\left(S_{t}D_{t}\right)\right)^{T}\left(S_{t}-\frac{2}{t}S_{t}D_{t}-\frac{2}{t}D_{t}S_{t}-\frac{2}{t}\diag\left(S_{t}D_{t}\right)\right)\right]}\\
 & =\sqrt{\begin{array}{c}
\left\Vert S_{t}\right\Vert _{F}^{2}-\frac{8}{t}\Tr\left(S_{t}S_{t}D_{t}\right)+\frac{8}{t^{2}}\Tr\left(S_{t}D_{t}S_{t}D_{t}\right)+\frac{8}{t^{2}}\left\Vert S_{t}D_{t}\right\Vert _{F}^{2}\\
-\frac{2}{t}\Tr\left(S_{t}\diag\left(S_{t}D_{t}\right)\right)+\frac{8}{t^{2}}\Tr\left(S_{t}D_{t}\diag\left(S_{t}D_{t}\right)\right)+\frac{4}{t^{2}}\left\Vert \diag\left(S_{t}D_{t}\right)\right\Vert _{F}^{2}
\end{array}}\\
 & =\sqrt{\begin{array}{c}
\left\Vert S_{t}\right\Vert _{F}^{2}-\frac{8}{t}\Tr\left(S_{t}S_{t}D_{t}\right)+\frac{8}{t^{2}}\Tr\left(S_{t}D_{t}S_{t}D_{t}\right)+\frac{8}{t^{2}}\left\Vert S_{t}D_{t}\right\Vert _{F}^{2}\\
-\frac{2}{t}\Tr\left(\diag\left(S_{t}\right)\diag\left(S_{t}D_{t}\right)\right)+\frac{12}{t^{2}}\left\Vert \diag\left(S_{t}D_{t}\right)\right\Vert _{F}^{2}
\end{array}}\\
 & \overset{\left(a\right)}{\le}\sqrt{\left\Vert S_{t}\right\Vert _{F}^{2}+\frac{16}{t^{2}}\left\Vert S_{t}\right\Vert _{F}^{2}\left\Vert D_{t}\right\Vert _{F}^{2}+\frac{12}{t^{2}}\left\Vert \diag\left(S_{t}D_{t}\right)\right\Vert _{F}^{2}}\\
 & \le\sqrt{\left\Vert S_{t}\right\Vert _{F}^{2}+\frac{16}{t^{2}}\left\Vert S_{t}\right\Vert _{F}^{2}\left\Vert D_{t}\right\Vert _{F}^{2}+\frac{12}{t^{2}}\left\Vert S_{t}D_{t}\right\Vert _{F}^{2}}\\
 & \le\left\Vert S_{t}\right\Vert _{F}\sqrt{1+\frac{28}{t^{2}}\sum_{l=1}^{d+1}\sum_{r=1}^{d+1}\eta_{l}\eta_{r}\left\Vert \delta_{j_{l}^{\left(t\right)}}\right\Vert ^{2}\left\Vert \delta_{j_{r}^{\left(t\right)}}\right\Vert ^{2}}\\
 & \overset{\left(b\right)}{\le}\left\Vert S_{t}\right\Vert _{F}\sqrt{1+\frac{448r_{x}^{4}}{t^{2}}\sum_{l=1}^{d+1}\sum_{r=1}^{d+1}\eta_{l}\eta_{r}}\\
 & =\left\Vert S_{t}\right\Vert _{F}\sqrt{1+\frac{448r_{x}^{4}}{t^{2}}},\label{eq:first grad bound}
\end{align}
where $(a)$ follows from $\Tr(S_{t}S_{t}D_{t})=\sum_{k=1}^{d+1}\eta_{k}\Tr(\delta_{j_{k}^{(t)}}^{T}S_{t}S_{t}\delta_{j_{k}^{(t)}})\ge0$,
$(b)$ follows from $\|\delta\|\le2r_{x}$, and $(c)$ follows since
$\sum_{l=1}^{d+1}\eta_{l}=1$.

Therefore, we can bound recursively,
\begin{align}
\left\Vert S_{t+1}\right\Vert _{F} & \le\left\Vert S_{t}-\frac{1}{t}\dsym\left(2S_{t}D_{t}\right)\right\Vert _{F}+\frac{1}{\lambda t}\left\Vert V_{t}\right\Vert _{F}\\
 & \le\left\Vert S_{t}\right\Vert _{F}\sqrt{1+\frac{448r_{x}^{4}}{t^{2}}}+\frac{1}{\lambda t}\left\Vert V_{t}\right\Vert _{F}\\
 & \le\left(\left\Vert S_{t-1}-\frac{1}{t-1}\dsym\left(2S_{t-1}D_{t-1}\right)\right\Vert _{F}+\frac{1}{\lambda\left(t-1\right)}\left\Vert V_{t-1}\right\Vert _{F}\right)\sqrt{1+\frac{448r_{x}^{4}}{t^{2}}}+\frac{1}{\lambda t}\left\Vert V_{t}\right\Vert _{F}\\
 & \le\sum_{i=1}^{t}\frac{1}{\lambda i}\left\Vert V_{i}\right\Vert _{F}\prod_{j=i+1}^{t}\sqrt{1+\frac{448r_{x}^{4}}{j^{2}}}\\
 & \le\frac{B}{\lambda}\sum_{i=1}^{t}\frac{1}{i}\prod_{j=i+1}^{t}\sqrt{1+\frac{448r_{x}^{4}}{j^{2}}.}
\end{align}
We bound the product by
\begin{align}
\sqrt{\prod_{j=i+1}^{t}1+\frac{448r_{x}^{4}}{j^{2}}} & =\sqrt{\prod_{j=i+1}^{t}\left(\left(1+\frac{448r_{x}^{4}}{j^{2}}\right)^{j^{2}}\right)^{j^{-2}}}\\
 & \le\sqrt{\prod_{j=i+1}^{t}\left(e^{448r_{x}^{4}}\right)^{j^{-2}}}\\
 & =\prod_{j=i+1}^{t}e^{\frac{448r_{x}^{4}}{2j^{2}}}\\
 & =e^{\sum_{j=i+1}^{t}\frac{224r_{x}^{4}}{j^{2}}}\\
 & \le e^{224r_{x}^{4}\int_{j=i+1}^{t}\frac{1}{x^{2}}dx}\\
 & =e^{224r_{x}^{4}\left(\frac{1}{i+1}-\frac{1}{t}\right)},\label{eq:prod bound}
\end{align}
where the first inequality is due to $1+x\le e^{x}$, and get
\begin{equation}
\left\Vert S_{t+1}\right\Vert _{F}\le\frac{B}{\lambda}\sum_{i=1}^{t}\frac{1}{i}e^{224r_{x}^{4}\left(\frac{1}{i+1}-\frac{1}{t}\right)}.\label{eq: a bound on the Frobenious norm of the iteration matrix}
\end{equation}
The sum is bounded as follows
\begin{align}
\sum_{i=1}^{t}\frac{1}{i}e^{\frac{224r_{x}^{4}}{i+1}} & \overset{\left(a\right)}{\le}\sum_{i=1}^{t}\frac{1}{i}\left[\frac{1}{i+1}\left(e^{224r_{x}^{4}}-1\right)+1\right]\\
 & =H_{t}+\left(e^{224r_{x}^{4}}-1\right)\sum_{i=1}^{t}\frac{1}{i\left(i+1\right)}\\
 & \le H_{t}+\left(e^{224r_{x}^{4}}-1\right)\sum_{i=1}^{\infty}\frac{1}{i^{2}}\\
 & \le2\ln\left(t\right)+2\left(e^{224r_{x}^{4}}-1\right)\\
 & \le2\ln\left(t\right)+2e^{224r_{x}^{4}},\label{eq:sum bound grad proof}
\end{align}
where $(a)$ is due to $\exp(x)\le\frac{x}{x_{0}}(\exp(x_{0})-1)+1$
for all $x\in(0,x_{0})$ and where $H_{t}\triangleq\sum_{i=1}^{t}\frac{1}{i}$
is the harmonic series. With this bound we conclude from (\ref{eq: a bound on the Frobenious norm of the iteration matrix})
that
\[
\left\Vert S_{t+1}\right\Vert _{F}\le\frac{B}{\lambda}\left(2\ln\left(t\right)+2e^{224r_{x}^{4}}\right)e^{-\frac{224r_{x}^{4}}{t}}.
\]
Finally, we bound the sub-gradient by
\begin{align}
\left\Vert \nabla_{t}\right\Vert  & \le\left\Vert \lambda\dsym\left(S_{t}D_{t}\right)-V_{t}\right\Vert _{F}\\
 & \le\lambda\left\Vert \dsym\left(\sum_{k=1}^{d+1}\eta_{k}S_{t}\delta_{j_{k}^{\left(t\right)}}\delta_{j_{k}^{\left(t\right)}}^{T}\right)\right\Vert _{F}+B\\
 & \le\lambda\sum_{k=1}^{d+1}\eta_{k}\left\Vert \dsym\left(S_{t}\delta_{j_{k}^{\left(t\right)}}\delta_{j_{k}^{\left(t\right)}}^{T}\right)\right\Vert _{F}+B\\
 & \overset{(a)}{\le}\lambda\sum_{k=1}^{d+1}2\eta_{k}\left\Vert S_{t}\delta_{j_{k}^{\left(t\right)}}\right\Vert \left\Vert \delta_{j_{k}^{\left(t\right)}}\right\Vert +B\\
 & \le2\lambda\cdot4r_{x}^{2}\cdot\frac{B}{\lambda}\left(2\ln\left(t\right)+2e^{224r_{x}^{4}}\right)e^{-\frac{224r_{x}^{4}}{t}}+B\\
 & \le16Br_{x}^{2}\left(\ln\left(t\right)+e^{224r_{x}^{4}}\right)+B\\
 & \le32Br_{x}^{2}\left(\ln\left(t\right)+e^{224r_{x}^{4}}\right),
\end{align}
where $(a)$ follows from Claim \ref{claim:dsym norm bound}. This
is the claimed bound.
\end{IEEEproof}
Next, we use the above lemma to bound the sub-gradient of additive
noise channel RLM (\ref{eq:stable RLM}).
\begin{cor}
\label{additive subgradient bound}The Frobenius norm of the sub-gradient
(\ref{eq:objective subgradient}) is bounded by
\[
\|\nabla_{t}\|_{F}\leq G\triangleq64r_{x}^{3}\sqrt{r_{x}^{2}+4r_{x}r_{z}+3r_{z}^{2}}\left(\ln\left(t\right)+e^{224r_{x}^{4}}\right).
\]
\end{cor}
\begin{IEEEproof}
We identify $V_{t}$ in the sub-gradient with
\[
V_{t}\triangleq\frac{1}{c}\sum_{i\in\boldsymbol{A}_{t}}\mathbb{I}\left[a_{i}^{T}S_{t}\delta_{a_{i}}<1\right]\dsym\left(a_{i}\delta_{a_{i}}^{T}\right),
\]
and bound its Frobenius norm by
\begin{align}
\left\Vert V_{t}\right\Vert _{F}^{2} & \le\max_{i\in\boldsymbol{A}_{t}}2\|a_{i}\|^{2}\|\delta_{a_{i}}\|^{2}+2\left\langle a_{i},\delta_{a_{i}}\right\rangle ^{2}\\
 & \le4\left(2r_{x}r_{z}\right)\left(2r_{x}^{2}\right)+2\left(2r_{x}r_{z}\right)^{2}+2r_{z}^{2}\cdot2r_{x}^{2}+4r_{x}^{4}\\
 & =16r_{x}^{3}r_{z}+12r_{x}^{2}r_{z}^{2}+4r_{x}^{4}\\
 & =4r_{x}^{2}\left(r_{x}^{2}+4r_{x}r_{z}+3r_{z}^{2}\right),
\end{align}
where the first inequality follows from Claim \ref{claim:dsym norm bound}
and the second from (\ref{eq:bound on a delta norm}). Note that for
a symmetric matrix$\|S\|_{F}=\sqrt{\sum_{i=1}^{d}\lambda_{i}^{2}}$,
therefore, the projection (\ref{eq:PSD projection}) satisfies for
condition from Lemma (\ref{lem:general grad bound}). Using this Lemma
concludes the proof.
\end{IEEEproof}
Now we turn to bound the Frobenius norm of the sub-gradient of (\ref{eq:stable RLM-1}).
In order to do so, we will need to establish some results first. The
following is a well known result on the projection on convex sets
in a Hilbert space (e.g., \cite[Prop. 2.2.1]{bertsekas2003convex}
for a proof in the Euclidean space). We provide here a short proof
for completeness.
\begin{prop}
\label{prop:pythagorean identity}Consider a Hilbert space ${\cal H}$
and a closed convex set $S$. Let $f\in{\cal H}$ be given, and let
$f_{0}$ be its projection on $S$ that is 
\[
f_{0}=\argmin_{h\in S}\|f-h\|^{2}.
\]
Let $h\in S$ be arbitrary. Then, the angle between $f-f_{0}$ and
$h-f_{0}$ is obtuse, that is 
\begin{equation}
\langle f-f_{0},h-f_{0}\rangle\leq0.\label{eq: obtuse angle}
\end{equation}
Thus, the following Pythagorean identity holds
\[
\|f-h\|^{2}\geq\|f-f_{0}\|^{2}+\|f_{0}-h\|^{2}.
\]
\end{prop}
\begin{IEEEproof}
By the definition of projection, for any $\alpha\in(0,1]$ it holds
that 
\begin{align}
0 & \leq\|f-(1-\alpha)f_{0}-\alpha h\|^{2}-\|f-f_{0}\|^{2}\\
 & =\|f-f_{0}+\alpha(f_{0}-h)\|^{2}-\|f-f_{0}\|^{2}\\
 & =2\alpha\langle f-f_{0},f_{0}-h\rangle+\alpha^{2}\|f_{0}-h\|^{2}.
\end{align}
Dividing by $\alpha$ we get 
\[
\langle f-f_{0},f_{0}-h\rangle+\alpha\|f_{0}-h\|^{2}\geq0
\]
and taking $\alpha\downarrow0$ implies the required claim. Arranging
the terms of the Pythagorean identity, it is seen to be equivalent
to (\ref{eq: obtuse angle}).
\end{IEEEproof}
Next, we show that the solution set of the non-linear channel RLM,
is indeed convex, and by this also prove Claim \ref{claim:projection claim}.
\begin{claim}
\label{claim:convex set claim}The set $\{(H,S):H^{T}H-S\preceq0\}$
is convex.
\end{claim}
\begin{IEEEproof}
Let $(H_{1},S_{1}),(H_{2},S_{2})\in\{(H,S):H^{T}H-S\preceq0\}$ and
$\lambda\in[0,1]$. Then, for any $x\in\mathbb{R}^{d_{x}}$ it holds
that 
\[
x^{T}\left(H_{1}^{T}H_{1}-S_{1}\right)x\le0,
\]
and
\[
x^{T}\left(H_{2}^{T}H_{2}-S_{2}\right)x\le0.
\]
Then,
\begin{align}
 & x^{T}\left[\left(\lambda H_{1}+\left(1-\lambda\right)H_{2}\right)^{T}\left(\lambda H_{1}+\left(1-\lambda\right)H_{2}\right)-\left(\lambda S_{1}+\left(1-\lambda\right)S_{2}\right)\right]x\nonumber \\
 & =x^{T}\left[\lambda^{2}H_{1}^{T}H_{1}+\lambda\left(1-\lambda\right)H_{1}^{T}H_{2}+\lambda\left(1-\lambda\right)H_{2}^{T}H_{1}+\left(1-\lambda\right)^{2}H_{2}^{T}H_{2}\right]x\nonumber \\
 & \phantom{=}-\lambda x^{T}S_{1}x-\left(1-\lambda\right)x^{T}S_{2}x\\
 & =\lambda x^{T}\left(\lambda H_{1}^{T}H_{1}-S_{1}\right)x+\left(1-\lambda\right)x^{T}\left(\left(1-\lambda\right)H_{2}^{T}H_{2}-S_{2}\right)x\nonumber \\
 & \phantom{=}+\lambda\left(1-\lambda\right)\left(H_{1}x\right)^{T}\left(H_{2}x\right)+\lambda\left(1-\lambda\right)\left(H_{2}x\right)^{T}\left(H_{1}x\right)\\
 & =\lambda x^{T}\left(H_{1}^{T}H_{1}-S_{1}\right)x-\lambda\left(1-\lambda\right)x^{T}H_{1}^{T}H_{1}x\nonumber \\
 & \phantom{=}+\left(1-\lambda\right)x^{T}\left(H_{2}^{T}H_{2}-S_{2}\right)x-\lambda\left(1-\lambda\right)x^{T}H_{2}^{T}H_{2}x\nonumber \\
 & \phantom{=}+2\lambda\left(1-\lambda\right)x^{T}H_{1}^{T}H_{2}x\\
 & \le2\lambda\left(1-\lambda\right)x^{T}H_{1}^{T}H_{2}x-\lambda\left(1-\lambda\right)x^{T}H_{1}^{T}H_{1}x-\lambda\left(1-\lambda\right)x^{T}H_{2}^{T}H_{2}x\\
 & =-\lambda\left(1-\lambda\right)\left\Vert H_{1}x-H_{2}x\right\Vert _{2}^{2}\\
 & \le0.
\end{align}
\end{IEEEproof}
Now we are ready to bound the sub-gradient with components (\ref{eq:subgrad H})
and (\ref{eq:subgrad K}).
\begin{lem}
\label{lem:non-linear subgradient bound}The Frobenius norm of the
sub-gradient with components (\ref{eq:subgrad H}) and (\ref{eq:subgrad K})
is bounded by
\[
\left\Vert \nabla_{t}\right\Vert _{F}=O\left(\max\left\{ 1,\lambda\right\} \cdot\ln\left(t\right)\right).
\]
\end{lem}
\begin{IEEEproof}
First,
\begin{align}
\left\Vert H_{t+1}\right\Vert _{F}^{2}+\left\Vert K_{t+1}\right\Vert _{F}^{2} & =\left\Vert \left(H_{t+1},K_{t+1}\right)\right\Vert ^{2}\\
 & =\left\Vert \Pi_{H^{T}H\preceq K}\left(H_{t}-\frac{1}{\lambda t}\nabla_{t}^{(1)},K_{t}+\frac{1}{\lambda t}\nabla_{t}^{(2)}\right)\right\Vert ^{2}\\
 & \le\left\Vert \left(H_{t}-\frac{1}{\lambda t}\nabla_{t}^{(1)},K_{t}+\frac{1}{\lambda t}\nabla_{t}^{(2)}\right)\right\Vert ^{2}\\
 & =\left\Vert H_{t}-\frac{1}{\lambda t}\nabla_{t}^{(1)}\right\Vert _{F}^{2}+\left\Vert K_{t}+\frac{1}{\lambda t}\nabla_{t}^{(2)}\right\Vert _{F}^{2},
\end{align}
where the first inequality is due to Prop. \ref{prop:pythagorean identity},
when taking into account that the set $\{(H,S):H^{T}H-S\preceq0\}$
is convex, by Claim \ref{claim:convex set claim}, and that the zero
point is in the set. Now, denote for readability the following:

\[
D_{t}^{(i)}\triangleq\sum_{k=1}^{d+1}\eta_{k}\delta_{j_{t,k}^{\left(i\right)}}\delta_{j_{t,k}^{\left(i\right)}}^{T},
\]
and
\[
V_{t}^{(1)}=\frac{1}{\left|\boldsymbol{A}_{t}\right|}\sum_{i\in\boldsymbol{A}_{t}}\frac{1}{m-1}\sum_{j'\in\left[m\right]\backslash\left\{ j_{i}\right\} }\mathbb{I}\left[y_{i}^{T}H\delta_{j_{i}j'}-\frac{1}{2}\left(x_{j_{i}}+x_{j'}\right)^{T}K\delta_{j_{i}j'}<1\right]y_{i}\delta_{j_{i}j'}^{T},
\]
as well as
\begin{multline}
V_{t}^{(2)}=\frac{1}{\left|\boldsymbol{A}_{t}\right|}\sum_{i\in\boldsymbol{A}_{t}}\frac{1}{m-1}\sum_{j'\in\left[m\right]\backslash\left\{ j_{i}\right\} }\mathbb{I}\left[y_{i}^{T}H_{t}\delta_{j_{i}j'}-\frac{1}{2}\left(x_{j_{i}}+x_{j'}\right)^{T}K_{t}\delta_{j_{i}j'}<1\right]\dsym\left(\frac{1}{2}\left(x_{j_{i}}+x_{j'}\right)\delta_{j_{i}j'}^{T}\right).
\end{multline}
With these notations the update rule is written as 
\[
H_{t}-\frac{1}{\lambda t}\nabla_{t}^{(1)}=H_{t}-\frac{2}{t}H_{t}D_{t}^{(1)}+\frac{1}{\lambda t}V_{t}^{(1)},
\]
\[
K_{t}-\frac{1}{\lambda t}\nabla_{t}^{(2)}=K_{t}-\frac{1}{t}\dsym\left(2K_{t}D_{t}^{(2)}\right)-\frac{1}{\lambda t}V_{t}^{(2)}.
\]
Consequently, using the triangle inequality,
\begin{align}
\left\Vert H_{t+1}\right\Vert _{F}^{2}+\left\Vert K_{t+1}\right\Vert _{F}^{2} & =\left\Vert H_{t}-\frac{2}{t}H_{t}D_{t}^{(1)}+\frac{1}{\lambda t}V_{t}^{(1)}\right\Vert _{F}^{2}+\left\Vert K_{t}-\frac{1}{t}\dsym\left(2K_{t}D_{t}^{(2)}\right)-\frac{1}{\lambda t}V_{t}^{(2)}\right\Vert _{F}^{2}\\
 & =\left\Vert H_{t}-\frac{2}{t}H_{t}D_{t}^{(1)}\right\Vert _{F}^{2}+\frac{2}{\lambda t}\left\langle H_{t}-\frac{2}{t}H_{t}D_{t}^{(1)},V_{t}^{(1)}\right\rangle +\frac{1}{\lambda^{2}t^{2}}\left\Vert V_{t}^{(1)}\right\Vert _{F}^{2}+\frac{1}{\lambda^{2}t^{2}}\left\Vert V_{t}^{(2)}\right\Vert _{F}^{2}\nonumber \\
 & \phantom{=}+\left\Vert K_{t}-\frac{2}{t}\dsym\left(K_{t}D_{t}^{(2)}\right)\right\Vert _{F}^{2}-\frac{2}{\lambda t}\left\langle K_{t}-\frac{2}{t}\dsym\left(K_{t}D_{t}^{(2)}\right),V_{t}^{(2)}\right\rangle \\
 & \le\left\Vert H_{t}-\frac{2}{t}H_{t}D_{t}^{(1)}\right\Vert _{F}^{2}+\left\Vert K_{t}-\frac{2}{t}\dsym\left(K_{t}D_{t}^{(2)}\right)\right\Vert _{F}^{2}\nonumber \\
 & \phantom{=}+\frac{1}{\lambda^{2}t^{2}}\left\Vert V_{t}^{(1)}\right\Vert _{F}^{2}+\frac{1}{\lambda^{2}t^{2}}\left\Vert V_{t}^{(2)}\right\Vert _{F}^{2}+\frac{2}{\lambda t}\left(\left\langle H_{t},V_{t}^{(1)}\right\rangle -\left\langle K_{t},V_{t}^{(2)}\right\rangle \right)\nonumber \\
 & \phantom{=}-\frac{4}{\lambda t^{2}}\left(\left\Vert H_{t}\right\Vert _{F}\left\Vert D_{t}^{(1)}\right\Vert _{F}\left\Vert V_{t}^{(1)}\right\Vert _{F}-\left\Vert \dsym\left(K_{t}D_{t}^{(2)}\right)\right\Vert _{F}\left\Vert V_{t}^{(2)}\right\Vert _{F}\right).
\end{align}
Now, regarding the fifth term
\begin{align}
\left\langle H_{t},V_{t}^{(1)}\right\rangle -\left\langle K_{t},V_{t}^{(2)}\right\rangle  & =\frac{1}{\left|\boldsymbol{A}_{t}\right|}\sum_{i\in\boldsymbol{A}_{t}}\frac{1}{m-1}\sum_{j'\in\left[m\right]\backslash\left\{ j_{i}\right\} }\mathbb{I}\left[y_{i}^{T}H\delta_{j_{i}j'}-\frac{1}{2}\left(x_{j_{i}}+x_{j'}\right)^{T}K\delta_{j_{i}j'}<1\right]\nonumber \\
 & \phantom{=}\times\left[\left\langle H_{t},y_{i}\delta_{j_{i}j'}^{T}\right\rangle -\left\langle K_{t},\frac{1}{4}\left(\left(x_{j_{i}}+x_{j'}\right)\delta_{j_{i}j'}^{T}+\delta_{j_{i}j'}\left(x_{j_{i}}+x_{j'}\right)^{T}\right)\right\rangle \right]\\
 & =\frac{1}{\left|\boldsymbol{A}_{t}\right|}\sum_{i\in\boldsymbol{A}_{t}}\frac{1}{m-1}\sum_{j'\in\left[m\right]\backslash\left\{ j_{i}\right\} }\mathbb{I}\left[y_{i}^{T}H\delta_{j_{i}j'}-\frac{1}{2}\left(x_{j_{i}}+x_{j'}\right)^{T}K\delta_{j_{i}j'}<1\right]\nonumber \\
 & \phantom{=}\times\left[y_{i}^{T}H_{t}\delta_{j_{i}j'}-\frac{1}{2}\left(x_{j_{i}}+x_{j'}\right)^{T}K_{t}\delta_{j_{i}j'}\right]\\
 & \le1.
\end{align}
Plugging this back gives
\begin{align}
\left\Vert H_{t+1}\right\Vert _{F}^{2}+\left\Vert K_{t+1}\right\Vert _{F}^{2} & \le\left\Vert H_{t}-\frac{2}{t}H_{t}D_{t}^{(1)}\right\Vert _{F}^{2}+\frac{1}{\lambda^{2}t^{2}}\left\Vert V_{t}^{(1)}\right\Vert _{F}^{2}+\left\Vert K_{t}-\frac{2}{t}\dsym\left(K_{t}D_{t}^{(2)}\right)\right\Vert _{F}^{2}+\frac{1}{\lambda^{2}t^{2}}\left\Vert V_{t}^{(2)}\right\Vert _{F}^{2}\nonumber \\
 & \phantom{=}+\frac{2}{\lambda t}-\frac{4}{\lambda t^{2}}\left(\left\Vert H_{t}\right\Vert _{F}\left\Vert D_{t}^{(1)}\right\Vert _{F}\left\Vert V_{t}^{(1)}\right\Vert _{F}-\left\Vert \dsym\left(K_{t}D_{t}^{(2)}\right)\right\Vert _{F}\left\Vert V_{t}^{(2)}\right\Vert _{F}\right).
\end{align}
Similarly to (\ref{eq:first grad bound}) from Lemma \ref{lem:general grad bound}
we get that
\[
\left\Vert H_{t}-\frac{2}{t}H_{t}D_{t}^{(1)}\right\Vert _{F}\le\left\Vert H_{t}\right\Vert _{F}\sqrt{1+\frac{448r_{x}^{4}}{t^{2}}},
\]
and
\[
\left\Vert K_{t}-\frac{2}{t}\dsym\left(K_{t}D_{t}^{(2)}\right)\right\Vert _{F}\le\left\Vert K_{t}\right\Vert _{F}\sqrt{1+\frac{448r_{x}^{4}}{t^{2}}}.
\]
Therefore, we can bound,
\begin{align}
\left\Vert H_{t+1}\right\Vert _{F}^{2}+\left\Vert K_{t+1}\right\Vert _{F}^{2} & \le\left(\left\Vert H_{t}\right\Vert _{F}^{2}+\left\Vert K_{t}\right\Vert _{F}^{2}\right)\left(1+\frac{448r_{x}^{4}}{t^{2}}\right)+\frac{\left(2r_{x}\left(R_{x}+r_{z}\right)\right)^{2}}{\lambda^{2}t^{2}}+\frac{5r_{x}^{4}}{\lambda^{2}t^{2}}\nonumber \\
 & \phantom{\le}+\frac{2}{\lambda t}+\frac{32r_{x}^{3}}{\lambda t^{2}}\left(2R_{x}+2r_{z}+\sqrt{5}r_{x}\right)\left(\left\Vert H_{t}\right\Vert _{F}+\left\Vert K_{t}\right\Vert _{F}\right)\\
 & \le\left(\left\Vert H_{t}\right\Vert _{F}^{2}+\left\Vert K_{t}\right\Vert _{F}^{2}\right)\left(1+\frac{448r_{x}^{4}}{t^{2}}+\frac{32r_{x}^{3}}{\lambda t^{2}}\left(2R_{x}+2r_{z}+\sqrt{5}r_{x}\right)\right)\nonumber \\
 & \phantom{\le}+\frac{\left(2r_{x}\left(R_{x}+r_{z}\right)\right)^{2}}{\lambda^{2}t^{2}}+\frac{5r_{x}^{4}}{\lambda^{2}t^{2}}+\frac{2}{\lambda t}+\frac{64r_{x}^{3}}{\lambda t^{2}}\left(2R_{x}+2r_{z}+\sqrt{5}r_{x}\right)\\
 & \le\left(\left\Vert H_{t}\right\Vert _{F}^{2}+\left\Vert K_{t}\right\Vert _{F}^{2}\right)\left(1+\frac{448r_{x}^{4}}{t^{2}}+\frac{32r_{x}^{3}}{\lambda t^{2}}\left(2R_{x}+2r_{z}+\sqrt{5}r_{x}\right)\right)\nonumber \\
 & \phantom{\le}+\frac{1}{t}\left[\frac{\left[4r_{x}^{2}\left(R_{x}+r_{z}\right)^{2}+5r_{x}^{4}\right]}{\lambda^{2}}+\frac{\left[64r_{x}^{3}\left(2R_{x}+2r_{z}+\sqrt{5}r_{x}\right)+2\right]}{\lambda}\right].
\end{align}
Denote for readability,
\[
C_{1}\triangleq\frac{\left[4r_{x}^{2}\left(R_{x}+r_{z}\right)^{2}+5r_{x}^{4}\right]}{\lambda^{2}}+\frac{\left[64r_{x}^{3}\left(2R_{x}+2r_{z}+\sqrt{5}r_{x}\right)+2\right]}{\lambda}
\]
and
\[
C_{2}\triangleq448r_{x}^{4}+\frac{32r_{x}^{3}}{\lambda}\left(2R_{x}+2r_{z}+\sqrt{5}r_{x}\right).
\]
Applying this bound recursively we get
\[
\left\Vert H_{t+1}\right\Vert _{F}^{2}+\left\Vert K_{t+1}\right\Vert _{F}^{2}\le C_{1}\sum_{i=1}^{t}\frac{1}{i}\prod_{j=i+1}^{t}\left(1+\frac{C_{2}}{j^{2}}\right).
\]
We bound this product, similarly to (\ref{eq:prod bound}) from the
proof of Lemma \ref{lem:general grad bound}, by
\[
\prod_{j=i+1}^{t}\left(1+\frac{C_{2}}{j^{2}}\right)\le e^{\frac{C_{2}}{i+1}}.
\]
Then, we remain with
\[
\left\Vert H_{t+1}\right\Vert _{F}^{2}+\left\Vert K_{t+1}\right\Vert _{F}^{2}\le C_{1}\sum_{i=1}^{t}\frac{1}{i}e^{\frac{C_{2}}{i+1}}.
\]
The sum is bounded, similarly to (\ref{eq:sum bound grad proof})
from the proof of Lemma \ref{lem:general grad bound}, by
\[
\sum_{i=1}^{t}\frac{1}{i}e^{\frac{C_{2}}{i+1}}\le2\ln\left(t\right)+2e^{C_{2}}.
\]
With this we conclude that
\[
\left\Vert H_{t+1}\right\Vert _{F}^{2}+\left\Vert K_{t+1}\right\Vert _{F}^{2}\le2C_{1}\left(\ln\left(t\right)+e^{C_{2}}\right).
\]
Finally, we bound the sub-gradient by
\begin{align}
\left\Vert \nabla_{t}\right\Vert ^{2} & \le\left\Vert \nabla_{t}^{(1)}\right\Vert ^{2}+\left\Vert \nabla_{t}^{(1)}\right\Vert ^{2}\\
 & =\left\Vert 2\lambda H_{t}D_{t}^{(1)}-V_{t}^{(1)}\right\Vert _{F}^{2}+\left\Vert \lambda\dsym\left(2K_{t}D_{t}^{(2)}\right)+V_{t}^{(2)}\right\Vert _{F}^{2}\\
 & \le4\lambda^{2}\left\Vert H_{t}\right\Vert _{F}^{2}\left\Vert D_{t}^{(1)}\right\Vert _{F}^{2}+4\lambda\left\Vert H_{t}\right\Vert _{F}\left\Vert D_{t}^{(1)}\right\Vert _{F}\left\Vert V_{t}^{(1)}\right\Vert _{F}+\left\Vert V_{t}^{(1)}\right\Vert _{F}^{2}\nonumber \\
 & \phantom{\le}+4\lambda^{2}\left\Vert \dsym\left(K_{t}D_{t}^{(2)}\right)\right\Vert _{F}^{2}+4\lambda\left\Vert \dsym\left(K_{t}D_{t}^{(2)}\right)\right\Vert _{F}\left\Vert V_{t}^{(2)}\right\Vert _{F}+\left\Vert V_{t}^{(2)}\right\Vert _{F}^{2}\\
 & \le4\lambda^{2}\cdot16r_{x}^{4}\left\Vert H_{t}\right\Vert _{F}^{2}+4\lambda\cdot4r_{x}^{2}\cdot2r_{x}\left(R_{x}+r_{z}\right)\left\Vert H_{t}\right\Vert _{F}+4r_{x}^{2}\left(R_{x}+r_{z}\right)^{2}\nonumber \\
 & \phantom{\le}+4\lambda^{2}\cdot64r_{x}^{4}\left\Vert K_{t}\right\Vert _{F}^{2}+4\lambda\cdot8r_{x}^{2}\cdot\sqrt{5}r_{x}^{2}\left\Vert K_{t}\right\Vert _{F}+5r_{x}^{4}\\
 & \le320\lambda^{2}r_{x}^{4}\left(\left\Vert H_{t}\right\Vert _{F}^{2}+\left\Vert K_{t}\right\Vert _{F}^{2}\right)+32\lambda r_{x}^{3}\left(R_{x}+r_{z}+\sqrt{5}r_{x}\right)\left(\left\Vert H_{t}\right\Vert _{F}+\left\Vert K_{t}\right\Vert _{F}\right)+4r_{x}^{2}\left(R_{x}+r_{z}\right)^{2}+5r_{x}^{4},
\end{align}
which leads to the claimed bound.
\end{IEEEproof}
Finally, we prove Theorem \ref{thm: optimization bound}.
\begin{IEEEproof}[Proof of Th. \ref{thm: optimization bound}]
First, note that the PSD cone is convex, as well as $\{(H,S):H^{T}H-S\preceq0\}$,
by Claim \ref{claim:convex set claim}. Denote by $\Gamma_{t}$ the
hypothesis from round $t$ of Algorithm \ref{subgradient alg}, i.e.,
$S_{t}$ for the additive noise channel and $(H_{t},K_{t})$ for the
non-linear channel. Now, by using \cite[Lemma 1]{pegasos} we get
that for every $\Gamma\in\boldsymbol{\Gamma}$
\[
\text{Reg}_{T}\triangleq\sum_{t=1}^{T}f\left(\Gamma_{t};\boldsymbol{A}_{t}\right)-\sum_{t=1}^{T}f\left(\Gamma;\boldsymbol{A}_{t}\right)\le\frac{G^{2}\left(1+\ln T\right)}{4\lambda\gamma},
\]
where $\gamma$ is the strong convexity constant either from Lemma
\ref{strong convexity lemma} or from Lemma \ref{strong convexity lemma-1},
respectively, $\lambda$ is the regularization parameter, and $G$
is the bound for the Frobenius norm of the sub-gradient from either
Corollary \ref{additive subgradient bound} or Lemma \ref{lem:non-linear subgradient bound},
respectively. Combining this with \cite[Th. 2]{kakade2009generalization},
we get, with probability of at least $1-4\delta\ln(T)$, that
\begin{align}
f\left(\bar{\Gamma}\right)-f\left(\Gamma^{*}\right) & \le\frac{\text{Reg}_{T}}{T}+4\sqrt{\frac{G^{2}\ln\left(1/\delta\right)}{2\lambda\gamma}}\frac{\sqrt{\text{Reg}_{T}}}{T}+\max\left\{ \frac{16G^{2}}{2\lambda\gamma},6B\right\} \frac{\ln\left(1/\delta\right)}{T}\\
 & \le\frac{G^{2}\ln T}{2\lambda\gamma T}+\frac{4G^{2}\sqrt{\ln T}}{\lambda\gamma T}\sqrt{\ln\left(\frac{1}{\delta}\right)}+\max\left\{ \frac{16G^{2}}{2\lambda\gamma},6B\right\} \frac{\ln\left(1/\delta\right)}{T},
\end{align}
where $\bar{\Gamma}\triangleq\frac{1}{T}\sum_{t=1}^{T}\Gamma_{t}$.
The objective of (\ref{eq:stable RLM}) is bounded by
\begin{align}
B & =\max_{S\in\mathbb{S}_{+}}f\left(S\right)\\
 & \le1+\max_{S\in\mathbb{S}_{+}}\max_{1\le p<q\le m}\max_{i\in\boldsymbol{D}_{p}\bigcup\boldsymbol{D}_{q}}\left|a_{pqi}^{T}S\delta_{pq}\right|+\lambda\max_{S\in\mathbb{S}_{+}}\max_{1\le p<q\le m}\left\Vert S\delta_{pq}\right\Vert _{F}^{2}\\
 & \le1+2r_{x}\left\Vert S\right\Vert _{F}\left(r_{x}+r_{z}\right)+4\lambda r_{x}^{2}\left\Vert S\right\Vert _{F}^{2}\\
 & =O\left(\frac{1}{\lambda}\ln^{2}\left(T\right)\right),
\end{align}
and the objective of (\ref{eq:stable RLM-1}) is bounded by
\begin{align}
B & =\max_{\left(H,K\right)\in\mathcal{H\times K}}f\left(H,K\right)\\
 & \le1+\max_{i\in\left[n\right]}\max_{j'\in\left[m\right]\backslash\left\{ j_{i}\right\} }\left|y_{i}^{T}H\delta_{j_{i}j'}-\frac{1}{2}\left(x_{j_{i}}+x_{j'}\right)^{T}K\delta_{j_{i}j'}\right|\nonumber \\
 & \phantom{\le}+\lambda\left(\max_{H\in\mathcal{H}}\max_{1\le p<q\le m}\left\Vert H\delta_{pq}\right\Vert _{F}^{2}+\max_{K\in\mathcal{K}}\max_{1\le p<q\le m}\left\Vert K\delta_{pq}\right\Vert _{F}^{2}\right)\\
 & \le1+2r_{x}\left(R_{x}+r_{z}\right)\left\Vert H\right\Vert _{F}+r_{x}^{2}\left\Vert K\right\Vert _{F}+4\lambda r_{x}^{2}\left(\left\Vert H\right\Vert _{F}^{2}+\left\Vert K\right\Vert _{F}^{2}\right)\\
 & =O\left(\max\left\{ 1,\frac{1}{\lambda}\right\} \cdot\ln\left(T\right)\right).
\end{align}
Plugging $G=O(\ln(T))$ from Corollary \ref{additive subgradient bound}
or $G=O(\max\{1,\lambda\}\cdot\ln(T))$ from Lemma \ref{lem:non-linear subgradient bound}
we conclude that
\[
f\left(\bar{\Gamma}\right)-f\left(\Gamma^{*}\right)=O\left(\frac{\ln^{3}\left(T\right)\ln\left(1/\delta\right)}{\lambda T}\right).
\]
At least half of the hypothesis satisfy the previous bound (as argued
in \cite[Lemma 3]{pegasos}). We conclude that the previous result
holds for $S_{t}$, with probability of at least $\frac{1-4\delta\ln\left(T\right)}{2}$.
\end{IEEEproof}
\bibliographystyle{ieeetr}
\bibliography{Decoding_SVM}

\begin{thebibliography}{10}

\bibitem{o2017introduction}
T.~O'Shea and J.~Hoydis, ``An introduction to deep learning for the physical
  layer,'' {\em IEEE Transactions on Cognitive Communications and Networking},
  vol.~3, no.~4, pp.~563--575, 2017.

\bibitem{ye2019deep}
H.~Ye, L.~Liang, G.~Y. Li, and B.-H.~F. Juang, ``Deep learning based end-to-end
  wireless communication systems with conditional {GAN} as unknown channel,''
  {\em arXiv preprint arXiv:1903.02551}, 2019.

\bibitem{nachmani2018deep}
E.~Nachmani, E.~Marciano, L.~Lugosch, W.~J. Gross, D.~Burshtein, and Y.~Be'ery,
  ``Deep learning methods for improved decoding of linear codes,'' {\em IEEE
  Journal of Selected Topics in Signal Processing}, vol.~12, no.~1,
  pp.~119--131, 2018.

\bibitem{sahai2019learning}
A.~Sahai, J.~Sanz, V.~Subramanian, C.~Tran, and K.~Vodrahalli, ``Learning to
  communicate in a noisy environment,'' {\em arXiv preprint arXiv:1910.09630},
  2019.

\bibitem{park2019meta}
S.~Park, O.~Simeone, and J.~Kang, ``Meta-learning to communicate: {F}ast
  end-to-end training for fading channels,'' {\em arXiv preprint
  arXiv:1910.09945}, 2019.

\bibitem{shlezinger2019viterbinet}
N.~Shlezinger, Y.~C. Eldar, N.~Farsad, and A.~J. Goldsmith, ``Viterbi{N}et:
  {S}ymbol detection using a deep learning based {V}iterbi algorithm,'' in {\em
  2019 IEEE 20th International Workshop on Signal Processing Advances in
  Wireless Communications (SPAWC)}, pp.~1--5, IEEE, 2019.

\bibitem{liao2021doubly}
S.~Liao, C.~Deng, M.~Yin, and B.~Yuan, ``Doubly residual neural decoder:
  Towards low-complexity high-performance channel decoding,'' {\em arXiv
  preprint arXiv:2102.03959}, 2021.

\bibitem{gruber2017deep}
T.~Gruber, S.~Cammerer, J.~Hoydis, and S.~ten Brink, ``On deep learning-based
  channel decoding,'' in {\em 2017 51st Annual Conference on Information
  Sciences and Systems (CISS)}, pp.~1--6, IEEE, 2017.

\bibitem{shlezinger2020data}
N.~Shlezinger, N.~Farsad, Y.~C. Eldar, and A.~J. Goldsmith, ``Data-driven
  factor graphs for deep symbol detection,'' in {\em 2020 IEEE International
  Symposium on Information Theory (ISIT)}, pp.~2682--2687, 2020.

\bibitem{watanabe2021deep}
T.~Watanabe, T.~Ohseki, and K.~Yamazaki, ``Deep learning-based bit reliability
  based decoding for non-binary {LDPC} codes,'' in {\em 2021 IEEE International
  Symposium on Information Theory (ISIT)}, pp.~1451--1456, 2021.

\bibitem{lian2019learned}
M.~Lian, F.~Carpi, C.~H{\"a}ger, and H.~D. Pfister, ``Learned
  belief-propagation decoding with simple scaling and {SNR} adaptation,'' in
  {\em 2019 IEEE International Symposium on Information Theory (ISIT)},
  pp.~161--165, 2019.

\bibitem{askri2019dnn}
A.~Askri and G.~R.-B. Othman, ``{DNN} assisted sphere decoder,'' in {\em 2019
  IEEE International Symposium on Information Theory (ISIT)}, pp.~1172--1176,
  2019.

\bibitem{wadayama2019deep}
T.~Wadayama and S.~Takabe, ``Deep learning-aided trainable projected gradient
  decoding for {LDPC} codes,'' in {\em 2019 IEEE International Symposium on
  Information Theory (ISIT)}, pp.~2444--2448, 2019.

\bibitem{jiang2019mind}
Y.~Jiang, H.~Kim, H.~Asnani, and S.~Kannan, ``Mind: Model independent neural
  decoder,'' in {\em 2019 IEEE 20th International Workshop on Signal Processing
  Advances in Wireless Communications ({SPAWC})}, pp.~1--5, IEEE, 2019.

\bibitem{kim2018communication}
H.~Kim, Y.~Jiang, R.~Rana, S.~Kannan, S.~Oh, and P.~Viswanath, ``Communication
  algorithms via deep learning,'' {\em arXiv preprint arXiv:1805.09317}, 2018.

\bibitem{berner2021modern}
J.~Berner, P.~Grohs, G.~Kutyniok, and P.~Petersen, ``The modern mathematics of
  deep learning,'' {\em arXiv preprint arXiv:2105.04026}, 2021.

\bibitem{weinberger2021generalization}
N.~Weinberger, ``Generalization bounds and algorithms for learning to
  communicate over additive noise channels,'' {\em IEEE Transactions on
  Information Theory}, 2021.

\bibitem{marzetta2015massive}
T.~L. Marzetta, ``Massive {MIMO}: {A}n introduction,'' {\em Bell Labs Technical
  Journal}, vol.~20, pp.~11--22, 2015.

\bibitem{sybis2016channel}
M.~Sybis, K.~Wesolowski, K.~Jayasinghe, V.~Venkatasubramanian, and
  V.~Vukadinovic, ``Channel coding for ultra-reliable low-latency communication
  in {5G} systems,'' in {\em 2016 IEEE 84th vehicular technology conference
  (VTC-Fall)}, pp.~1--5, IEEE, 2016.

\bibitem{barry2012digital}
J.~R. Barry, E.~A. Lee, and D.~G. Messerschmitt, {\em Digital communication}.
\newblock Springer Science \& Business Media, 2012.

\bibitem{shalev2014understanding}
S.~Shalev-Shwartz and S.~Ben-David, {\em Understanding machine learning: {F}rom
  theory to algorithms}.
\newblock Cambridge university press, 2014.

\bibitem{viterbi2013principles}
A.~J. Viterbi and J.~K. Omura, {\em Principles of digital communication and
  coding}.
\newblock Courier Corporation, 2013.

\bibitem{crammer2002learnability}
K.~Crammer and Y.~Singer, ``On the learnability and design of output codes for
  multiclass problems,'' {\em Machine learning}, vol.~47, no.~2, pp.~201--233,
  2002.

\bibitem{bottou2007support}
L.~Bottou and C.-J. Lin, ``Support vector machine solvers,'' {\em Large scale
  kernel machines}, vol.~3, no.~1, pp.~301--320, 2007.

\bibitem{pegasos}
S.~Shalev-Shwartz, Y.~Singer, and N.~Srebro, ``Pegasos: Primal estimated
  sub-{G}radient {SO}lver for {SVM},'' 2007.
\newblock A fast online algorithm for solving the linear svm in primal using
  sub-gradients.

\bibitem{Goodfellow-et-al-2016}
I.~Goodfellow, Y.~Bengio, and A.~Courville, {\em Deep Learning}.
\newblock MIT Press, 2016.

\bibitem{wang2017deep}
T.~Wang, C.-K. Wen, H.~Wang, F.~Gao, T.~Jiang, and S.~Jin, ``Deep learning for
  wireless physical layer: Opportunities and challenges,'' {\em China
  Communications}, vol.~14, no.~11, pp.~92--111, 2017.

\bibitem{caciularu2020unsupervised}
A.~Caciularu and D.~Burshtein, ``Unsupervised linear and nonlinear channel
  equalization and decoding using variational autoencoders,'' {\em IEEE
  Transactions on Cognitive Communications and Networking}, vol.~6, no.~3,
  pp.~1003--1018, 2020.

\bibitem{li2021knowledge}
D.~Li, Y.~Xu, M.~Zhao, S.~Zhang, and J.~Zhu, ``Knowledge-driven machine
  learning-based channel estimation in massive {MIMO} system,'' in {\em 2021
  IEEE Wireless Communications and Networking Conference Workshops (WCNCW)},
  pp.~1--6, IEEE, 2021.

\bibitem{ma2021model}
X.~Ma, Z.~Gao, F.~Gao, and M.~Di~Renzo, ``Model-driven deep learning based
  channel estimation and feedback for millimeter-wave massive hybrid {MIMO}
  systems,'' {\em IEEE Journal on Selected Areas in Communications}, 2021.

\bibitem{zhao2021deep}
S.~Zhao, Y.~Fang, and L.~Qiu, ``Deep learning-based channel estimation with
  {SRGAN} in {OFDM} systems,'' in {\em 2021 IEEE Wireless Communications and
  Networking Conference (WCNC)}, pp.~1--6, IEEE, 2021.

\bibitem{carpi2019reinforcement}
F.~Carpi, C.~H{\"a}ger, M.~Martal{\`o}, R.~Raheli, and H.~D. Pfister,
  ``Reinforcement learning for channel coding: Learned bit-flipping decoding,''
  in {\em 2019 57th Annual Allerton Conference on Communication, Control, and
  Computing (Allerton)}, pp.~922--929, IEEE, 2019.

\bibitem{akin2020joint}
S.~Ak{\i}n, M.~Penner, and J.~Peissig, ``Joint channel estimation and data
  decoding using {SVM}-based receivers,'' {\em arXiv preprint
  arXiv:2012.02523}, 2020.

\bibitem{charrada2017analyzing}
A.~Charrada and A.~Samet, ``Analyzing performance of joint {SVR} interpolation
  for {LTE} system with 64-{QAM} modulation under 500 {K}m/h mobile velocity,''
  in {\em 2017 Sixth International Conference on Communications and Networking
  (ComNet)}, pp.~1--6, IEEE, 2017.

\bibitem{sanchez2004svm}
M.~S{\'a}nchez-Fern{\'a}ndez, M.~de~Prado-Cumplido, J.~Arenas-Garc{\'\i}a, and
  F.~P{\'e}rez-Cruz, ``{SVM} multiregression for nonlinear channel estimation
  in multiple-input multiple-output systems,'' {\em IEEE transactions on signal
  processing}, vol.~52, no.~8, pp.~2298--2307, 2004.

\bibitem{lee2019learning}
J.~Lee and M.~Raginsky, ``Learning finite-dimensional coding schemes with
  nonlinear reconstruction maps,'' {\em SIAM Journal on Mathematics of Data
  Science}, vol.~1, no.~3, pp.~617--642, 2019.

\bibitem{levrard2013fast}
C.~Levrard, ``Fast rates for empirical vector quantization,'' {\em Electronic
  Journal of Statistics}, vol.~7, pp.~1716--1746, 2013.

\bibitem{antos2005improved}
A.~Antos, ``Improved minimax bounds on the test and training distortion of
  empirically designed vector quantizers,'' {\em IEEE Transactions on
  Information Theory}, vol.~51, no.~11, pp.~4022--4032, 2005.

\bibitem{antos2005individual}
A.~Antos, L.~Gyorfi, and A.~Gyorgy, ``Individual convergence rates in empirical
  vector quantizer design,'' {\em IEEE Transactions on Information Theory},
  vol.~51, no.~11, pp.~4013--4022, 2005.

\bibitem{linder2000training}
T.~Linder, ``On the training distortion of vector quantizers,'' {\em IEEE
  Transactions on Information Theory}, vol.~46, no.~4, pp.~1617--1623, 2000.

\bibitem{chen1990adaptive}
S.~Chen, G.~Gibson, C.~Cowan, and P.~Grant, ``Adaptive equalization of finite
  non-linear channels using multilayer perceptrons,'' {\em Signal processing},
  vol.~20, no.~2, pp.~107--119, 1990.

\bibitem{kaleh1994joint}
G.~K. Kaleh and R.~Vallet, ``Joint parameter estimation and symbol detection
  for linear or nonlinear unknown channels,'' {\em IEEE Transactions on
  Communications}, vol.~42, no.~7, pp.~2406--2413, 1994.

\bibitem{xenoulis2012new}
K.~Xenoulis, N.~Kalouptsidis, and I.~Sason, ``New achievable rates for
  nonlinear {V}olterra channels via martingale inequalities,'' in {\em 2012
  IEEE International Symposium on Information Theory Proceedings},
  pp.~1425--1429, 2012.

\bibitem{roberts2016convex}
I.~Roberts, J.~M. Kahn, and D.~Boertjes, ``Convex channel power optimization in
  nonlinear {WDM} systems using gaussian noise model,'' {\em Journal of
  Lightwave Technology}, vol.~34, no.~13, pp.~3212--3222, 2016.

\bibitem{essiambre2010capacity}
R.-J. Essiambre, G.~Kramer, P.~J. Winzer, G.~J. Foschini, and B.~Goebel,
  ``Capacity limits of optical fiber networks,'' {\em Journal of Lightwave
  Technology}, vol.~28, no.~4, pp.~662--701, 2010.

\bibitem{mecozzi2012nonlinear}
A.~Mecozzi and R.-J. Essiambre, ``Nonlinear {S}hannon limit in pseudolinear
  coherent systems,'' {\em Journal of Lightwave Technology}, vol.~30, no.~12,
  pp.~2011--2024, 2012.

\bibitem{mitra2001nonlinear}
P.~P. Mitra and J.~B. Stark, ``Nonlinear limits to the information capacity of
  optical fibre communications,'' {\em Nature}, vol.~411, no.~6841,
  pp.~1027--1030, 2001.

\bibitem{dar2014shaping}
R.~Dar, M.~Feder, A.~Mecozzi, and M.~Shtaif, ``On shaping gain in the nonlinear
  fiber-optic channel,'' in {\em 2014 IEEE International Symposium on
  Information Theory}, pp.~2794--2798, IEEE, 2014.

\bibitem{shalev2008svm}
S.~Shalev-Shwartz and N.~Srebro, ``{SVM} optimization: {I}nverse dependence on
  training set size,'' in {\em Proceedings of the 25th international conference
  on Machine learning}, pp.~928--935, 2008.

\bibitem{mcculloch1982symmetric}
C.~E. McCulloch, ``Symmetric matrix derivatives with applications,'' {\em
  Journal of the American Statistical Association}, vol.~77, no.~379,
  pp.~679--682, 1982.

\bibitem{Boyd}
S.~P. Boyd and L.~Vandenberghe, {\em Convex Optimization}.
\newblock Cambridge university press, 2004.

\bibitem{batshon2008proposal}
H.~G. Batshon, I.~B. Djordjevic, L.~L. Minkov, L.~Xu, T.~Wang, and M.~Cvijetic,
  ``Proposal to achieve 1 {T}b/s per wavelength transmission using
  three-dimensional {LDPC}-coded modulation,'' {\em IEEE Photonics Technology
  Letters}, vol.~20, no.~9, pp.~721--723, 2008.

\bibitem{gronlund2020near}
A.~Gr{\o}nlund, L.~Kamma, and K.~G. Larsen, ``Near-tight margin-based
  generalization bounds for support vector machines,'' in {\em International
  Conference on Machine Learning}, pp.~3779--3788, PMLR, 2020.

\bibitem{guo2021recent}
Y.~Guo and C.~Zhang, ``Recent advances in large margin learning,'' {\em IEEE
  Transactions on Pattern Analysis and Machine Intelligence}, 2021.

\bibitem{rigollet2015high}
P.~Rigollet and J.-C. H{\"u}tter, ``High dimensional statistics,'' tech. rep.,
  Massachusetts Institute of Technology, 2019.

\bibitem{vershynin2018high}
R.~Vershynin, {\em High-dimensional probability: {A}n introduction with
  applications in data science}, vol.~47.
\newblock Cambridge University Press, 2018.

\bibitem{rakhlin2012statistical}
A.~Rakhlin and K.~Sridharan, ``Statistical learning theory and sequential
  prediction,'' {\em Lecture Notes in University of Pennsyvania}, 2012.

\bibitem{bartlett2002rademacher}
P.~L. Bartlett and S.~Mendelson, ``{R}ademacher and {G}aussian complexities:
  {R}isk bounds and structural results,'' {\em Journal of Machine Learning
  Research}, vol.~3, no.~Nov, pp.~463--482, 2002.

\bibitem{bertsekas2003convex}
D.~Bertsekas, A.~Nedic, and A.~Ozdaglar, {\em Convex analysis and
  optimization}, vol.~1.
\newblock Athena Scientific, 2003.

\bibitem{kakade2009generalization}
S.~M. Kakade and A.~Tewari, ``On the generalization ability of online strongly
  convex programming algorithms,'' in {\em Advances in Neural Information
  Processing Systems}, pp.~801--808, 2009.

\end{thebibliography}

\end{document}